\documentclass[%
 twocolumn,superscriptaddress,aps,prx,amssymb,amsfonts,amsmath,floatfix
longbibliography,
]{revtex4-2}

\usepackage[utf8]{inputenc}

\usepackage{amsmath}
\usepackage{amssymb}
\usepackage{xcolor}
\usepackage{graphicx}
\usepackage{comment}
\usepackage{algorithm}
\usepackage[noend]{algpseudocode}
\usepackage[normalem]{ulem}
\usepackage{here}

\makeatletter
\def\BState{\State\hskip-\ALG@thistlm}
\makeatother

\begin{document}

\title{Wavelet Conditional Renormalization Group}

\author{Tanguy Marchand}
\affiliation{D\'partement d'Informatique de l'Ecole Normale Sup\'erieure, ENS,CNRS, Universit\'e PSL}

\author{Misaki Ozawa}
\affiliation{Laboratoire de Physique de l'Ecole Normale Sup\'erieure, ENS, Universit\'e PSL, CNRS, Sorbonne Universit\'e, Universit\'e Paris-Diderot, Sorbonne Paris Cit\'e, Paris, France}

\author{Giulio Biroli}
\affiliation{Laboratoire de Physique de l'Ecole Normale Sup\'erieure, ENS, Universit\'e PSL, CNRS, Sorbonne Universit\'e, Universit\'e Paris-Diderot, Sorbonne Paris Cit\'e, Paris, France}

\author{Stéphane Mallat}
\affiliation{Coll\`ege de France, DI ENS,CNRS, Universit\'e PSL, CCM Flatiron Institute, New York}




\date{\today}

\newtheorem{theorem}{Theorem}[section]
\newtheorem{lemma}{Lemma}[section]
\newtheorem{definition}{Definition}[section]
\newtheorem{proposition}{Proposition}[section]
\newtheorem{corollary}{Corollary}[section]

\newcommand{\Cov} {{\rm Cov}}
\newcommand{\R} {{\mathbb R}}
\newcommand{\Z} {{\mathbb Z}}
\newcommand{\la} {{\lambda}}
\newcommand{\ttt} {{t}}
\newcommand{\HH} {{E}}
\newcommand{\wHH} {\widetilde {E}}
\newcommand{\bHH} {\overline {E}}
\newcommand{\wK} {\widetilde {\theta}}
\newcommand{\wbK} {\widetilde {\bf \theta}}
\newcommand{\oHH} {{E}}
\newcommand{\bFF} {\overline {F}}
\newcommand{\FF} {{F}}
\newcommand{\wFF} {\widetilde {F}}
\newcommand{\wU} {\widetilde {U}}
\newcommand{\ZZ} {{Z}}
\newcommand{\wG} {{\overline G}}
\newcommand{\G} {{\overline G}}
\newcommand{\NN} {{\cal N}}
\newcommand{\NGG} {{\cal N G}}
\newcommand{\C} {{\bf C}}
\newcommand{\K} {{\theta}}
\newcommand{\bK} {\overline {\theta}}
\newcommand{\bbK} {\overline {K}}
\newcommand{\LLL} {{\bf G}}
\newcommand{\LL} {{G}}
\newcommand{\LD} {{\bf L}[0,1]^d}
\newcommand{\V} {{\bf V}}
\newcommand{\VV} {{V}}
\newcommand{\DKL} {{D_{\rm KL}}}
\newcommand{\dphi} {{\overline \varphi}}
\newcommand{\cphi} {{f}}
\newcommand{\aphi} {{\varphi}}
\newcommand{\ophi} {{\overline \varphi}}
\newcommand{\oK} {{\Delta {\bf K}}}
\newcommand{\U} {{U}}
\newcommand{\bU} {\overline {U}}
\newcommand{\lB} {{\big\langle}}
\newcommand{\rB} {{\big\rangle}}
\newcommand{\lb} {{\langle}}
\newcommand{\rb} {{\rangle}}
\newcommand{\ReLU} {{\rm ReLU}}
\newcommand{\E} {{\mathbb E}}
\newcommand{\N} {{\mathbb N}}
\newcommand{\W} {{\bf W}}
\newcommand{\WW} {{\cal W}}
\newcommand{\om} {{k}}
\newcommand{\sm}[1]{\textcolor{blue}{#1}}
\newcommand{\tm}[1]{\textcolor{green}{#1}}
\newcommand{\mo}[1]{\textcolor{magenta}{#1}}
\newcommand{\gb}[1]{\textcolor{red}{#1}}
\renewcommand{\algorithmicrequire}{\textbf{Input:}}
\renewcommand{\algorithmicensure}{\textbf{Output:}}

\begin{abstract}
We develop a multiscale approach to estimate high-dimensional probability distributions from a dataset of physical fields or configurations observed in experiments or simulations. In this way we can estimate energy functions (or Hamiltonians) and efficiently generate new samples of many-body systems in various domains, from statistical physics to cosmology. 
Our method -- the Wavelet Conditional Renormalization Group (WC-RG) -- proceeds scale by scale, estimating models for the conditional probabilities of "fast degrees of freedom" conditioned by coarse-grained fields. These probability distributions are modeled by energy functions associated with scale interactions, and are represented 
in an orthogonal wavelet basis.
WC-RG decomposes the microscopic energy function as a sum of interaction energies at
all scales and can efficiently generate new samples by going from coarse to fine scales.
 Near phase transitions, it avoids the "critical slowing down" of direct estimation and sampling algorithms.
This is explained theoretically by combining results from RG and wavelet theories, and verified numerically for the Gaussian and $\varphi^4$ field theories. 
We show that multiscale WC-RG energy-based models are more general than local potential models and can capture the physics of complex many-body interacting systems at all length scales. This is demonstrated for weak-gravitational-lensing fields reflecting dark matter distributions in cosmology, which include long-range interactions with long-tail probability distributions. WC-RG has a large number of potential applications in non-equilibrium systems, where the underlying distribution is not known {\it a priori}. Finally, we discuss the connection between WC-RG and deep network architectures.
\\
\end{abstract}

\maketitle

\section{\label{sec:introduction}Introduction}

For a long time, physicists have determined the energy function (or Hamiltonian) for a given problem by following a bottom-up approach starting from physical principles. Nowadays, thanks to a large amount of data available,  it has become possible to estimate the energy function statistically, i.e., by inferring it from a large dataset of observations from experiments or simulations. 
The majority of physical systems are characterized by a probability distribution.
At equilibrium, this distribution has the Boltzmann-Gibbs form. Estimating such probability distributions and generating new samples efficently is now a major endeavor at the center of intense research activity. For instance,  generative models have been used in cosmology to constrain the values of the cosmological constants \cite{perraudin2019cosmological}, and in condensed matter theory and statistical physics to characterize phase transitions \cite{carrasquilla2017machine}. This problem has important connections with the topic of sampling in scientific computing, which is witnessing a surge of activity due to progress in machine learning, see, e.g., Refs.~\cite{noe2019boltzmann,gabrie2021efficient}.

\begin{figure}[h]
\includegraphics[width=\linewidth]{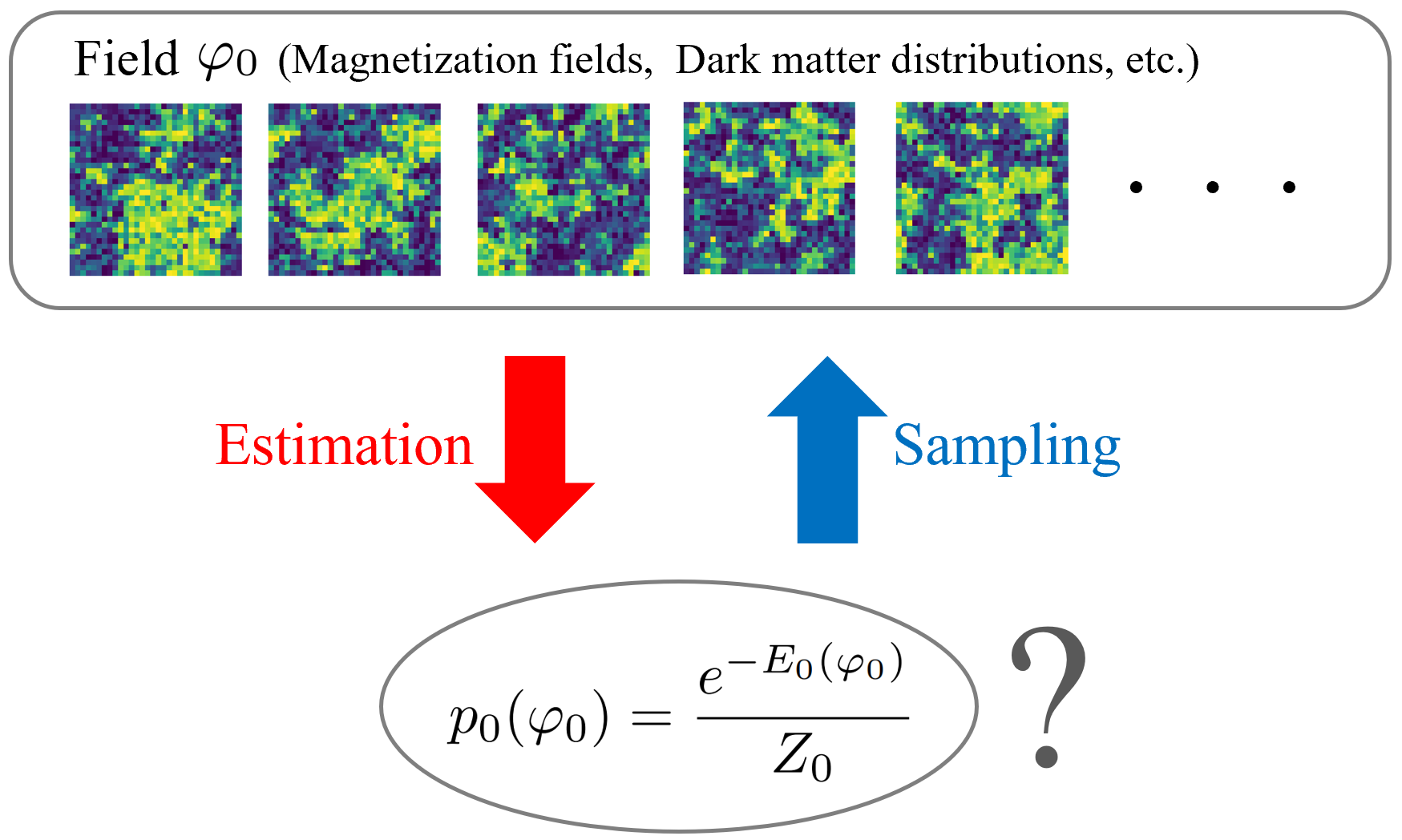}
\caption{A major challenge in physics and machine learning is to estimate the probability distribution $p_0$ of a field $\varphi_0$ and its microscopic energy function $E_0$, from examples of fields or configurations. New fields can then be generated by sampling this probability distribution. 
The Wavelet-Conditional Renormalization Group (WC-RG) is a fast multiscale approach which eliminates
"critical slowing down" phenomena near phase transitions.} 
\label{fig:concept}
\end{figure}

Estimating the probability distribution of a very large number of variables from datasets of examples is a major challenge that goes well beyond physics. It is called
"unsupervised learning" in data sciences and has considerable applications in all sciences, including neuroscience \cite{schneidman2006weak} and protein sequence data analysis \cite{cocco2018inverse}. 
In image processing, such problems were introduced in Ref.~\cite{geman1984stochastic}, to synthesize and discriminate image textures with maximum entropy models
\cite{portilla2000parametric,zhu1997minimax}.
In recent years generative models based on deep neural networks have obtained impressive results for images \cite{vae,Chintala}. Yet, their theoretical understanding still remains limited.  Developing a theory able to explain the success, the limitations, and the scalability of the algorithms used to estimate high-dimensional distributions from a large dataset is a central and widely open scientific question.

The problem, illustrated in Fig.~\ref{fig:concept}, can be described in a nutshell as follows. One has a dataset of microscopic many-body configurations represented as spatial fields $\aphi_0$,   provided by multiple experimental observations or numerical simulations.  
One can think of $\aphi_0$ as the magnetization field in a model of ferromagnetism, e.g., the $\aphi^4$ model \cite{zinn2002quantum},
or as a density field in a fluid, or in a dark matter distribution in cosmology. 
Thus $\aphi_0$ can be either in thermal equilibrium (e.g., the $\varphi^4$ model) or inherently non-equilibrium (dark matter).
The underlying probability distribution, $p_0 (\aphi_0)=e^{-E_0(\aphi_0)}/Z_0$, is unknown {\it a priori} and must
be estimated from the dataset, 
to generate new samples efficiently and determine the energy function $E_0(\aphi_0)$ (we have set $\beta=(k_BT)^{-1}=1$, i.e., we measure the energy in units of $k_BT$). 
For non-equilibrium systems, $E_0(\aphi_0)$ can be considered as an effective Hamiltonian that allows us to represent the high dimensional (and non-trivial) probability distribution in a compact way.
The tasks mentioned above are usually very difficult for two reasons. First, general high-dimensional distributions can only be estimated if the number of examples in the dataset grows exponentially with the system size \cite{bellman1959mathematical} -- an impossible requirement in practice. This curse of dimensionality can be avoided with prior information, which specifies suitable families of models for the probability distribution. 
Second, for systems that have fluctuations on a wide range of length scales, particularly near critical points, the estimation of
model parameters is usually badly conditioned. 
It requires a large number of iterations, 
which grows as a power law of the system size and leads to large estimation errors. Furthermore, estimating the parameters and generating new samples requires performing Monte Carlo simulations~\cite{krauth2006statistical}. 
Close to or at a critical point, Monte Carlo computations are hampered by  a critical slowing down ~\cite{chaikin1995principles,sethna2021statistical}, which produces a strong divergence of the mixing or decorrelation timescales when the system size increases.

This work develops a new approach to estimating energy functions and
generate new samples by addressing both problems.
It leverages and combines ideas from the theory of renormalization group (RG) and wavelet theory to define new energy-based models of
probability distributions, which ensure that the estimation is well-conditioned, 
with no critical slowing down at phase transitions. Similarly to
the renormalization group~\cite{delamotte2012introduction}, 
we connect effective theories at different scales.
In the context of phase transitions, the RG theory assumes that the microscopic energy function is known {\it a priori}, and RG computes the evolution of energy functions at progressively larger scales. 
In our problem, the microscopic
energy is instead {\it unknown} and must be estimated from the dataset.

Our method, the Wavelet Conditional Renormalization Group (WC-RG), focuses on the probabilities of the "fast degrees of freedom" (or high wave-vector fluctuations), given the configuration at the coarser scale. These conditional probabilities connect the effective theories at different scales.
From the conditional probability at each scale and going progressively from the coarsest to the finest scales, one can obtain a novel multiscale representation of the probability distribution of $\varphi_0$. WC-RG is based on this representation and estimates the probability distribution $p_0(\varphi_0)$ by the product of conditional probability models defined over orthogonal wavelet fields. 
Although WC-RG is inspired by the standard forward RG, it performs RG {\it inversely} from coarse to fine, dealing with the conditional probabilities.
We develop models of conditional probabilities across scales by introducing energy functions associated with scale interactions. This leads to more flexible models than the ones based on estimating the energy function directly at the microscopic scale. 
Remarkably, our approach also guarantees that the method remains well-conditioned even at phase transitions. Thus it requires only a relatively small number of data, and it can generate new samples very fast. For instance, it allows generating new samples on times of order one, even at the critical point of statistical physics models. Our numerical applications to two-dimensional Gaussian and $\varphi^4$ field theories confirm that, indeed, WC-RG circumvents the problem of critical slowing down completely. Its successful application to  weak-gravitational-lensing fields in cosmology shows that WC-RG is a powerful approach to model complex many-body interactions. This is a highly non-trivial case study because the system is inherently out-of-equilibrium and is characterized by long-range interactions due to gravity.


The use of renormalization group (RG) techniques and ideas in data sciences and machine learning has attracted a lot of attention over the years. All these interesting applications have elements in common with our approach but also differ in aims and content. In particular, RG  has been used recently as an inspiring analogy to study image generative models
with deep neural networks in Refs.~\cite{tanaka2015inverse,shiina2021inverse,li2018neural,mehta2014exact}.
However, the complexity of these architectures does not yet allow any mathematical analysis of these models, and thus do not provide explicit representations of the energy. Several works \cite{evenbly2016entanglement,mallatReview} also demonstrated close relations between deep convolutional network representations and iterated wavelet transforms with non-linearities. They
capture scale interactions but without conditional probabilities. As we shall discuss in conclusion, WC-RG provides a framework to analyze convolutional neural networks by modeling scale interactions through conditional probabilities, contrary to simpler methods that are based on energy models at a given scale. 



\subsection{A Brief Introduction to WC-RG}

\begin{figure}[ht]
\includegraphics[width=\linewidth]{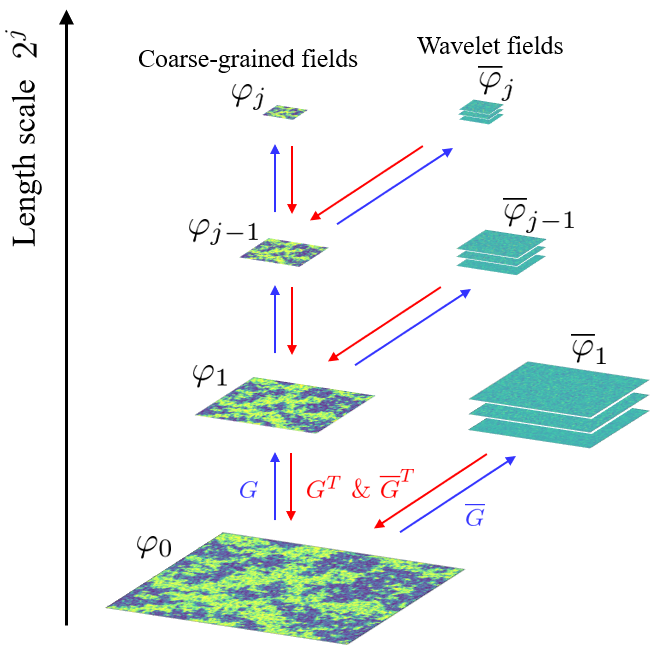}
\caption{
Coarse-grained fields $\aphi_j$ and wavelet fields $\dphi_j$ at length scales $2^j$ are iteratively computed from a field $\aphi_{j-1}$ at a finer scale $2^{j-1}$. It is implemented with orthogonal convolutional and subsampling 
operators $G$ and $\overline G$. A wavelet field
$\dphi_j$ represents the "fast degrees of freedom" of $\aphi_{j-1}$ which have disappeared in $\aphi_j$. It contains three sub-fields in dimension $d=2$, corresponding to spatial fluctuations along different orientations.
The inverse wavelet transform reconstructs $\aphi_{j-1}$ from $\aphi_j$ and $\dphi_j$ with the adjoint operators $\LL^T$ and $\wG^T$.
Near the phase transition, a coarse-grained field $\aphi_j$ has long-range spatial correlations, whereas  $\dphi_j$ has short range correlations.
}
\label{fig:wavelet_concept}
\end{figure}

One of the key elements of RG is {\it coarse-graining}. From an input field $\aphi_0$ in $d$ spatial dimensions, which may be an image ($d=2$), one computes coarse-grained versions $\aphi_j$ at each length scale $2^j a$, where $a$ is the unit of length that is set to one. The coarse grained fields $\aphi_j$ are illustrated in Fig.~\ref{fig:wavelet_concept}, together with wavelet fields $\dphi_j$ corresponding to "fast-degrees of freedom"  discussed below.  
The probability distribution $p_j (\aphi_j)$ is defined at each scale $2^j$ 
by an energy function $E_j(\aphi_j)$. This energy
is parameterized by a vector of coupling parameters $\theta_j$,
with $E_j(\aphi_j)=\theta_j^T U_j(\aphi_j)$, where $U_j$ is the basis functions (or "operators" in the quantum field theory jargon). RG procedures specify the evolution of
$p_j (\aphi_j)$ across scales through the RG flow of coupling parameters $\theta_j$. In an RG step, one computes $p_j (\aphi_j)$ and $\theta_j$ from the finer scale $p_{j-1}(\aphi_{j-1})$ and $\theta_{j-1}$, 
by integrating over the fast degrees of freedom of $\aphi_{j-1}$ which have been eliminated in $\aphi_j$. These degrees of freedom represent the short scale or highest wave-vector variations of $\aphi_{j-1}$. 
The theories describing large-scale critical fluctuations at a phase transition correspond to a non-trivial fixed point of the RG transformation.
To compute this fixed point for the $\aphi^4$ model, Wilson originally represented the fast degrees of freedom over a basis of functions that are well localized in the Fourier basis \cite{wilson1971renormalization}. He did so by defining what was later called a Shannon wavelet basis \cite{wilson1971renormalization}.
A considerable body of work has then been devoted to RG calculations in Fourier bases
\cite{delamotte2012introduction}, and also in wavelet bases \cite{battle1999wavelets,altaisky2016unifying,evenbly2016entanglement}, particularly 
with perturbative methods in dimension $d = d_{U}-\epsilon$ (where $d_U$ is the upper critical dimension, which is equal to $4$ for the $\varphi^4$ model). 

The Wavelet Conditional Renormalization Group does not focus on the probability distribution $p_j(\varphi_j)$, nor on $E_j(\varphi_j)$ at a given scale or on $\theta_j$. It is instead based on a different
representation of microscopic probability distributions. It relies on conditional probabilities, which specify interactions across scales. These are functions of the wavelet fields $\dphi_j$ representing the fast degrees of freedom or high wave-number fluctuations. 
The importance of scale interactions is well known
for multiscale physics phenomena such as fluid turbulence, producing energy cascades across scales
\cite{kolmogorov_1962,FrischParisi80,frisch1991global}. Wavelet transform has been used to analyze these interactions \cite{muzy1994multifractal} and to introduce specific parametric models such as Multifractal Random Walks
\cite{mandelbrotmultifractals,MRW01}. Low-dimensional wavelet scale interaction models have been shown to be
sufficient to synthesize complex turbulent fluids and mass density fields in 
astrophysics~\cite{Allys_2020,zhang2021maximum,Cheng_2021}. However, such
approaches do not provide a general numerical and theoretical framework that enables us to recover the
energy functions of probability distributions.
Energy functions are a very useful output of our method as, besides their fundamental character, they provide a compact and interpretable representation of high-dimensional probability distributions.

Within WC-RG, the connection between scales is studied as follows: at each scale $2^j$, a wavelet transform decomposes a field $\aphi_{j-1}$ into 
wavelet fields $\dphi_j$, which capture its high wave-vector variations,
and a coarse-grained field $\aphi_j$~\cite{stephane1999wavelet}.
In spatial dimension $d=2$, $\dphi_j$ has $3$ wavelet sub-fields corresponding to different spatial orientations that we call channels. This is shown in Fig.~\ref{fig:wavelet_concept} for the two dimensional $\varphi^4$ field theory just above the critical point. In spatial dimension $d$ there are $2^d-1$ channels. The WC-RG focuses on the conditional probability $\overline p_j (\dphi_j | \aphi_j)$ of each
wavelet field $\dphi_j$ given the coarse-grained field $\aphi_j$.
The microscopic probability $p_0 (\aphi_0)$ is factorized as a product of these
conditional probabilities up to a maximum scale which (without loss of generality) is $L=2^J$:
\[
p_0 (\aphi_0) = \alpha\, p_J (\aphi_J)\, \prod_{j=1}^J \overline p_j (\dphi_{j} | \aphi_j),
\]
where $\alpha$ is a Jacobian constant.
Each conditional probability $\overline p_j (\dphi_{j} | \aphi_j)$ is represented by a Gibbs energy function $\overline E_j$ which specifies scale interactions. The energy $\overline E_j=\bK_j^T \overline U_j$ is represented as a linear combination of local operators $\overline U_j$ weighted by conditional coupling parameters $\bK_j$. The microscopic probability $p_0 (\varphi_0)$
is specified by the WC-RG coupling parameters of all $\overline p_j$ and of the maximum scale 
$p_J$:
\[
\Big\{ \K_J ~,~\bK_j \Big\}_{1 \leq j \leq J } .
\]
Using this representation WC-RG {\it estimates} from data the probability distribution, the energy function and {\it samples} new configurations. 
The coupling parameters $\Big\{ \K_J ~,~\bK_j \Big\}_{1 \leq j \leq J }$ are obtained from the training dataset by maximum likelihood estimation, which
is equivalent to minimize a Kullback–Leibler divergence. They are used to generate new samples $\varphi_0$
of the microscopic field distribution $p_0(\varphi_0)$. 
The theories describing large-scale fluctuations at critical points correspond to fixed points  $\bK_j=\bK_{j-1}$. WC-RG parametrization remains stable even at phase transitions because
it relies on a parametrization $\bK_j$ of interaction energies $\overline E_j$ 
instead of a parametrisization of $E_j$.
Indeed, although each $\varphi_j$ has long-range correlations and a singular
covariance matrix at phase transitions,  wavelet
fields $\dphi_j$ still have short-range correlations, as proved by
the theorem in Sec.~\ref{precondHamilsec}. This is illustrated by the decomposition 
of $\varphi^4$ field near phase transition in Fig.~\ref{fig:wavelet_concept}.
As a result, 
WC-RG coupling parameters does not suffer from any critical slowing down, 
for Gaussian fields, $\varphi^4$ models and simulated cosmological data, as we shall show numerically.\\
To recover the microscopic energy function $E_0(\varphi_0)$,
we regress a parametrised model of the free energy $\overline F_j(\aphi_j)$ associated to
each interaction energy $\overline E_j$. It is the normalization factor of the conditional probability $\overline p_j (\dphi_{j} | \aphi_j)$. The microscopic energy is then obtained 
as a sum of all interaction
energies parameterised by $\overline \theta_j$, together with their
free energies. As opposed to microscopic energies $E_0$ with local potentials, such multiscale
models can also capture long-range interactions.

The WC-RG sampling proceeds from
coarse to fine scales by iteratively sampling 
$\overline p_j (\dphi_j | \aphi_j)$. It computes a new field $\dphi_j$ given $\aphi_j$ from, which we recover the finer scale field $\aphi_{j-1}$ with an inverse wavelet
transform. It does not require to compute the free energies $\overline F_j$, which are
normalisation factors. As opposed to 
inverse RG algorithms \cite{ron2002inverse,bachtis2022inverse}, which also proceed from 
coarse to fine scales, it does not sample directly the distributions $p_j$ to recover each $\aphi_j$. Indeed, sampling fast degrees of freedom $\dphi_j$ avoids critical slowing down phenomena, because they have a short range correlation.

In summary, WC-RG estimates and samples conditional probability distributions of wavelet fields connecting fluctuations across scales. It relies
on the mathematics and physics of RG, which focuses on interactions across scales and
fast degrees of freedom ~\cite{delamotte2012introduction}.
The manuscript is organized as follows. 
Section~\ref{waverensec} reviews the properties of wavelet transforms and their connection with
the RG formalism. Section~\ref{wavecondRGsec} introduces the Wavelet Conditional Renormalization Group
representation of probabilities and energy functions. Section~\ref{precondHamilsec} presents maximum likelihood
estimations of WC-RG coupling parameters. It shows that 
the resulting algorithms are well-conditioned and have a fast convergence that is not 
affected by critical slowing down.
Section~\ref{Recovermicrosec} relates WC-RG coupling parameters to microscopic energy functions through
free-energy calculations, which are performed with thermodynamic integrations. 
Sec.~\ref{numericalSec} gives numerical applications of WC-RG to two-dimensional systems, Gaussian theory, 
$\varphi^4$ field theory, and weak lensing cosmological data, demonstrating the fast convergence of
estimation and sampling algorithms even at critical points. Section VII provides a conclusion with discussions and perspectives.

\section{Renormalization Group and Wavelets}
\label{waverensec}

The pioneering works on RG by Kadanoff \cite{kadanoff}, Wilson \cite{wilson1971renormalization}, Fisher \cite{wilson1972critical} and others \cite{fisher1974renormalization,delamotte2012introduction} rely on scale separation. They characterize the evolution of the energy function from fine to coarse scales by progressively integrating out "fast degrees of freedom". Critical phenomena such as phase transitions are identified as fixed points of the renormalization group.
In the following, we set up the formalism needed to perform the renormalization group transformation over orthogonal wavelet coordinates. This will provide the basic tools to then 
introduce a conditional renormalization based on conditional probabilities across scales.

\subsection{RG in a Wavelet Orthogonal Basis}
\label{RGwavebasis}
We shall focus on $d$-dimensional scalar field theories, but our framework can be straightforwardly extended to vector fields. We consider theories that are translational invariant. In order to provide a concrete example of our theory, we will focus on the $\varphi^4$-field theory, which is a central model in the theory of second-order phase transitions. Henceforth, we will denote by $\aphi_0$ a many-body configuration field. The site index $i$ in $\aphi_0(i)$ belongs to a $d$-dimensional lattice of lattice spacing $a=1$, linear size $L$, and hence with $L^d$ sites, e.g., 
$d = 1$ (linear), $d=2$ (square) or $d = 3$ (cube).
The probability distribution of $\aphi_0$ is given by the Boltzmann law $p_0 (\aphi_0) = {\ZZ_0^{-1}} \, e^{-\HH_0 (\aphi_0)}$, where $\HH_0$ is the (configurational) energy function of the field $\aphi_0$. We have set $\beta=(k_BT)^{-1}=1$, i.e., we measure the energy in units of $k_BT$. The notation $\langle \cdot \rangle_p$ will be used to denote statistical averages over the probability distribution $p$.
Throughout the paper, we consider the microscopic scalar field $\aphi_0$, which is normalized such that $\langle \aphi_0(i) \rangle_{p_0}=0$ and $\langle | \aphi_0(i)|^2 \rangle_{p_0}=1$. Note also that we will not use the field theory terminology "connected correlation function" but the more standard one in probability theory "covariance".

\subsubsection{Coarse graining and renormalization}
Let us denote $\aphi_j$ the coarse-grained version of $\aphi_0$ at a length scale $2^j$ defined over a coarser grid (lattice) with intervals $2^j$ and hence $( L2^{-j})^d$ sites. In this coarse-graining procedure, the lattice spacing is each time reduced by two. The coarser field $\aphi_j$ is iteratively computed from  $\aphi_{j-1}$ by applying a coarse-graining operator, which acts as a scaling filter $\LL$ eliminating high wave-vector contributions and subsamples the field \cite{Mallat:89}:
\begin{equation}
  \label{low-pass20}
(\LL \aphi_{j-1}) (i) =  \sum_{i'} \aphi_{j-1} (i')\, \LL (2i-i')\,\, .
\end{equation}
The site index $i$ on the left-hand side runs on the coarser lattice, whereas $i'$ on the finer one. 
More detailed properties of $G$, together with a brief introduction to wavelet theory, are described in App.~\ref{app1}. In order to guarantee that the fluctuations of $\aphi_{j}$ remains of the order of one, a standard RG procedure normalizes $\LL$ with a factor $\gamma_j$ which guarantees that $\langle |\aphi_{j}(i)|^2 \rangle_{p_j}-\langle \aphi_{j}(i) \rangle_{p_j}^2=1$:
\[
\aphi_j = \gamma_j^{-1}\, \LL\, \aphi_{j-1} .
\]
A different normalization condition will be chosen for WC-RG later. 
The Kadanoff block averaging scheme is a particular example
of the coarse-graining procedure in Eq.~(\ref{low-pass20}). 
In the $d=1$ case, it sums pairs of consecutive samples:
\begin{equation}
\label{blockav}
\aphi_j(i)=  \frac{\aphi_{j-1}(2i)+\aphi_{j-1}(2i-1)}{\sqrt{2}\,\gamma_j}.
\end{equation}
In the renormalization group procedure, one computes scale by scale 
the energy function $\HH_j$ of the marginal probability distribution restricted to the coarse-grained field
$\aphi_j$, as given by
\begin{equation}
  \label{probaj}
p_j (\aphi_j) = \frac{1}{\ZZ_j} \, e^{-\HH_j (\aphi_j)} ~~\mbox{with}~~\HH_j (0) := 0~,
\end{equation}
where $\ZZ_j = \int e^{-\HH_j (\aphi_j)}\,d \aphi_j$. 
RG proceeds from fine (microscopic) 
to coarse (macroscopic) scales, by calculating $\HH_{j}$ from $\HH_{j-1}$. The marginal probability
distribution $p_j(\aphi_j)$ is computed by
integrating $p_{j-1} (\aphi_{j-1})$ along the degrees of freedom of $\aphi_{j-1}$ that are not in  $\aphi_{j}$~\cite{delamotte2012introduction}. These "fast degrees of freedom" correspond to high wave-vector fluctuations, which
can be computed by using Fourier or wavelet bases. Wavelet bases provide localized representations of large classes of energy functions, which leads to an efficient calculation of
marginal integrations, as we shall discuss below.

\subsubsection{Fast wavelet transform}
We now introduce the orthogonal decomposition, which allows us to represent a microscopic field in terms of its coarse-grained version and "fast degrees of freedom" contributions. This is done with the fast wavelet transform algorithm introduced in Ref.~\cite{Mallat:89}. 
It computes an orthogonal representation 
with a cascade of filtering and subsampling illustrated in Fig.~\ref{fig:wavelet_concept}, which separates the field fluctuations at different scales. For completeness we recall below and in the appendix A some properties of wavelet theory. 

The degrees of freedom of $\aphi_{j-1}$ that are not in  $\aphi_{j}$ are encoded in orthogonal wavelet field $\dphi_j$.
The representation $(\aphi_j , \dphi_j)$ 
is an orthogonal change of basis calculated from $\aphi_{j-1}$. The coarse-grained field
$\aphi_j$ is calculated in Eq.~(\ref{low-pass20}) with a low-pass scaling filter $\LL$ and a subsampling. In spatial dimension $d$,
the wavelet field $\dphi_j$ has $2^d-1$ sub-fields (or channels) computed with a convolution and subsampling operator $\G$. 
By including the normalization factor
$\gamma_j$, we get 
\begin{equation}
  \label{fastdec3}
\aphi_j = \gamma_j^{-1}\, \LL\, \aphi_{j-1} ~~\mbox{and}~~
\dphi_j = \gamma_j^{-1}\, \G\, \aphi_{j-1} .
\end{equation}
In spatial dimension $d$, we define a separable low wave-number filter $\LL$ in Eq.~(\ref{low-passfi}), which computes a
coarse-grained field $\aphi_j (i)$ which has $ (L2^{-j})^d$ sites. Instead, the wavelet filter
$\G$ computes $2^d-1$ wavelet sub fields $\dphi_j (m, i)$ indexed by $1 \leq m \leq 2^d-1$, 
with separable high-pass filters $\G_m$ specified in Eq.~(\ref{high-passfi}):
\[
\dphi_{j} (m, i) =  \gamma_j^{-1} \sum_{i'} \aphi_{j-1} (i')\, \G_m (2i-i').
\]
For each $m$, $\dphi_{j} (m, i)$ carries the fast fluctuations 
of $\aphi_{j-1}$ along a particular spatial orientation.
Cascading Eq.~(\ref{fastdec3}) for $1 \leq j \leq J$ 
computes the decomposition of the microscopic field $\aphi_0$ into its orthogonal wavelet representation
over $J$ scales, denoted as
\begin{equation}
\label{orthowave}
\left\{ \aphi_J\, ,\, \dphi_j \right\}_{1 \leq j \leq J} .
\end{equation}

In $d=1$, there is a single wavelet field in $\dphi_j$.
The Kadanoff scheme is computed with a block averaging filter $\LL$ in Eq.~(\ref{blockav}).
The corresponding wavelet filter $\wG$ in Eq.~(\ref{quadrature}) computes the
wavelet field with normalized increments:
\begin{equation}
    \label{Harwnsdf}
\dphi_j(i)= \frac{\aphi_{j-1}(2i-1)-\aphi_{j-1}(2i)}{\sqrt{2}\,\gamma_j}.
\end{equation}
If $d = 2$ then there are $3$ channels, as  illustrated in Fig.~\ref{fig:wavelet_concept}.

The wavelet orthonormal filters $\LL$ and $\G$ define
a unitary transformation, which satisfies
\begin{equation}
  \label{consdsf}
\G\,\LL = \LL\,\G = 0~~\mbox{and}~~\LL^T \LL + \G^T \G = Id~,
\end{equation}
where $Id$ is the identity. Appendix~\ref{wavefiltdesign} gives in
Eqs.~(\ref{quadrature0}) and (\ref{quadrature}) a condition on the Fourier transform on $\LL$ and on $\G$ to build such filters.
The filtering equations in Eq.~(\ref{fastdec3})
can be inverted with the adjoint operators, $\LL^T$ and $\G^T$, as given by
\begin{equation}
  \label{fastrec3}
  \aphi_{j-1} =  \gamma_j \LL^T \aphi_j + \gamma_j \G^T \dphi_j~.
\end{equation}
The adjoint $\LL^T$ enlarges the field size $\aphi_j$ by inserting a zero between coefficients and filters the output, as
\[
(\LL^T \aphi_j)(i) = \sum_{i'} \aphi_j (i' )\, \LL (2i'-i) .
\]
The adjoint $\G^T$ performs the same operations over the $2^d-1$ channels and adds them,
\[
(\G^T \dphi_j )(i) = \sum_{m=1}^{2^d-1} \sum_{i'} \dphi_j (m, i' )\, \G_m (2i'-i) .
\]
The fast inverse wavelet transform \cite{Mallat:89} reconstructs 
$\aphi_0$ from its wavelet representation in Eq.~(\ref{orthowave}) 
by progressively recovering $\aphi_{j-1}$ from $\aphi_j$ and $\dphi_j$
with Eq.~(\ref{fastrec3}), for $j$ going from $J$ to $1$.

The fast wavelet transform computes each $\dphi_j$ from $\aphi_0$ by applying $j-1$ times
the convolutional operator $G$ and then applying $\wG$. This cascade of convolutional 
operators defines a single convolutional operator with an 
equivalent filter which is called a wavelet. When $j$ increases, 
the wavelet theory \cite{Mallat:89b,Meyer:92c} reviewed in App.~\ref{app1.2} proves that
wavelet filters converges to wavelet functions of a continuous spatial variable $x \in \R^d$. Wavelet functions
belong to the space ${\bf L^2} (\R^d)$
of square-integrable functions, $\int |f(x)|^2 dx < \infty$. 
Dilating and translating these $2^d -1$ wavelet functions 
defines a wavelet orthonormal basis of ${\bf L^2} (\R^d)$.
Wavelet fields $\dphi_j$ can
be rewritten in Eq.~(\ref{waveletbasis}) as
decomposition coefficients in this wavelet orthonormal basis. 
The wavelet basis plays an important role in understanding the properties
of wavelet fields, depending upon the representation of operators involved in energy functions. 
Actually, Wilson, in his seminal paper~\cite{wilson1971renormalization}, computes
such a wavelet decomposition by using a specific wavelet that is now called Shannon wavelet.
Kadanoff RG scheme instead can be interpreted as a decomposition in a Haar wavelet basis.
Figure~\ref{fig20} shows three representative wavelets that we discuss in this paper, Haar (a, b), Shannon (c, d), and Daubechies (e, f). 
 We shall see that results are improved by using wavelets having a better localization both in the spatial and
Fourier domains, as a Daubechies wavelet \cite{daubechies1992ten} in Fig.~\ref{fig20}(e, f).

For latter convenience, we define a projection operator, $P=G^T G$, and its complement,  $\overline P = \overline G^T \overline G$, where $P+\overline P = Id$ because of Eq.~(\ref{consdsf}). $P$ and $\overline P$ play a role in projecting a field $\varphi_{j-1}$ onto the low and high wave-vector components, respectively,
\[
P \varphi_{j-1}=\gamma_j G^T \varphi_j~~\mbox{and}~~\overline P \varphi_{j-1}=\gamma_j \overline G^T \overline \varphi_j. 
\]

\subsubsection{Wavelet renormalization}
 
The standard RG scheme begins from a microscopic energy function $\HH_0$,
and iteratively computes $\HH_j$ from $\HH_{j-1}$. 
To compute $p_j (\aphi_j) = \ZZ_j^{-1} e^{-\HH_{j} (\varphi_j)}$
from
$p_{j-1} (\aphi_{j-1}) = \ZZ_{j-1}^{-1}  e^{-\HH_{j-1} (\varphi_{j-1})}$,
one can represent $\aphi_{j-1}$ as orthogonal wavelet coordinates
$(\aphi_j , \dphi_j)$ by Eq.~(\ref{fastrec3}),
and perform a marginal integration along the wavelet field $\dphi_j$, as 
\begin{equation}
    \label{marn08sdf00}
\HH_{j} (\varphi_j) = -\log
\int e^{-\HH_{j-1}(\varphi_{j-1})}\, d \dphi_j + c_j,
\end{equation}
where $c_j$ is a constant chosen such that $\HH_{j} (0)=0$, thus $c_j=\log
\int e^{-\HH_{j-1}(\overline P \varphi_{j-1})}\, d \dphi_j$.
Eq. \ref{marn08sdf00} plays the same role as a momentum shell integration in the standard RG, in which 
a similar marginal integration is calculated in Fourier
basis. 

\begin{figure}
\includegraphics[width=0.5\linewidth]{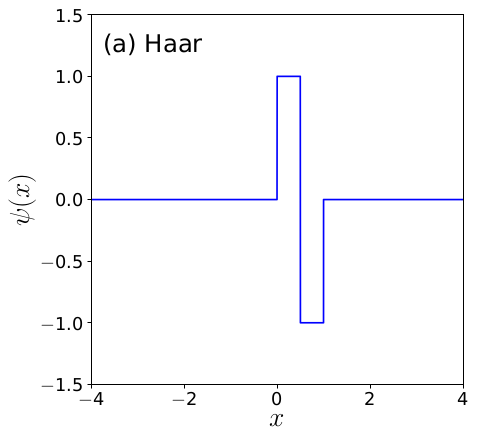}
\includegraphics[width=0.48\linewidth]{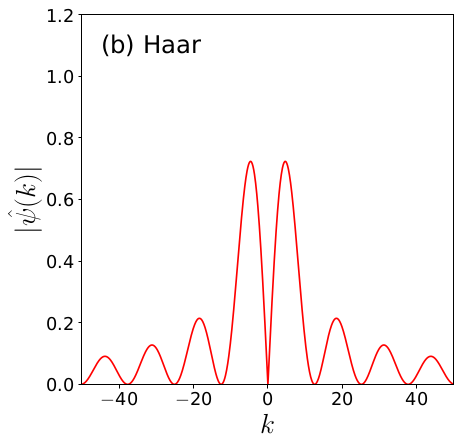}
\includegraphics[width=0.5\linewidth]{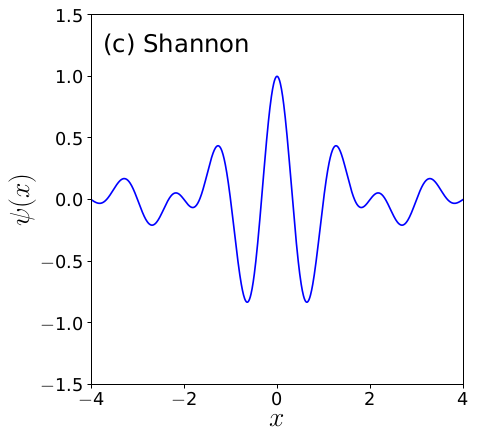}
\includegraphics[width=0.48\linewidth]{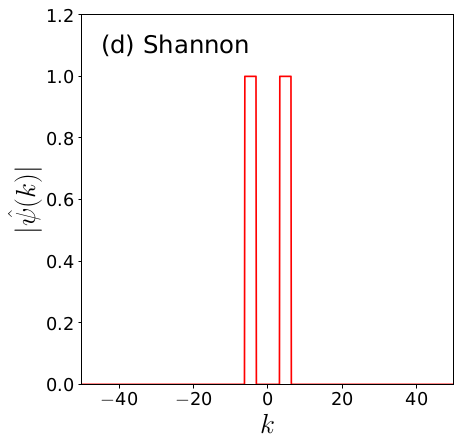}
\includegraphics[width=0.5\linewidth]{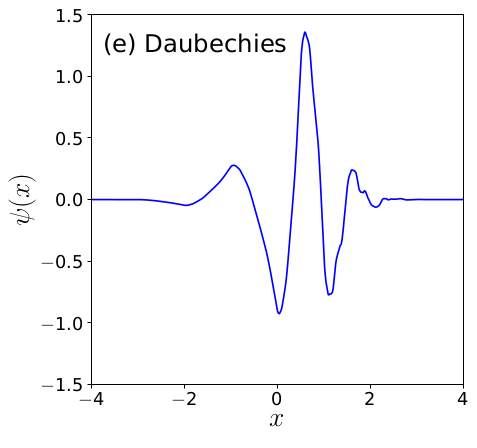}
\includegraphics[width=0.48\linewidth]{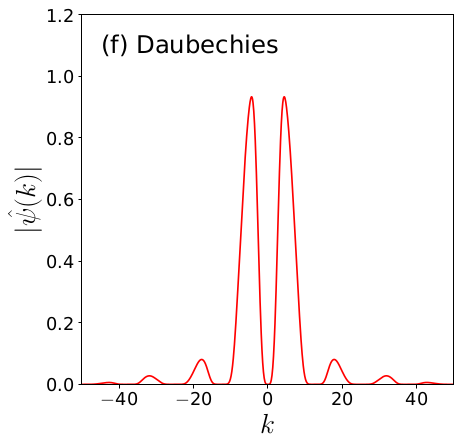}
\caption{The left graphs show one-dimensional wavelets $\psi(x)$ in real space, 
and the right graphs give their Fourier transform amplitude $|\widehat \psi(k)|$.
(a, b): Haar wavelet. (c, d): Shannon wavelet. (e, f): Daubechies wavelet with $q=4$ vanishing moments ~\cite{daubechies1992ten}.}
 \label{fig20}
\end{figure}

Large classes of energy functions involve differential operators such as Laplacians and gradients, which are
diagonal in a Fourier basis. However, they often also involve local (pointwise) non-linear
operators as, e.g., in the $\varphi^4$ model, whose representations are delocalized on a Fourier basis. 
For appropriate wavelets,  which are well localized in both
real-space and Fourier domains, energy functions $E_{j-1}(\varphi_{j-1})$ depend upon local interactions
of the wavelet fields $\dphi_\ell$ for $\ell \geq j$. This is very useful both for numerical and analytical approaches, as it provides a compact short-range representation of the theory both in real and Fourier space. 
In his original calculation, Wilson used such "diagonal" approximations in a Shannon wavelet basis to analyze the properties of the $\varphi^4$ model at the phase transition \cite{wilson1971renormalization}.
Indeed, if $\HH_{j-1}$ can be approximated by a sum over sites $i$ of terms that are functions of the local value of the wavelet field $\dphi_j (m,i)$, then the multidimensional marginal integral in Eq.~(\ref{marn08sdf00}) becomes a product of one-dimensional integrals, and the RG flow can be easily analyzed. Yet, the Shannon wavelet is not well adapted to the $\varphi^4$ energy function because it has a slow spatial decay in real-space, as shown in Fig.~\ref{fig20}(c).

We stress that there are several studies of RG in wavelet bases~\cite{battle1999wavelets,altaisky2016unifying}, but they remain on a formal level or are focused on $d = d_{U}-\epsilon$ perturbation theory. Our interest is instead developing a non-perturbative method both for forward RG and estimation by WC-RG. 

\subsubsection{The $\aphi^4$ Model}
\label{sec:intro_phi4}

This small section introduces 
a discrete lattice version of the $\aphi^4$ field theory, as a specific model, which provides a reader a concrete example of energy functions.
The $\aphi^4$ field theory played a central role in the theory of critical phenomena \cite{zinn2002quantum}, because it captures the essential 
properties of standard second-order phase transitions. 
Its microscophic energy on a lattice reads
\begin{equation}
  \label{nsodufsbis01}
\HH_{0} (\aphi_0) = -\frac \beta 2 \aphi^T_0 \Delta \aphi_0  + 
\sum_i \Big(\aphi^4_0(i) - (1+2\beta) \aphi^2_0(i) \Big) ,
\end{equation}
where $\Delta$ is a discrete Laplacian: 
\begin{equation}\label{discretizedlaplacian}
   -\aphi^T_0 \Delta \aphi_0 = \sum_i \sum_{i'\in {\cal N}(i)} \frac{1}{2} \left(\aphi_0(i) - \aphi_0(i')\right)^2 , 
\end{equation}
where ${\cal N}(i)$ is a set of the nearest neighbor sites of $i$.
It contains $2d$ neighbor sites in the $d$-dimensional cubic lattice.
The first term in Eq.~(\ref{nsodufsbis01}) disfavors spatial fluctuations.  
The second term is an even local potential with a double-well shape that favors two opposite values of the field (corresponding to the two wells). The parameter $\beta$ specifies the relative magnitude between the two terms. For $\beta \ll 1$, fluctuations are entropically favored, and hence, the system is in a disordered phase. For $\beta \gg 1$, the two terms compete and lead to a ferromagnetic phase. 
The $\aphi^4$ model in Eq.~(\ref{nsodufsbis01}) undergoes a phase transition in the thermodynamics limit for $\beta=\beta_c\simeq 0.68$~\cite{kaupuvzs2016corrections}. For $\beta > \beta_c$ the system is in the ordered broken symmetry phase whereas for $\beta < \beta_c$ the system is in a disordered phase.  

When expressed on Fourier basis, the Laplacian operator in the energy function in Eq.~(\ref{nsodufsbis01}) becomes diagonal, but the $\aphi_0^4(i)$ local potential term instead produces global interactions between all wave-vectors in the Fourier domain. On the other hand,
appropriate wavelet bases (see below) provide nearly diagonal representations for the Laplacian operators \cite{Meyer:92c} and the $\aphi_0^4(i)$ term because wavelets are well localized in both spatial and Fourier domains. The nearly diagonal representations by such wavelet bases are proven in App.~\ref{Theorem1Proof}.

Kadanoff and Wilson's RG approaches can be interpreted as different choices of wavelet basis to represent high wave-number fluctuations.
The Shannon wavelet shown in Fig.~\ref{fig20}(c, d) is well localized in the Fourier domain, but it has a slow spatial 
decay in real space. It thus does not provide good local approximations of 
pointwise polynomial non-linearities in the $\varphi^4$ model.
Instead, Kadanoff block averaging~\cite{kadanoff} is equivalent to decomposition in a wavelet basis generated by a 
Haar wavelet presented in Fig.~\ref{fig20}(a, b). It has a narrow compact support in real space, but it is discontinuous and therefore
extended in the Fourier space. As a consequence, the Haar wavelet is unable to nearly diagonalize the Laplacian term. The most important consequence is that Kadanoff block averaging RG is not able to describe the Gaussian fixed point (associated with pure Laplacian energy), and hence it is incorrect in high spatial dimensions and unable to capture the existence of the upper critical dimension.
Appendix~\ref{app1.2} introduces Daubechies wavelets \cite{daubechies1992ten} which have a compact support wider than the Haar
wavelet but are well localized in Fourier space. They provide nearly diagonalized differential operators such as a Laplacian as well as pointwise non-linearities of local potentials. 
Figure~\ref{fig20}(e, f) shows the Daubechies wavelet used in our numerical calculations.

\subsection{Energy Ansatz}
\label{HmitAnsec}

The RG coarse-graining in Eq.~(\ref{marn08sdf00}) cannot be performed exactly in general. A strategy developed both for real space \cite{kadanoff} and non-perturbative RG \cite{delamotte2012introduction} is to 
approximate each $\HH_j$ as a linear combination of a few terms, each one determined by coupling parameters and basis functions (or operators) of the field. 
We will proceed in a similar way. In the following, we review the 
energy Ansatz that we will use as approximation models.  
We write such an Ansatz in terms of coupling parameters and basis functions, so that one can express the wavelet RG flow equation as an evolution of the coupling parameters.

\subsubsection{Local Potentials}

The energy Ansatz, i.e., the approximate form of the energy function for $\HH_{j} (\aphi_j)$, is defined at each scale $2^j$ as the sum of bilinear terms characterized by 
a two-point symmetric positive coupling matrix $K_j$ and a non-linear local potential composed of a vector of basis functions $V$ and the associated coupling parameters $C_j$:
\begin{equation}
  \label{nsodufsbis0}
\HH_{j} (\aphi_j) = \frac 1 2 \aphi_j^T K_{j} \aphi_j  + C_j^T V (\aphi_j),
\end{equation}
where $V (\aphi)$ is given by
\begin{equation}
    \label{ensdfiweef00}
V(\aphi) = \left( V_1 (\aphi), V_2 (\aphi), \cdots, V_s (\aphi) \right),
\end{equation}
which do not depend on $j$.

Local potentials are defined as the sum of independent contributions from local field values. For each $n$ ($1 \leq n \leq s$), they are given by
\begin{equation}
    \label{scnasdf8sd}
\VV_n (\aphi_j) = \sum_i v_n (\aphi_j (i)) ,
\end{equation}
where $v_n$ is, by definition, the local potential for each site $i$, which will be described below.
Thus the non-linear potential in Eq.~(\ref{nsodufsbis0}) is written by
\begin{eqnarray}
      \label{ensdfiweef}
C_j^T V (\aphi_j) &=& \sum_{n} C_{j,n}\, V_{n} (\aphi_j) = \sum_i \sum_{n} C_{j,n}\,  v_{n} (\aphi_j(i)) \nonumber \\
&=& \sum_i C_j^T v(\aphi_j(i)).
\end{eqnarray}

The microscopic energy $\HH_0$ of the $\varphi^4$ model in Eq.~(\ref{nsodufsbis01}) has
such a local potential, and thus it can be written as the form in Eq.~(\ref{nsodufsbis0}). 
The energy Ansatz for the $\varphi^4$ model assumes that the expression in Eq.~(\ref{nsodufsbis0}) holds at all scales. Numerical applications in this paper are based on local potentials, but this is not a necessary condition
for the theory and algorithms that are introduced.

We may further impose that two-point interactions have a finite range, i.e., they are defined over neighborhoods ${\cal N}(i)$
of constant size $s'$, which does not depend upon the scale $2^j$:
\[
\aphi_j^T K_j \aphi_j = \sum_i \sum_{i' \in {\cal N}(i)} K_{j}(i,i')\, \varphi_j (i)\, \varphi_j (i') .
\]
The energy Ansatz in Eq.~(\ref{nsodufsbis0}) can be rewritten as an inner product,
\begin{equation}
  \label{Hamilnansdfs00}
\HH_{j} (\aphi_j) = \K_{j} ^T \U_j(\aphi_j) ,
\end{equation}
between a vector of coupling parameters, 
\begin{equation}
  \label{nsodufs700}
\K_j = \left(\frac 1 2 \, K_j \,,\, C_j \right),
\end{equation}
of dimension $s+s'$, and the vector of the basis functions,
\begin{equation}
  \label{nsodufs70}
\U_j (\aphi_j) = \Big({\aphi_j\, \aphi_j^T\,},\, V (\aphi_j)\Big) ,
\end{equation}
where $\aphi_j \aphi_j^T = \{ \aphi_j (i)\, \aphi_j (i') \}_{i,i' \in {\cal N}(i)}$ is a band matrix
of width $s'$. 
For stationary fields, $K_j$ is a convolution kernel whose support size is $s'$ and the matrix $\aphi_j \aphi_j^T$
can be replaced by a translation invariant vector
$\left\{ \sum_i \aphi_j (i)\, \aphi_j (i-i') \right\}_{|i-i'| \leq s'/2}$. This will be the case for all applications
in this paper, but we shall keep the notation $\aphi_j \aphi_j^T$ for simplicity. 
Having an energy Ansatz whose functional form is the same at all scales allows one to capture the self-similarity emerging at phase transitions. One expects that the critical point is described by a scale-invariant theory, i.e., the coupling vectors $\K_j$ do not depend upon the scale $2^j$ for $1\ll j \ll J$. Henceforth, we will call {\it local} an Ansatz 
such as Eq.~(\ref{nsodufs70}), as it is the sum over the sites $i$ of functions of the field evaluated in sites at a finite bounded (local) distance from $i$.  

\begin{figure}
\includegraphics[width=0.75\linewidth]{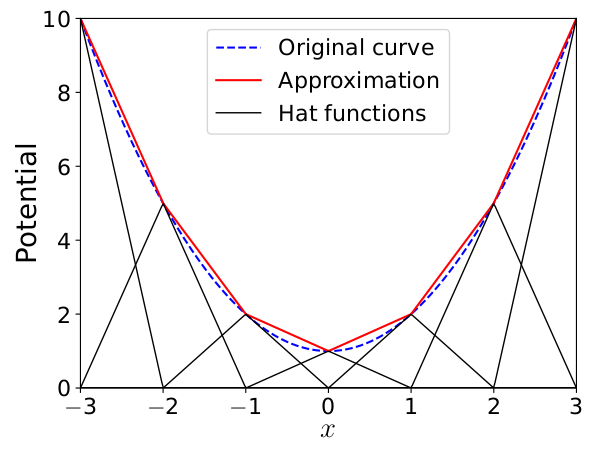}
\caption{
A quadratic function (original curve, dashed) is represented by a piecewise linear approximation (red) given by a linear combination of hat functions (black). 
}
\label{fig:hat_functions}
\end{figure}

We now describe the local potential $v_n$. In the theory of phase transitions, $v_n$
are usually chosen to be polynomials, e.g., quadratic and quartic terms, because it leads to easier analytic computations. However, the fast growth of high-degree polynomials may lead to numerical instabilities. As a consequence, in the following, we employ a piecewise linear approximation of $C_j^T v(\aphi_j(i))$. This is done by considering a basis of the "hat" functions,
$v_n (x) = h(x - n a)$, which are uniform translations of a linear box spline $h(x)$, 
which is a linear combination of $3$ linear rectifiers:
\[
h (x) = \max \{ x + a, 0 \} - 2 \max\{ x,0\} + \max\{x - a,0\},
\]
where $2a$ is the width of the hat function.
Figure~\ref{fig:hat_functions} shows an example of the piecewise linear approximation described above with $a=1$ for a quadratic function by hat functions.

We guarantee that our approximation of the probability distribution at scale $2^j$ is normalizable by 
imposing that the symmetric matrix $K_j$ is strictly
positive. The other coupling vector $C_{j}$ does not affect this convergence because $C_j^T V(\aphi_j)$ has at best a linear growth at infinity by construction due to the linear growth of the hat function. Note that we do not aim to reproduce the true growth of the potential at infinity, as this is completely determined by rare events and out of the scope of the present work. The choice of having a linear growth of $C_j^T V(\aphi_j)$ is dictated by the requirement of having a stable algorithm in which normalization of the probability distribution is guaranteed by construction. 
For stationary processes,
$K_j$ is a translation-invariant Toeplitz matrix that is diagonalized in the Fourier basis.
Normalizability amounts to imposing that its spectrum is strictly positive.

\subsubsection{RG Flow Equation}

The coupling parameters of energy functions at successive scales are related by an equation
derived from the marginal integration in Eq.~(\ref{marn08sdf00}).
Inserting the Ansatz in Eq.~(\ref{Hamilnansdfs00}) for $j$ and
$j-1$ in the RG equation, Eq.~(\ref{marn08sdf00}), gives
\begin{equation}
  \label{partionsdf}
- \log \int e^{-\K_{j-1} ^T \U_{j-1} (\aphi_{j-1})}\, d \dphi_j + c_j \approx \K_{j} ^T \U_j(\aphi_j) . 
\end{equation}
In practice, this is not an equality since the marginal integration produces extra-terms with respect to the energy Ansatz in Eq.~(\ref{Hamilnansdfs00}). In order to establish an approximate RG treatment, one has to define a procedure to obtain $\theta_j$ from the left-hand side of Eq.~(\ref{partionsdf}). Several ways have been introduced in the literature: truncations such as in the Migdal-Kadanoff real-space RG \cite{kadanoff}, projections as done in Non-Perturbative RG \cite{delamotte2012introduction}. Another possibility, that we will use later in Sec.~\ref{FreeEnerCalc}, is to linearly regress $\K_j$ over $\aphi_j$. This also allows us to estimate the approximation error as well.

Note that approximating the RG equation does not lead to instabilities (or divergencies) at the critical point, contrary to the case of perturbation theory directly done on the microscopic scale. The reason is that the integration in Eq.~(\ref{partionsdf}) is performed over $\dphi_j$ which only impacts the interaction terms between high and low wave-vectors and within high wave-vectors. See Ref.~\cite{delamotte2012introduction} for a general discussion of the stability of the RG procedure. 

In this paper, the usual (forward) RG is introduced only as a preamble to the wavelet conditional (inverse) RG. Since we do not need to develop an approximate wavelet RG scheme for our purposes, this issue will be analyzed in a future publication.

\section{Wavelet Conditional RG}
\label{wavecondRGsec}

Wavelet Conditional RG estimates an efficient representation of the probability distribution of the microscopic field from several examples measured in experiments or obtained by simulations. The aim of WC-RG is to obtain form data: (i) a model of the probability distribution that allows us to generate new samples, (ii) a generative method that is fast even in the presence of long-range correlations, e.g., at criticality, and (iii) an estimate of the underlying microscopic energy function $\HH_0$. 


\subsection{Conditional Probability Representation}

The conditional wavelet RG defines a representation of the microscopic energy $\HH_0$ into multiscale interaction energies, which represent interactions between high and low wave-vectors at each scale.
We introduce this new representation below. 

Since $(\aphi_j , \dphi_j)$ is an orthogonal representation of $\aphi_{j-1}$,
computing $p_{j-1}(\aphi_{j-1})$ from $p_j(\aphi_{j})$ amounts to compute
\begin{equation}
  \label{condproba}
p_{j-1} (\aphi_{j-1}) =  \alpha_j\,  \overline p_j ( \dphi_j | \aphi_j)\, p_j (\aphi_j)\, ,
\end{equation}
where $\alpha_j = \gamma_j^{(L/2^{j-1})^d}$ is the Jacobian factor resulting from the normalization in Eq.~(\ref{fastrec3}). This procedure can be iterated to express $p_j(\aphi_j)$ as a telescopic product: 
\begin{equation}
    p_j(\aphi_j) = p_J(\aphi_J) \, \prod_{l=j+1}^J \alpha_l \overline p_l (\dphi_l | \aphi_l)\, .
\end{equation}
At the microscopic scale ($j=0$), we can express $p_0(\varphi_0)$ as 
\begin{equation}
\label{teles}
    p_0(\aphi_0) = \alpha\, p_J(\aphi_J) \, \prod_{j=1}^J \overline p_j (\dphi_j | \aphi_j)\, . 
\end{equation}
Thus, the probability distribution $p_0$ can be represented in terms of all the conditional probabilities $\overline p_j$ and the coarsest probability $p_J$.
Each conditional probability distribution can be exactly rewritten as 
\begin{equation}
  \label{conosdf2}
\overline p_j (\dphi_j | \aphi_j) = \frac{1}{ \overline \ZZ_j} \exp\Big(-\bHH_{j} (\aphi_{j-1})+ \bFF_j(\aphi_j)\Big) ,
\end{equation}
where $\bHH_j$ and $\bFF_j$ are the energy associated with scale interactions between 
$\dphi_j$ and $\aphi_j$, and the free energy, respectively. 
More formally, the scale interaction energy $\bHH_j(\aphi_{j-1})$ is constructed such that $\bHH_j(P\aphi_{j-1}) = \left. \bHH_j(\varphi_{j-1}) \right|_{\overline \varphi_j=0} =0$ for all $\aphi_{j-1}$.
Then the free energy $\bFF_j$ in Eq.~(\ref{conosdf2}) is defined as 
\begin{equation}
  \label{freeEne}
\bFF_j(\aphi_j)  = -  \log \int  \, e^{-\bHH_{j}(\aphi_{j-1})}\, d \dphi_j + c'_j  \,,
\end{equation}
where $c'_j$ is a constant chosen such that $ \bFF_j(0) = 0$. Hence, $c'_j=\log \int  \, e^{-\bHH_{j}(\overline P\aphi_{j-1})}\, d \dphi_j$. The normalization constant in Eq.~(\ref{conosdf2}) is defined as
\[\overline \ZZ_j=  \int \, e^{-\bHH_{j}(\overline P \aphi_{j-1})}\, d \dphi_j.
\]
It results from Eq.~(\ref{condproba}) that
\begin{equation}
 \label{partionsdf9sdf9sdf00}
    \HH_{j-1}  = \HH_{j} + \bHH_{j} - \bFF_{j}.
\end{equation}
Using the representation in Eq.~(\ref{teles}) of $p_0$, and cascading by Eq.~(\ref{partionsdf9sdf9sdf00}), one finds 
\begin{equation}
  \label{partionsdf9sdf9sdf00981}
\HH_0 = \HH_J + 
\sum_{j=1}^{J} \left( \bHH_{j}  - \bFF_{j} \right).
\end{equation}
These results show that instead of working with $p_0$ directly, one can work with its alternative representation in Eq.~(\ref{teles}). Sampling a field $\aphi_0$ can thus be achieved by successive sampling of $\dphi_j$ using the multiscale interaction energy functions,
\begin{equation}
  \label{partionsdf9sdf9sdf00982}
\left\{ \HH_J~,~ \bHH_{j} \right\}_{1 \leq j \leq J} .
\end{equation}
Those energy functions specify completely $\HH_0$, since each free energy $\bFF_j$ is calculated from $\bHH_j$ with
Eq.~(\ref{freeEne}). In conclusion, the multiscale interaction energy representation in Eq.~(\ref{partionsdf9sdf9sdf00982}) offers an alternative way to sample and to define the microscopic probability distribution $p_0$. 
It is a way to represent probability distributions that describes fluctuations progressively from coarse to fine scales. 

We have shown above how to obtain all the energy functions $\HH_j$ from the multiscale interaction energies $\left\{ \HH_J~,~ \bHH_{j} \right\}_{1 \leq j \leq J}$. One can also obtain the inverse relationship that constructs the scale interaction energies and free energies from $\HH_j$. By subtracting Eq.~(\ref{partionsdf9sdf9sdf00}) to its counterpart evaluated for $\dphi_j=0$, one obtains 
\begin{equation}
 \label{partionsdf9sdf9sdf001}
\bHH_j (\aphi_{j-1})   = \HH_{j-1}(\aphi_{j-1}) - \HH_{j-1}(P \aphi_{j-1}) .
\end{equation}
Evaluating Eq.~(\ref{partionsdf9sdf9sdf00}) 
for $\dphi_j = 0$ gives
\begin{equation}
 \label{partionsdf9sdf9sdf0012}
\bFF_j(\aphi_j)  = \HH_{j}(\aphi_j)  - \HH_{j-1}(P\aphi_{j-1}).
\end{equation}
In a standard RG, the fields $\aphi_j$ are normalized with a multiplicative factor $\gamma_j$ in Eq.~(\ref{fastdec3}),
so that at second-order phase transitions, the probability distribution is scale-invariant under RG transformations. It results that $\overline p_j(\dphi_j | \aphi_j)$ is also scale-invariant. In order to find a fixed point, one has to impose a normalization condition for the fields. In our case, we choose the normalization condition such that 
\begin{equation}
    \label{normalisation}
    \sum_{m=1}^{2^d-1} 
    \langle |\dphi_j(m,i)|^2 \rangle_{p_{j-1}} = 1 .
\end{equation}
This normalization condition has an important role even when the system is not at a critical point. It leads to coupling parameters associated with scale interaction energies of order one at all scales. 

\subsection{Conditional Couplings Parameters}
\label{energycouplsec}

We now introduce a model (or Ansatz) for conditional probabilities
$\overline p_j (\dphi_j | \aphi_j)$ with associated coupling parameters. This is at the core of our parametric representation of the conditional probabilities.
In the next section, we show
that the estimation of such conditional coupling parameters remains stable at phase transitions,
as opposed to coupling parameters $\theta_j$ of $p_j (\aphi_j)$. Moreover, we shall see that it defines
richer multiscale models of the microscopic probability distribution $p_0(\aphi_0)$. This is crucial for systems with long-range interactions such as cosmological fields.


\subsubsection{Scale Interactions}
For a Gaussian field, the conditional probability $\overline p_j (\dphi_j | \aphi_j)$ contains a quadratic interaction between 
$\dphi_j $ and $\aphi_j$. However, this is not the case for non-Gaussian processes, which often have non-linear dependencies across scales. 
To simplify notations, we shall write $\aphi_J = \dphi_{J+1}$.
Since $\aphi_j$ is specified by its orthogonal wavelet coefficients $\{ \dphi_\ell \}_{j < \ell \leq J+1}$, one finds:
\[
\overline p_j (\dphi_j | \aphi_j) = \overline p_j (\dphi_j |  \{ \dphi_\ell \}_{j < \ell \leq J+1} ). 
\]
Several studies have shown that
such conditional dependencies can indeed be represented with low-dimensional models 
taking advantage of the
sparsity of wavelet fields \cite{portilla2000parametric}.
For instance, wavelet scattering and phase harmonic models 
are conditioned 
by the correlation of the amplitude and the phase of $\dphi_j$ relatively
to $\dphi_\ell$ for $j < \ell \leq J+1$. They have been shown to
approximate well the distribution of complex physical fields such 
as fluid and gas turbulence \cite{bruna2019multiscale,zhang2021maximum}.  

\subsubsection{WC-RG Conditional Coupling Model}
We define a conditional probability model by approximating the scale interaction energy
$\bHH_j$ of $\overline p_j$ with a conditional coupling vector $\bK_j$
and a vector of basis functions $\bU_j$ such that
\begin{equation}
    \label{noiusdfaa0}
\bHH_{j} (\aphi_{j-1}) = \bK_j ^T \bU_j (\aphi_{j-1}) ~~\mbox{and}~~
\bU_j(P\aphi_{j-1}) = 0 . 
\end{equation}
The last requirement means that $\left.\bU_j(\aphi_{j-1})\right|_{\dphi_j = 0} = 0$. It guaranties that $\bHH_j(P\aphi_{j-1}) = 0$ and hence that the model taken for $\bHH_j$ contains the interactions between the wavelet field $\dphi_j$ and the coarse-grained field $\aphi_j$, and not the interactions of $\aphi_j$ with itself.

These models of scale interaction energies $\bHH_j$ define models of the microscopic energy $\HH_0$ through Eq.~(\ref{partionsdf9sdf9sdf00981}).
Free energies are defined by
\begin{equation}
  \label{freeEne2}
\bFF_j(\aphi_j)  = -  \log \int  e^{-\bK_j ^T \bU_j(\aphi_{j-1})} \, d \dphi_j 
+  c''_j, 
\end{equation}
where $c''_j$ is chosen such that $\bFF_j(0)=0$.
According to Eq.~(\ref{partionsdf9sdf9sdf00981}), we get
\begin{equation}
  \label{partionsdf9sdf9sdf009800}
\HH_0 = \K_J ^T \U_J + 
\sum_{j=1}^{J} \left(  \bK_j ^T \bU_j - \bFF_{j} \right),
\end{equation}
which proves that the microscopic energy function is entirely specified by the
WC-RG conditional coupling model, given by 
\begin{equation}
  \label{partio800}
\Big\{ \K_J\,,\, \bK_j \Big\}_{1 \leq j \leq J}  .
\end{equation}

These models are more general than the energy models discussed in Sec.~\ref{HmitAnsec}, because
they may include different local potentials at all scales, as we describe in more detail below.

\subsubsection{From local to multiscale  potential models}


Equation (\ref{partionsdf9sdf9sdf001}) implies that a model for the energy functions $\HH_j$ translates into a model for the scale interaction energies, with
\begin{equation}
    \label{noiusdfaa}
\bU_j(\aphi_{j-1}) = \U_{j-1}(\aphi_{j-1}) - \U_{j-1}(P \aphi_{j-1}) ,
\end{equation}
and 
\begin{equation}
    \label{noiusdfaa2}
\bK_j ^T \bU_j  = \K_{j-1} ^T \bU_{j}  ~.
\end{equation}
The conditional coupling parameter $\bK_j$ is uniquely defined as the orthogonal projection of the coupling vector $\K_{j-1}$ 
of $\HH_{j-1}$ over the space generated by the interaction potential vector $\bU_j$.

Let us consider models of $\HH_j$ with a single local potential at the scale $2^j$, 
defined by $\U_j (\aphi_j) = \left({\aphi_j \aphi_j^T} , V (\aphi_j)\right)$.
Appendix~\ref{interintsce} shows that the corresponding scale interaction energy model obtained from Eq.~(\ref{noiusdfaa}) can be written as
\begin{align}
\label{Ingssdcouprs}
\bK_j ^T \bU_j(\aphi_{j-1}) = & \frac{1}{2} \sum_{\ell=j}^{J+1} \dphi_j^T \bbK_{j,\ell}\, \dphi_{\ell} \\
\nonumber
& +  \overline C_j^T \Big(V (\aphi_{j-1}) - V (P \aphi_{j-1}) \Big), 
\end{align}
with a scale interaction potential,
\begin{equation}
    \label{interaposnit2}
\bU_j = \Big( \{ \dphi_{\ell} \dphi_j^T \}_{j \leq \ell \leq J+1}\,,\,V (\aphi_{j-1}) - V (P \aphi_{j-1}) \Big) ,
\end{equation}
and the associated coupling vector
\begin{equation}
\label{multiscouplinsdf}
\bK_j = \left( \frac{1}{2} \{ \bbK_{j,\ell} \}_{j \leq \ell \leq J+1}\,,\,  \overline C_j\right). 
\end{equation}
Each $\bbK_{j,\ell}$ is a convolution operator and hence depends upon the wave-number overlap between wavelets
at length scales $2^j$ and $2^{\ell}$. It is large for $\ell = j$, small for $\ell = j+1$, and negligible beyond if the wavelet is sufficiently well localized in the Fourier domain. 
We typically
only estimate $\bbK_{j,\ell}$ for $\ell = j$ and $j+1$ , and we set to zero all matrices for $\ell > j+1$ (see below).

Local scalar energy models $\HH_0$ have a single local potential at the finest scale $2^0$, as in the $\varphi^4$ field theory. Section \ref{free-energ-sec} shows that their expansion in Eq.~(\ref{partionsdf9sdf9sdf009800})
corresponds to particular cases where local potential parameters $\overline C_j$ cancel with
the free energy terms $\bFF_j$ at all scales $2^j$ but the finest and coarsest ones, thus leading indeed to a single local potential defined on the microscopic scale. However, 
general scale interaction models in Eq~(\ref{Ingssdcouprs}) may have arbitrary conditional coupling parameters $\overline C_j$ which do not disappear. The resulting microscopic energy $\HH_0$ then has
a different local potential at each scale, which can capture long-range interactions.

\subsection{Coarse to Fine WC-RG Sampling}
\label{Coafinsec}

We now discuss the first important outcome of WC-RG: sampling independent fields or configurations in an efficient way, which fully circumvents the problem of critical slowing down for second-order phase transitions and numerical instabilities associated with metastable states.     
In order to sample by WC-RG, one has to first determine the model of scale interaction energy functions by estimating the associated conditional coupling parameters in Eq.~(\ref{noiusdfaa0}) from the training dataset. This can be done efficiently (again in a way that is not affected by critical slowing down in the presence of long-range correlations) and is explained in the next section. Once the value of the coupling parameters has been learned, the WC-RG representation of the probability distribution is known as explained above.  

Sampling is then performed as follows. 
At a coarsest scale $2^J$, which is of the order of the system size $L$,
we draw a sample $\aphi_J$ of the probability distribution whose energy is $\HH_J = \K_J ^T \U_J$. This is done with an MCMC Metropolis sampling algorithm and is fast 
because $\aphi_J$ is reduced to a single point if
$2^J = L$ or a few grid points if $2^J$ is of the order of $L$. The corresponding distribution is simple: it is a Gaussian in the disordered phase of second-order phase transition (if $L$ is large enough).  
We then iteratively draw $\dphi_j$ given $\aphi_j$ by sampling with an MCMC Metropolis algorithm using
the conditional probability distribution in Eq.~(\ref{conosdf2}).
Note that we do not need to compute the free energy $\bFF_j$ because 
it is a normalization
factor and thus not needed in the MCMC algorithm. As we explain and show in the next sections, this sampling is also fast.
Indeed, the MCMC Metropolis algorithm used here has a decorrelation time that
does not increase near phase transitions. It does not suffer from any critical slowing down. In consequence, at scale $2^j$, the number of operations needed to draw a sample of $\dphi_j$ is just of the order of the number of degrees of freedom, i.e., $(2^d-1)(L2^{-j})^d$. We only update the highest wave-numbers through local interactions.
Since $2^{Jd}=L^d$, by summing up all scales from $J$ to $1$, we find that the number of MC steps
to sample a new configuration is proportional to
$\sum_{j=1}^J (2^d -1)(L 2^{-j})^d=L^d (1-2^{-Jd}) = \mathcal{O}(L^d)$.
The unit of time of an MC step is $L^d$.
The resulting characteristic MCMC time to sample using WC-RG is therefore
{\it independent} of the system size $L$, namely $\tau_{\rm MC} \sim \mathcal{O}(1)$, where $\tau_{\rm MC}$ is a MCMC decorrelation (or mixing) timescale.
This is an important improvement relative to state of the art. Indeed,
standard MCMC simulations for models with long-range correlations, such as Ising or $\aphi^4$ models at criticality, require an MCMC time $\tau_{\rm MC} \sim L^z$, where $z$ is a dynamic critical exponent ($z\simeq 2$ in two dimensions). More efficient cluster algorithms such as the Swendsen–Wang algorithm and the Wolff algorithm have an exponent $z \simeq 0.2-0.3$~\cite{swendsen1987nonuniversal,wolff1989comparison}. 
The WC-RG sampling is thus faster. It is done in "one-shot" because the number of operations needed to draw a sample is of the order of the number of degrees of freedom - the best possible in terms of computational complexity since to draw $L^d$ variables, one needs to do at least ${\mathcal O}(L^d)$ operations. It can be applied to generic systems, including {\it non-equilibrium} fields, as long as the conditional probability represents the underlying distribution well, whereas cluster algorithms are only used for simple systems such as the Ising model and its variants.

We note that various Monte-Carlo algorithms and techniques have been developed in order to achieve faster sampling, such as Hamilton Monte-Carlo~\cite{duane1987hybrid}, the multi-grid Monte-Carlo~\cite{goodman1989multigrid}, umbrella sampling~\cite{torrie1977nonphysical}, and parallel tempering~\cite{marinari1992simulated} to name but a few.
Hamilton Monte-Carlo~\cite{duane1987hybrid} uses proposal updates along the direction of the gradient of the Hamiltonian of the system instead of random updates used in the standard MCMC. Although it leads to collective movements, which might be more efficient than the standard MCMC in some cases, the algorithm still remains within the paradigm of physical dynamics, and hence does not lead to drastic improvements in terms of efficiency. 
The multi-grid Monte-Carlo methods~\cite{goodman1989multigrid}, which is a stochastic version of the multi-grid method in the field of numerical analysis, uses proposal updates that are 
performed in a scale-by-scale manner. They are not renormalization group algorithms because they do not use the conditional probability distribution at each scale, but directly project the microscopic energy on coarse-grained fields
\cite{goodman1989multigrid}. Although multi-grid Monte-Carlo can completely eliminate the critical slowing down for the Gaussian model, it still produces finite dynamic critical exponent for non-Gaussian systems. For example, the multi-grid MC method for the $\varphi^4$ field model shows the same exponent as the standard MC ($z \simeq 2$).

\section{Preconditioned Estimation of Conditional Couplings and absence of critical slowing down}
\label{precondHamilsec}
This section discusses the determination of the coupling parameters from the training dataset.
When inferring coupling vectors from data using the maximum likelihood estimation, there are two crucial difficulties that can make the method unstable and unfeasible. First, a dramatic slowing down of the gradient decent because we must use an extremely small step size in order to avoid instabilities. Second, a large increase of the decorrelation (or mixing) time $\tau_{\rm MC}$ of the MCMC, which must be computed at each gradient decent step. These two issues are actually facets of the same phenomenon originates from critical slowing down.

Let us focus on the case of second-order phase transitions (the $\varphi^4$ field theory in particular) to explain intuitively what the problem is, which in this context is related to the phenomenon called critical slowing down. Let us consider the setting in which the estimation is performed directly on the microscopic lattice, i.e., on the finest scale. We will call it a {\it direct coupling estimation} method. For instance, imagine estimating an energy function formed by two terms corresponding to a local potential and a discrete Laplacian, respectively, as in Eq.~(\ref{nsodufsbis0}).

Close to a critical point, a very tiny change in the coupling parameters induces a dramatic change in the spatial correlations of the fields. In consequence, in order to estimate an energy function that correctly reproduces the training data, the method has to be extremely precise, and hence a very small integration step in the gradient descent is required. At the critical point, the susceptibility to changes in the coupling parameters diverges, and hence the required step size vanishes, thus making the gradient descent unfeasible. 
Moreover, for each gradient descent step leading to the maximum likelihood, one has to perform an MCMC to evaluate the gradient term. If the estimated probability distribution is a faithful representation of the original one, as it should be if the method works, then the MCMC has a very long decorrelation time $\tau_{\rm MC}$ close to the second-order phase transition and diverges as a power law of $L$ at the critical point, $\tau_{\rm MC} \sim \mathcal{O}(L^z)$ with $z>0$. 

In summary, the long-range correlations that emerge near the second-order phase transition make a direct coupling estimation of the probability distribution problematic. In the following, 
we first show in detail why standard algorithms to infer the coupling vectors suffer from a critical slowing down near the critical point. We then present the theory which shows that WC-RG circumvents this problem and provides preconditioning of the maximum
likelihood gradient estimation and the MCMC, leading to much faster convergence, overcoming the critical slowing down completely.

\subsection{Direct coupling estimation by gradient descent}
\label{finegridestimat}

We estimate a model $p_{\K_j}= \ZZ_j^{-1} e^{-\K_j^T \U_j}$ of
$p_j$ by maximizing the likelihood $\langle \log p_{\K_j} \rangle_{p_j}$ \cite{wasserman2021}.
This is equivalent to minimize the Kullback L eibler (KL) divergence,
\[
\DKL (p_j \| p_{\K_j}) =  \int p_j(\varphi_j)\, \log p_j(\varphi_j)\, d\varphi_j- \langle \log p_{\K_j} \rangle_{p_j} ,
\]
which measures the information loss when approximating $p_j$ by $p_{\K_j}$.
The gradient of $\langle \log p_{\K_j} \rangle_{p_j}$ or $-\DKL (p_j \| p_{\K_j})$ relatively to $\K_j$ is
  \[
 \nabla_{{\K_j}} \langle \log p_{{\K_j}} \rangle_{p_j} = - \nabla_{{\K_j}} \DKL (p_j \| p_{\K_j})= 
 \langle \U_j  \rangle_{p_{\K_j} } - 
 \langle \U_j \rangle_{p_j},
 \]
where $\U_j$ is obtained from Eq.~(\ref{nsodufs70}):
\begin{equation}
\label{eq:mu_j}
\U_j (\aphi_j) = \Big({\aphi_j\, \aphi_j^T\,},\, \mu_j^{-1} V (\aphi_j)\Big),
\end{equation}
where $\mu_j$ is a normalization factor that will be determined below.
The optimal $\K_j^{\star}$, which satisfies $ \langle \U_j \rangle_{p_{\K_j^{\star}}} = \langle \U_j \rangle_{p_j}$, is searched by a gradient descent with a step size (or learning rate) $\epsilon$,
\begin{equation}
\label{gradse2}
 {\K}_{j}^{(\ttt+1)} - {\K}_{j}^{(\ttt)} = \epsilon \,\Big(
 \left\langle \U_j  \right\rangle_{p_{\K_{j}^{(\ttt)}}} - 
 \left\langle \U_j \right\rangle_{p_j} \Big), 
  \end{equation}
where $t$ represents discrete time step.
Since  $\DKL (p_j \| p_{\K_j})$ is a convex function of $\K_j$, as it can be easily checked, the gradient descent is guaranteed to converges if $\epsilon$ is smaller than the inverse of the largest eigenvalue 
of the Hessian of $\DKL (p_j \| p_{\K_j})$ relatively to $\K_j$.
The first right hand-side term $\left\langle \U_j  \right\rangle_{p_{\K_{j}^{(\ttt)}}}$
of Eq.~(\ref{gradse2})
is computed from the energy ${\K_j^T \U_j}$ of $p_{\K_{j}^{(\ttt)}}$ with an MCMC Metropolis algorithm. The
second term $\left\langle \U_j \right\rangle_{p_j}$ is estimated with an empirical
average of $U_j (\varphi_j)$ over the training dataset. 
Note that this coupling estimation by maximum likelihood or minimization of the KL divergence is equivalent to an application of the maximum entropy principle~\cite{zhu1997minimax,batou2013calculation}.

The above gradient descent algorithm is simple but it converges extremely slowly for multiscale processes or near critical points. The convergence rate of the gradient descent depends 
upon the condition number $\kappa$ of the Hessian $H_{\K_j}$ of $\DKL (p_j \| p_{\K_j})$ relatively to $\K_j$.
If the eigenvalues of $H_{\K_j}$ are between $\lambda_{\min}$ and $\lambda_{\max}$ then the gradient descent converges for $\epsilon \leq \lambda^{-1}_{\max}$ with a rate $\kappa^{-1} = \lambda_{\min} / \lambda_{\max}$. The convergence is thus very slow if $\kappa$ is large. 
A direct calculation shows that $H_{\K_j}$ is equal to the covariance of $\U_j$ 
  \begin{equation}
     \label{Hessian1}
 H_{\K_j}   = {\rm Cov} (\U_j)_{p_{\K_j}} =
 \langle \U_j \U_j^T \rangle_{p_{\K_j}} - \langle \U_j \rangle_{p_{\K_j}} \langle \U_j \rangle_{p_{\K_j}}^T . 
 \end{equation}
 When $p_{\K_j}$ becomes close to $p_j$ this Hessian becomes close to ${\rm Cov} (\U_j)_{p_j}$, which must be
 well conditioned to have a fast convergence rate.
 At phase transitions, $\aphi_0(i)$ is a stationary field whose correlation length 
 $\xi$ grows up to the system size $L$. As we shall explain below, this is the key reason responsible for the bad-conditioning of the Hessian. 
We determine the normalization factor $\mu_j$ in Eq.~(\ref{eq:mu_j}) such that ${\rm Cov}(\mu_j^{-1}V)_{p_j}$ is order one. To do so we fix $\mu_j$ by
 \begin{equation}
 \label{eq:normalization_mu_j}
     \mu_j^2 = {\rm Tr} \ {\rm Cov}(V)_{p_j} = \sum_{n=1}^s \left\{ \left\langle V_n^2 \right\rangle_{p_j} - \left\langle V_n \right\rangle_{p_j}^2  \right \}.
 \end{equation}
 The covariance of $U_j(\aphi_j) = ( \aphi_j \aphi_j^T , \mu_j^{-1} V(\aphi_j))$
 is an operator formed by different connected correlation functions (covariances)
 of $\aphi_j \aphi_j^T$, $V(\aphi_j)$ and between them. 
  Since 
 the probability distribution of $\aphi_0$ is invariant by translation (it is stationary), all covariances are also invariant by translation, and hence diagonal in Fourier. In order to understand intuitively why long-range correlations
 lead to very large maximum eigenvalues for these covariances, let us focus first on the covariance of the continuous field $\aphi(x)$ itself (this covariance does not appear in the Hessian, but it plays an essential role). The connection between continuous and discrete lattice fields, $\aphi(x)$ and $\aphi_j(i)$, is summarized in App.~\ref{app1}. Approaching the phase transition, this covariance  displays a power-law behavior in real space as \cite{cardy1996scaling}: $$\langle \aphi(x) \aphi(x')\rangle-\langle \aphi(x)\rangle \langle  \aphi(x')\rangle \sim \frac{1}{|x-x'|^{d-2+\eta}}$$ for 
 $1 \ll|x-x'|\ll \xi$, where $\xi$ is the correlation length.
Accordingly, the eigenvalue (Fourier transform) of the covariance associated with the 
field $\aphi$, $\lambda_\aphi(k)$,  has a power-law decay in Fourier space as $|k|^{-(2-\eta)}$ with $2-\eta>0$ for $2\pi/\xi \ll |k|$.  Approaching critical points, the
correlation length $\xi$ grows up to the order of the system size $L$. This leads to a very large ratio between minimum and maximum eigenvalue for the covariance of the field. 

The Hessian does not only contain the covariance of the field but also more complicated covariances between $\aphi \aphi^T$ and $V(\aphi)$. However, the same reasoning holds. The long-range correlation of the field $\aphi(x)$ also induces a power law decay for $1 \ll|x-x'|\ll \xi$ for these covariances, although with possibly different critical exponents \cite{cardy1996scaling}. As in the previous case, this behavior in real space leads to a power law decay in Fourier space for  $2\pi/\xi \ll |k| $. The blow-up at small $k$ produces a very large ratio $\kappa = \lambda_{\max} / \lambda_{\min}$ and a small $\lambda_{\max}^{-1}$, and thus a bad-conditioning. 
In the Gaussian case the computation can be done explicitly since the covariance of $U_j$ is only composed of  $4^{\rm th}$ order moments that can be computed using Wick theorem. This gives an explicit confirmation of the general argument explained above. 
 
The emergence of long-range correlations also leads to a very large decorrelation time for the MCMC. This effect, called critical slowing down~\cite{chaikin1995principles,sethna2021statistical}, is a consequence of spatial long-range correlations. An MCMC performing local updates has to break long-range correlated patterns in order to show decorrelation and hence reach ergodicity. This is further explained in Sec.~\ref{precondsec}.

In summary, long-range correlations in fields make the direct coupling estimation by gradient descent perform very poorly. In many cases, the requirement of a very small learning rate and very large decorrelation time is so stringent to make estimation practically unfeasible.

\subsection{Conditional Coupling Estimation}
\label{Hessian-ga}

The conditional coupling estimation for WC-RG is performed scale by scale, unlike what is done when estimating the full-probability distribution directly in the previous section (direct coupling estimation). One starts from the coarsest scale, $2^J$, of the order of the system size $L$, i.e., $2^J \approx L$. In this case, estimating the coupling parameters $\K_J$ of an energy model $\HH_J = \K_J ^T \U_J$ is easy and fast since $\aphi_J$ is reduced to a single or few values. It amounts to performing a low-dimensional estimation from a large training dataset. We then estimate all conditional probabilities $\overline p_j (\dphi_j | \aphi_j)$ needed to recover the probability distribution with Eq.~(\ref{teles}). 

For each given scale, the procedure is done as follows. First of all, the renormalization factor $\gamma_j$ in Eq.~(\ref{fastdec3}) is adjusted according to Eq.~(\ref{normalisation}).
In order to do this, the variances of wavelet fields are estimated from $R$ examples $\dphi_{j,r}$ calculated from
$R$ examples $\aphi_{0,r}$ of $\aphi_0$ in the dataset.
The parameter $\gamma_j$ is adjusted to normalize these variances. 
As we shall discuss in the next section, this renormalization has an important role in making the method well-conditioned. 

We wish to estimate the conditional probability model,
\begin{equation}
    \label{parmnosdfles0}
\overline p_{\bK_j} (\dphi_j | \aphi_j) = \frac{1}{\overline \ZZ_j}\, e^{- \bK_j ^T \bU_j (\aphi_{j-1}) + \bFF_j (\aphi_j)} 
\end{equation}
of $\overline p_j (\dphi_j | \aphi_j)$.
Given $p_j(\aphi_j)$, this conditional probability model
defines the following parameterized model of $p_{j-1} (\aphi_{j-1})$,
\begin{equation}
    \label{parmnosdfles}
p_{\bK_j}(\aphi_{j-1}) =  \overline p_{\bK_j}(\dphi_j | \aphi_j) \,p_j(\aphi_j).
\end{equation}
This model is optimized by maximizing
the log-likelihood $\lB\log p_{\bK_j} \rB_{p_{j-1}}$, which is equivalent to minimize
the Kullback-Leibler divergence $\DKL (p_{j-1} \| p_{\bK_j})$.
It is calculated with the same gradient descent algorithm as in 
the previous section, but we now only estimate high wave-vector conditional coupling parameters $\bK_j$. As we shall show later, this eliminates the 
ill-conditioning due to low wave-vectors.
The gradient of the log-likelihood or the KL divergence relatively to $\bK_j$ is 
\begin{eqnarray}
  \nabla_{\bK_j} \lB \log p_{\bK_j} \rB_{p_{j-1}} &=& - \nabla_{\bK_j} \DKL (p_{j-1} \| p_{\bK_j}) \nonumber \\ &=&
 \big\langle \bU_j   \big\rangle_{ p_{\bK_j}} - 
 \big\langle \bU_j  \big\rangle_{p_{j-1}}.
\nonumber
\end{eqnarray}
Therefore, the KL divergence gradient descent iteratively computes this gradient with a step size $\epsilon$:
 \begin{equation}
   \label{gradse20}
 \bK_j^{(\ttt+1)} - \bK_j^{(\ttt)} = \epsilon \,\Big( 
 \lB \bU_j  \rB_{p_{\bK_j^{(\ttt)}}} - 
 \lB \bU_j \rB_{p_{j-1}} \Big) .
  \end{equation}
 The second expected value $\lB \bU_j \rB_{p_{j-1}}$ is estimated with an empirical average
 of $\bU_j (\varphi_{j-1})$ over examples of $\varphi_{j-1}$ calculated from the training dataset. 
 %
For $\overline \theta_j = \overline \theta_j^{(\ttt)}$,
since $p_{\bK_j}(\aphi_{j-1}) =  \overline p_{\bK_j}(\dphi_j | \aphi_j)\,p_j(\aphi_j)$,
we compute samples $\varphi_{j,r}$ of $p_j$ by coarse graining
the samples $\aphi_{0,r}$ of $p_0$ in the dataset.
For each $\varphi_{j,r}$, a sample $\varphi_{j-1}$
of $p_{\bK_j}$ is calculated from 
a sample $\dphi_j$ of 
$\overline p_{\bK_j}(\dphi_j | \aphi_{j,r})$, with
\begin{equation}
    \label{phijsdf1}
\varphi_{j-1} = 
 \gamma_j \LL^T \aphi_{j, r} + \gamma_j \G^T \dphi_{j} .
\end{equation}
Samples $\dphi_j$ are calculated with an
  MCMC Metropolis algorithm, which performs random local updates of $\dphi_{j}$
  according to the interaction energy $\bK_j ^T \bU_j$
  of $\overline p_{\bK_j}(\dphi_j | \aphi_{j,r})$.
 We estimate $\lB \bU_j  \rB_{p_{\bK_j}}$ 
 by averaging $\bU_j (\aphi_{j-1})$ over the $\aphi_{j-1}$ in Eq. (\ref{phijsdf1}), obtained
  along all the MCMC chains which update $\dphi_j$ for all $\aphi_{j,r}$.
   As it will be shown in Sec.~\ref{numtheory} on the $\varphi^4$ model, the MCMC on wavelet fields have a fast mixing time (decorrelation) which is not affected by critical slowing down near phase transitions.
  
To accelerate computations, one can successively estimate the couplings from coarse to fine scales and initialize $\bK_j^{(\ttt)}$ for $\ttt = 0$ with the final gradient descent value $\bK_{j+1}^{(\infty)}$ estimated at the previous scale $2^{j+1}$. 

The gradient descent in Eq.~(\ref{gradse20}) converges if the gradient step $\epsilon$ is smaller than
the inverse of the largest eigenvalues of the Hessians $H_{\bK_j}$ of
$\DKL (p_{j-1} \| p_{\bK_j})$ relatively to $\bK_j$. 
Since $p_{\bK_j}(\aphi_{j-1})$ is factorized by Eq.~(\ref{parmnosdfles}) with the exponential conditional probability
distribution in Eq.~(\ref{parmnosdfles0}), one can verify that 
\begin{equation}
  H_{\bK_j} = \Big\langle \Cov(\bU_j)_{\overline p_{\bK_j}} \Big\rangle_{p_j} ,
  \label{h5}
\end{equation}
where $\Cov(\bU_j)_{\overline p_{\bK_j}}$ 
is the covariance of $\bU_j(\aphi_{j-1})$ relatively to $\overline p_{\bK_j}(\dphi_j | \aphi_j)$, given $\aphi_j$ from $p_j(\aphi_j)$.
The rate of convergence of the gradient descent in Eq.~(\ref{gradse2}) is equal to the inverse of the condition number of $H_{\bK_j}$.
As we shall show below, this Hessian is well conditioned even at phase transitions, so the
gradient descent does not suffer from critical slowing down. 

  
\subsection{Preconditioning by Renormalization}
\label{precondsec}

The estimation of coupling parameters and sampling a microscopic energy function
suffer from a "critical slowing down" near phase transitions. This section shows that there is
no such critical slowing down with a wavelet conditional renormalization group (WC-RG). 
The central result is a theorem for the Gaussian model which proves that although the covariance of the microscopic field $\aphi_0$ is
badly conditioned at phase transitions, 
the covariance of wavelet fields $\dphi_j$ remain uniformly well-conditioned at all scales. 
We then argue, based on this result, that WC-RG is well-conditioned and does not suffer from "critical slowing down". In the next section on numerical applications, we verify that this conclusion
indeed holds numerically for the two-dimensional Gaussian and $\varphi^4$ field theories and cosmological data. Constructing a fully rigorous proof is an open mathematical challenge that we leave for future works.

\subsubsection{Preconditioned Gradient descent}

As previously explained, the rate of convergence of the log-likelihood gradient descent
is defined by the condition number of the Hessian in Eq.~(\ref{h5}).
It computes the covariance of $\bU_j$ relatively to $\dphi_j$ conditioned by $\aphi_j$ 
according to $\overline p_{\bK_j}(\dphi_j|\varphi_j)$. This covariance is then 
averaged over $\aphi_j$ according to $p_j$.
The scale interaction potential $\bU_j $ that we use is given by
\begin{equation}
    \label{interaposnit3}
\bU_j = \Big( \{ \dphi_{\ell} \dphi_j^T \}_{j \leq \ell \leq j+1}\,,\,\nu_j^{-1} (V (\aphi_{j-1}) - V (P \aphi_{j-1})) \Big),
\end{equation}
where $\nu_j$ is an additional normalization factor.
The quadratic interactions are reduced to neighboring scales because
they are otherwise negligible, which amounts to replacing $J+1$ by $j+1$ in Eq.~(\ref{interaposnit2}). 

The Hessian in Eq.~(\ref{h5}) includes the cross-covariance of the matrices $ \dphi_{\ell} \dphi_j^T $  for $\ell = j,j+1$, the cross-covariance
 of $\dphi_{\ell} \dphi_j^T$ and 
$V(\aphi_{j-1})$, and the covariance of
$V(\aphi_{j-1})$. One can verify that the subtraction of $V (P \aphi_{j-1})$ 
does not modify the Hessian. 
We fix the the normalization constant $\nu_j$ similarly to  Eq.~(\ref{eq:normalization_mu_j}), by setting
\begin{eqnarray}
\nu_j^2 &=& 
{\rm Tr}\,  \Cov\Big(V(\aphi_{j-1}) - V(P\aphi_{j-1}) \Big)_{ p_{{j-1}}}.
\end{eqnarray}
As already discussed in the direct coupling estimation case, since the probability distribution is translation invariant, the covariances in eq. \ref{h5} are translation invariant and hence diagonal in Fourier space.
To understand why they do not have power-law decay, we first focus on the covariance of $\dphi_{j}$. We show that 
they are uniformly well-conditioned at all scales $2^j$, which plays a key role.

To understand the covariance properties of the wavelet fields $\dphi_j(m, i)$,
we relate wavelet fields defined over the discrete grid $i \in \Z^d$ to fields defined over continuous space $x \in \R^d$, 
decomposed over orthogonal wavelet bases. 
Appendix \ref{app1.2} explains that $\dphi_j$ are the decomposition coefficients 
of a field $\aphi(x)$ for $x \in \R^d$ in a wavelet orthonormal basis of the space of
finite energy functions of $x \in \R^d$.
The microscopic field $\aphi_0(i)$ on a discrete lattice is a projection of $\aphi(x)$ at a length scale $1$ (or $j=0$),
which amounts to setting to zero all wavelet coefficients at scales $2^j < 1$, and hence eliminate
the highest wave-numbers to sample the field at unit intervals. The microscopic scale $2^j = 1$ plays the role of
a cut-off scale. In order to analyze the wavelet field covariance at all scales $2^j$, we consider the asymptotic properties of the continuous
field $\aphi(x)$ over $\R^d$ (see App.~\ref{app1.2}). 
The following theorem shows that the covariance of the wavelet field $\dphi_j$ has a behavior very different from the one of $\varphi$. It states on a rigorous basis what it means for the wavelet field $\dphi_j$ to be a "fast degree of freedom".

\begin{theorem}
\label{Specsn}
Let $\aphi(x)$ be stationary field over $x \in \R^d$, whose covariance has eigenvalues
$\lambda_\aphi(k) = c\,|k|^{-\zeta}$ for $k \in \R^d$. 
If $\widehat G(k) = \sqrt{2} + \mathcal{O}(|k|^q)$ for $q \geq \zeta / 2$ then
there exists $A >0$ and $B$ such that for all $j \in \Z$ and $k \in [-\pi,\pi]^d$
\begin{equation}
    \label{condiont0}
A \leq \lambda_{\dphi_j}(k) \leq \,B .
\end{equation}
\end{theorem}

This theorem proves that the eigenvalues (i.e., the Fourier transform) $\lambda_{\dphi_j}$ of the covariance of $\dphi_j$  
vary by a bounded factor at all scales $2^j$, which means that in real space, the wavelet field has local fluctuations of order one with short-range correlation. 
There are two essential ingredients for the theorem to hold. 
The first one is to use 
wavelets which are sufficiently well localized in Fourier space, more precisely that the wavelets have a sufficiently large number $q$ of vanishing moments relatively to the exponent $\zeta$ associated with the phase transition, with $\zeta=2-\eta$ using the standard notation of phase transition literature. Appendix \ref{app1.2} explains that for this to hold, it is sufficient to impose that the wavelet filter $G$ has a compact support and a Fourier transform which satisfies 
$\widehat G(k) = \sqrt{2} + \mathcal{O}(|k|^q)$ with $q\geq \zeta / 2$.
The second important ingredient is that 
wavelet fields are normalized at each scale, i.e., $\dphi_j$ satisfies Eq.~(\ref{normalisation}). Conditional probabilities amount to slicing the estimation over multi-wave-number bands over 
which the scale interaction potentials have well-conditioned covariances. The wavelet field normalization has the flavor of a second-order method (Newton method), where the normalization of
the wavelet field plays a similar role as an inverse Hessian to adjust each gradient step.

The theorem proof is in App.~\ref{Theorem1Proof}. It is a particular case 
of a more general class of results concerning the representation of Calder\'on-Zygmund
operators in wavelet bases \cite{Meyer:92c}.
Such operators include differential and pseudo-differential operators.
This theorem considers a
phase transition where the covariance of $\aphi$ is a singular 
operator whose eigenvalues have a homogeneous and isotropic blow-up at low wave-numbers, as it happens at the critical point in second-order phase transitions \cite{cardy1996scaling}. 
For systems close to the critical point in which there is a finite but very large correlation length, a modification of the theorem proof leads to the same result in Eq.~(\ref{condiont0})
with constants that do not depend
upon $\xi$.
This theorem result can be extended to a wider class of covariances by only imposing that
$\lambda_\aphi(k)$ varies at most by a constant over any wave-number annulus,
$c_1 2^{-(j+1)} \leq |k| \leq c_2 2^{-j}$, where $c_1$ and $c_2$ are some constants.

To insure a fast convergence of the gradient descent in Eq.~(\ref{gradse20}), we must insure
the Hessian $H_{\overline \theta_j}$ in Eq.~(\ref{h5}) is well conditioned.
We give a qualitative argument which justifies this property,
but not a formal proof. The Hessian is equal to the covariance of
the scale interaction potential in Eq.~(\ref{interaposnit3}). 
For a direct coupling estimation, the long-range spatial correlations of $\varphi_0$ lead to bad conditioning. 
The Hessian associated to the WC-RG depends upon covariances relatively to fluctuations
of the wavelet fields $\dphi_j$, conditioned and
averaged over $\aphi_j$. Since the $\dphi_j$ have a short-range correlation and
the scale interaction energies $\overline U_j$ are local, the resulting covariances are also short-ranged. Moreover, all terms in $\overline U_j$ have fluctuations of order $1$ because we normalized the variance of $\dphi_{j}$ and of the potential term in $\overline U_j$. Local interactions of terms whose
fluctuations are of the order of one have
covariance matrices whose eigenvalues remain of the order of one. They do not blow up at small $k$ and remain of order one for all $k$. The
resulting Hessian is therefore well-conditioned. These are the reasons why WC-RG is a well-conditioned method.

Turning these arguments into a rigorous proof is an open mathematical challenge that we leave for future works. 
The first difficulties are related to the non-linear potential functions, although this seems within reach when expressed as polynomials because they transform local wavelets into localized functions.
Another difficulty is to prove that averaging covariances conditioned by $\aphi_j$
does not affect their condition number. It requires to further decompose
$\aphi_j$ into normalised larger scale wavelet fields $\dphi_\ell$ for $\ell > j$ and
evaluate cross-correlations with the terms in $\dphi_j$.
The proof of the purely Gaussian models relies on the fact that the homogeneous singular operators
whose Fourier eigenvalues   $\lambda_\aphi(k)$ have a power-law decay like $\left(|k|^{\zeta} + (2\pi/\xi)^\zeta \right)^{-1}$ 
are fully preconditioned by diagonal terms in a wavelet basis \cite{Meyer:92c}, not just scale per scale. In other words, the symmetric covariance matrix
of all normalized wavelet fields $\{ \dphi_j \}_{1 \leq j \leq J+1}$,
even when taking into accounts interactions across scales, has eigenvalues that remain of the
order of $1$. In this Gaussian case, the Hessian
$H_{\overline \theta_j}$ involves $4^{th}$ order moments which
can be explicitly related to this full covariance matrix. The
log-likelihood gradient descent has an exponential convergence with no critical slowing down.

\subsubsection{Preconditioned MCMC and Langevin dynamics}
\label{sec:Langevin}


In the previous section, we showed that if the Hessian $H_{\overline{\theta}_j}$ is well-conditioned, then the gradient descent 
converges on times of the order of one. The condition on the Hessian requires analyzing a specific set of correlation functions. 
The requirements on correlation functions to guarantee a short decorrelation time
of the MCMC performed on $\dphi_j$ conditioned by $\aphi_j$ are much more restrictive. 
As proven in Ref.~\cite{montanari2006rigorous}, 
a necessary and sufficient condition for the decorrelation time of MCMC to be finite (not diverging with the system size) is that the so-called point-to-set length for the field $\dphi_j$ conditioned by a typical $\aphi_j$ is finite
\cite{montanari2006rigorous}. Physically, this requirement imposes that {\it all possible spatial connected correlation functions} (for the measure $\overline p_{\bK_j}(\dphi_j|\varphi_j)$) have a short
spatial range (and not just a specific set as for the Hessian). Only in this case, one can be sure that $\dphi_j$ conditioned by a typical $\aphi_j$ is non-critical with respect to any kind of ordering, and hence that the MCMC decorrelates quickly. 

Although a general proof for WC-RG is a challenge, RG approaches to dynamical critical 
phenomena \cite{hohenberg1977theory} show that this result holds for the $\varphi^4$ model, and more generally at critical points. These techniques are well studied in theoretical physics and considered to be fully under control, but they are not rigorous. In fact, it is at the core of the dynamical RG procedure that the short-scale degrees of freedom (the wavelet field in our case) are short-ranged and relax quickly at each scale, hence the name "fast-degrees of freedom" as shown by perturbative methods (in $d_U-d$, where $d_U$ is the upper critical dimensions) \cite{hohenberg1977theory} and by 
non-perturbative approximations \cite{canet2007non}. 
As in the previous case, the Gaussian field theory provides a framework where these results can be worked out in detail, even rigorously. In this case, the field $\dphi_j$ is short-ranged if all eigenvalues $\lambda_{\dphi_j}$
of its covariance is of the order of $1$.
Theorem \ref{Specsn} proves that this is true even for critical Gaussian fields such as Ornstein-Uhlenbeck process and Fractional Brownian motion.
Therefore, even if the MCMC for the critical field has a decorrelation time $\tau_{\rm MC}$ that diverges as $L^{z}$ with $z=\zeta$, the MCMC for the field $\dphi_j$ conditioned by a typical $\aphi_j$ converges on times of the order of one. 
 
One can directly obtain this result by considering Langevin dynamics for the field $\dphi_j$. It is known physically and rigorously \cite{andreanov2006field,gelman1997weak} that MCMC tends to Langevin in the continuum time limit. Let us then focus on the latter since the analysis is more straightforward. 
For simplicity, we focus on $d=1$, where only one wavelet channel exists.
The Langevin dynamics associated to the field $\dphi_j$ is a multi-dimensional Ornstein-Uhlenbeck process:
\begin{equation}\label{lang}
   \frac{\partial  \dphi_j(i, t)}{\partial t}=-\sum_{i'} \overline K_{j,j} (i-i') \dphi_j(i', t)+\xi(i,t), 
\end{equation}
where $\xi(i,t)$ are i.i.d. Gaussian white noise of variance equal to two. The matrix $\overline K_{j,j}$ (see Eq.~(\ref{DecompAnA})) is diagonal in Fourier space. As a consequence, the above set of Langevin equations decouple in Fourier space. For each Fourier component $k$, we obtain an independent Ornstein-Uhlenbeck process whose decorrelation time is the inverse of the eigenvalues of $\overline K_{j,j}$, which are equal to the
eigenvalues $\lambda_{\dphi_j}$ of the covariance of $\dphi_j$, namely, ${\widehat {\overline K}}_{j,j}^{-1}(k) = \lambda_{\dphi_j}$ (where the hat denotes the Fourier transform). Theorem \ref{Specsn} proves that all these eigenvalues are of the order of $1$ and hence that 
the normalised Langevin process in Eq.~(\ref{lang}) converges to the equilibrium stationary measure on times of the order of one. 

In conclusion, the crucial condition that makes RG work is that
” fast degrees of freedom” are not critical, i.e.,
without long-range correlations in space and time. It 
is the same key requirement for WC-RG, thus strengthening
the connection between the standard RG framework and WC-RG.

\section{Recovering the Microscopic energy}
\label{Recovermicrosec}

In the previous sections, we showed how to estimate the multiscale representation of the probability distribution $p_0(\aphi_0)$ in Eq.~(\ref{teles}), allowing us to perform {\it sampling} and obtain new data. 
In this section, we show how to recover {\it the microscopic energy function} $E_0(\aphi_0)$ with a well-conditioned method. 
This is a second important outcome of the WC-RG, which can be useful for obtaining essential physical information, and which is out of reach of many other generative models. It is done by computing a functional representation of the scale interaction free energies at each scale.

\subsection{Scale Interaction Free-energy Models}
\label{free-energ-sec}
In order to compute the microscopic energy in Eq.~(\ref{partionsdf9sdf9sdf009800}), we must
construct a functional representation of scale interaction free energy $\bFF_j(\varphi_j)$ defined in Eq.~(\ref{freeEne2}).
Modeling
$\bFF_j(\aphi_j)$ amounts to define another basis functions $\widetilde \U_j(\aphi_j)$ and the associated coupling vector $\widetilde \K_j$ such that $\widetilde \K_j ^T \widetilde \U_j$ provides an accurate approximation of $\bFF_j(\aphi_j)$ up to an additive constant:
\begin{equation}
    \bFF_j(\aphi_j) = \widetilde \K_j ^T \widetilde \U_j(\aphi_j) + \tilde c_j . 
\end{equation}
We will show that it defines multiscale potential models of the microscopic energy function $\HH_0(\aphi_0)$, which are more general than local microscopic potential models. 

Replacing each $\bFF_j$ by $\widetilde \K_j ^T \wU_j + \tilde c_j$ in Eq.~(\ref{partionsdf9sdf9sdf009800}) gives
\begin{equation}
\label{minsdfasd}
\HH_0 = \K_J ^T \U_J + 
\sum_{j=1}^{J} \left(  \bK_j ^T \bU_j - \widetilde \K_j ^T \wU_j \right) + c_0,
\end{equation}
where $c_0$ is a constant.
We consider the scale interaction energy model 
in Eq.~(\ref{Ingssdcouprs}), 
\begin{align}
\label{Ingssdcouprs22}
\bK_j ^T \bU_j (\aphi_{j-1}) = & \frac{1}{2} \sum_{\ell=j}^{J+1} \dphi_j^T \bbK_{j,\ell}\, \dphi_{\ell} \\
\nonumber
& +  \overline C_j^T \Big(V (\aphi_{j-1}) - V (P \aphi_{j-1}) \Big).
\end{align}
The normalization constant $\mu_j$ introduced in Eq.~(\ref{interaposnit3}) is absorbed into
$\overline C_j$ for simplicity.
The term $\overline C_j^T V (P \aphi_{j-1})$
is not modified by the free energy integration in Eq.~(\ref{freeEne2}), because it does not depend on $\dphi_j$. We thus define a model of $\bFF_j(\aphi_j)$ which includes this term 
\begin{equation}
\label{Ingssdcouprs23}
{\widetilde \K_j} ^T \wU_j(\aphi_{j}) = \frac{1}{2} \aphi_j^T \ \widetilde K_j \aphi_j
+  \widetilde C_j^T V (\aphi_{j}) - \overline C_j^T V (P \aphi_{j-1}) ,
\end{equation}
with $\widetilde \K_j = \left(\frac{1}{2} \widetilde K_j \,,\,\widetilde C_j \right)$. 
It results from the expansion in Eq.~(\ref{minsdfasd}) that $\HH_0$ can be written by
\begin{equation}
\label{Ingssdcouprs223}
\HH_0 (\aphi_0) =  \frac{1}{2}\aphi_0^T K_0 \aphi_0 +  \sum_{j=0}^J\, {C'_j}^T V (\aphi_j),
\end{equation}
where $C'_0 = \overline C_1$, $C'_J = C_J - \widetilde C_J$  and for $1 \leq j \leq J-1$,
\begin{equation}
C'_j = \overline C_{j+1} - \widetilde C_j~,
\label{eq:C'_j}    
\end{equation}
with
\begin{equation}
\aphi_0^T K_0 \aphi_0  = \aphi_J^T K_J \aphi_J + \sum_{j=1}^J \Big(
\sum_{\ell=j}^{J+1} \dphi_j^T \bbK_{j,\ell}\, \dphi_{\ell}
- \aphi_{j}^T \ \widetilde K_j \aphi_j \Big) . 
\label{eq:reconstruction_gaussian}
\end{equation}
The multiscale coupling parameters ${C'_j}^T V (\aphi_j)$ give rise to {\it long-range potential interactions}.
Local microscopic energy models correspond to a particular case where 
$K_0$ is a convolution operator with a small support, and $C'_j = 0$ for all $1 \leq j \leq J$,
hence $\widetilde C_j = \overline C_{j+1}$, and $ \widetilde C_J=C_J$.
Such models are sufficient
to describe the $\varphi^4$ field theory. Instead, for cosmological weak lensing fields, we shall see that $C'_j \neq 0$ for all $0 \leq j \leq J$. In this case,
the microscopic energy includes local potentials at all scales, 
producing long-range interactions at the microscopic scale.

\subsection{Free Energy Regression}
\label{FreeEnerCalc}

\begin{table*}
\label{tab:summary}
\caption{Summary of multiscale WC-RG estimation. It decomposes the microscopic probability
$p_0 (\aphi_0) = \alpha\, p_J (\aphi_J)\, \prod_{j=1}^J \overline p_j (\dphi_{j} | \aphi_j)$
 and its energy function $\HH_0(\aphi_0) = 
\K_J ^T \U_J(\varphi_J) + 
\sum_{j=1}^{J} \left(  \bK_j ^T \bU_j(\varphi_{j-1}) - \widetilde \K_j ^T \wU_j(\varphi_j) \right) + c_0$. The coupling parameters $\bK_j$ of $\overline p_j$ are estimated
by minimizing a KL divergence. The parameters $\widetilde \theta_j$ of $\overline F_j$
are computed by linear regression.
}
\begin{ruledtabular}
\begin{tabular}{ccc}
 Estimation Method & Parameterized Models  & Inference
 \\ \hline \\ \vspace{0.2cm}
 Direct estimation of $p_0(\aphi_0)$ & $p_{\K_0}(\varphi_0)= \ZZ_0^{-1} \exp\left[-\K_0^T \U_0(\varphi_0)\right]$ &  $\displaystyle \min_{\K_0}  \DKL (p_0 \| p_{\K_0}) $ \\
\vspace{0.2cm}

 Conditional estimation of $\overline p_j (\dphi_j | \aphi_j)$ & $ \overline p_{\bK_j} (\dphi_j | \aphi_j) = \overline \ZZ_j^{-1} \exp\left[- \bK_j ^T \bU_j (\aphi_{j-1}) + \bFF_j (\aphi_j)\right]$ & $\displaystyle \min_{\bK_j} \DKL (p_{j-1} \| \overline p_{\bK_j}p_j) $ \\
\vspace{0.2cm}

 Estimation of free energies $\bFF_j(\varphi_j)$ & $\widetilde \K_j ^T \widetilde \U_j(\varphi_j)$ & $\displaystyle \min_{\widetilde \K_j} \Big\langle \left( \bFF_j  - \widetilde \K_j ^T \wU_j \right)^2  \Big\rangle_{p_j}  $ \\

\end{tabular}
\end{ruledtabular}
\end{table*}

A coupling model $\widetilde \K_j^T \wU_j$ of the scale interaction
free energy $\bFF_j$ is calculated
by minimizing the mean-square error,
\begin{equation}
\label{menasqaloss}
\ell (\widetilde \K_j) = \Big\langle \left( \bFF_j  - \widetilde \K_j ^T \wU_j \right)^2  \Big\rangle_{p_j} .
\end{equation}
This section together with App.~\ref{app:freeenergy}
explains how to estimate 
the coupling vector $\widetilde \K_j$,
from $R$ samples $\varphi_{0,r}$ of $p_0$, with a thermodynamic 
integration~\cite{frenkel2001understanding}. 
The free energy model is validated by verifying that the regression error is small:
$\ell(\widetilde \K_j) \ll \langle \bFF_j^2 \rangle_{p_j}$. 

The mean-square loss $\ell (\widetilde \K_j)$ is minimized for 
\begin{equation}
\label{thermoint0}
\widetilde \K_j^{\star} = \langle \wU_j \, \wU_j^T \rangle_{p_j}^{-1} \, \langle \bFF_j \, \wU_j \rangle_{p_j} ,
\end{equation}
and the resulting error is
\begin{equation}
\label{thermoint01}
{\ell (\widetilde \K_j^{\star})} = {\langle \bFF_j^2 \rangle_{p_j} } -
{\wK_j^{\star T} \langle \bFF_j \, \wU_j \rangle_{p_j}}  .
\end{equation}
To estimate the expected values, 
we compute $R$ samples $\aphi_{j,r}$ of $p_j$ by coarse graining
$R$ samples $\aphi_{0,r}$ in the training dataset.
We then estimate $\langle \wU_j\, \wU_j^T \rangle_{p_j}$ with an empirical average over all $\aphi_{j,r}$. 

The main difficulty is to estimate ${\langle \bFF_j^2 \rangle_{p_j} } $ and $\langle \bFF_j \, \wU_j \rangle_{p_j}$, where
$\bFF_j = - \log  \int e^{-\bK_j ^T \bU_j}\, d \dphi_j$ up to an additive constant.
We evaluate these moments with a thermodynamic integration \cite{frenkel2001understanding}. This is done by
introducing a family of
models $\bK_{j}^T \bU_{j,\la}$, which performs an interpolation between a Gaussian model and
the full scale interaction energy model. For this purpose,
the non-linear potential term in Eq.~(\ref{Ingssdcouprs22}) is multiplied by $\lambda \in [0,1]$:
\begin{equation}
    \label{interuqnadfimso}
\bK_{j}^T \bU_{j,\la} (\aphi_{j-1}) =  Q_{\aphi_j} (\dphi_j) + \lambda \, \overline C_j^T \,V (\aphi_{j-1})~, 
\end{equation}
with a first term which is quadratic in $\dphi_j$ for fixed $\aphi_j$,
\begin{equation}
    \label{interuqnadfimso2}
Q_{\aphi_j} (\dphi_j) = 
\frac{1}{2} \sum_{\ell=j}^{J+1} \dphi_j^T \bbK_{j,\ell}\, \dphi_{\ell} - \overline C_j^T \,V (P \aphi_{j-1}) .
\end{equation}
The thermodynamic integration interpolates linearly the Gaussian measure ($\lambda=0$) and the original one ($\lambda=1$), by estimating expected values with empirical averages.
Appendix \ref{app:freeenergy} shows that it involves
MCMC calculations on $\dphi_j$ for fixed $\aphi_j$, which is fast and not affected by the critical slowing down. As explained in the previous section, it mostly results from 
Theorem \ref{Specsn} which proves that wavelet fields
$\dphi_j$ have covariances that are well-conditioned. 

In summary, the free-energy estimation is made jointly by thermodynamic integration and linear regression. This method is numerically stable even close to critical points because it only requires sampling the
wavelet field, keeping the coarse-grained field frozen. The numerical experiments in Sec.~\ref{numericalSec} over cosmological data show that
small errors introduced by the free energy estimation induce small errors on the microscopic
energy $\HH_0$. However, these small microscopic energy
errors may lead to large errors when sampling the probability 
distribution model $p_0$ using $\HH_0$. This is due to the instabilities
of microscopic coupling parametrizations of $\HH_0$. For instance, it is inherent to the phenomenon of phase transition that a small change in the parameters of the energy (the coupling constants) can change the physical behavior drastically. WC-RG circumvents this problem by generating new samples using the representation in Eq.~(\ref{teles}), it avoids the instabilities by using the parametrization based on conditional coupling parameters $\bK_j$. We will discuss this issue further when applying WC-RG to cosmological data. 

Table~\ref{tab:summary} summarizes the different steps to compute the microscopic energy function, with a direct approach or with a WC-RG. A WC-RG estimates
the coupling parameters $\overline \theta_j$
of $\overline p_j$, and the parameters 
$\widetilde \theta_j$ of the free energy $\overline F_j$ at all scales $2^j$.

\section{Numerical Applications of WC-RG: Sampling, Estimation of Energy Function, and Absence of Critical Slowing Down}
\label{numericalSec}

This section presents three different numerical applications of WC-RG on two-dimensional ($d=2$) fields. For each example, we first show the results of WC-RG sampling and then the recovery of microscopic energy functions. 

We start with a Gaussian field theory associated with the Ornstein-Uhlenbeck stochastic process. It provides a simple illustration and a validation test of our numerical method over a system with long-range correlated fields. As a second application, we focus on the $\aphi^4$-model, which is central in the theory of phase transitions and provides a challenging estimation problem with non-linear potential. The third application concerns the characterization of weak-gravitational-lensing maps in cosmology associated with the distribution of dark matter. In this last example, the field is inherently {\it out-of-equilibrium}, and there is no explicit energy model to compare with. However, we show that the resulting model reproduces visually and statistically nearly identical fields. This highly non-trivial example shows the potentiality of our method.
The algorithmic aspects of our implementation of WC-RG  are in App.~\ref{app:numerical_methods}. 

\subsection{Gaussian Field Theory}
\label{Gaussiansec}

We consider a model in which the field $\aphi _0$ is Gaussian at the finest scale, and thus has
an energy function, 
\[\HH_0 (\aphi _0) =\frac 1 2 \, \aphi _0^T K_0  \aphi _0, \]
without non-linear potential energy term (i.e., $V_0 (\aphi _0) = 0$). In particular, $K_0$ is given by
\begin{equation}\label{geq}
    K_0 = \alpha \, Id - \beta \Delta ~~\mbox{with}~~ \beta > 0,
\end{equation}
where $\Delta$ is a discretized approximation of the Laplacian, as in Eq.~(\ref{discretizedlaplacian}).
The corresponding Gibbs-Measure is the equilibrium measure of the Ornstein-Uhlenbeck stochastic process on the field $\aphi _0$. 
The convolutional matrix $K_0$ is the inverse of the covariance of $\aphi_0$,
which is diagonal in Fourier space with eigenvalues,
\begin{equation}
\lambda_{\aphi_0} (k) = \widehat K_0^{-1}(k) = \frac{1}{\beta (|\om|^2 + \alpha/\beta )} ,
\end{equation}
where $\om$ is the Fourier wave-vector. In consequence, 
the ratio $\alpha/\beta$ defines a correlation length $\xi=2 \pi (\alpha/\beta)^{-\frac{1}{2}}$, which is the control parameter specifying the model. We normalize the overall factor $\beta$ such that $\langle |\aphi_{0}(i)|^2 \rangle_{p_0}=1$.
The field $\aphi_0$ displays critical fluctuations for $|\om| > 2\pi/\xi$, and thus for length-scale smaller than $\xi$.

\begin{figure}
\includegraphics[width=0.48\linewidth]{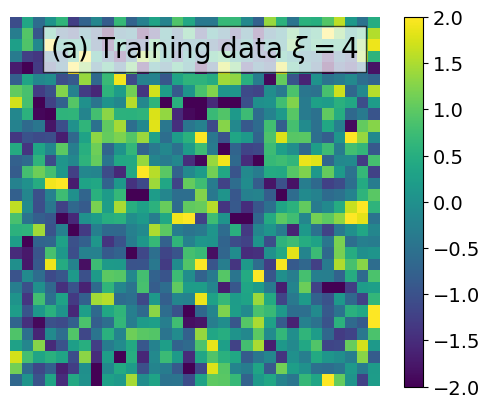}
\includegraphics[width=0.48\linewidth]{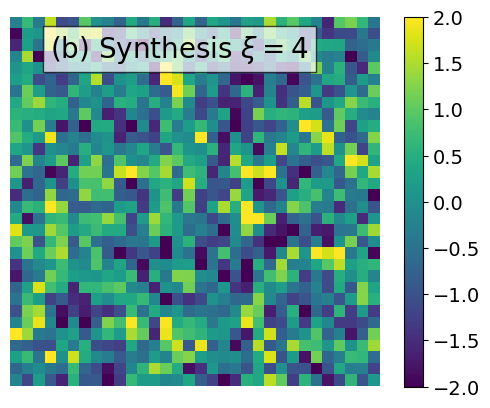}
\includegraphics[width=0.48\linewidth]{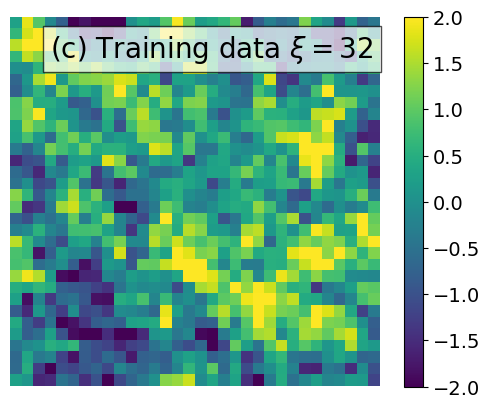}
\includegraphics[width=0.48\linewidth]{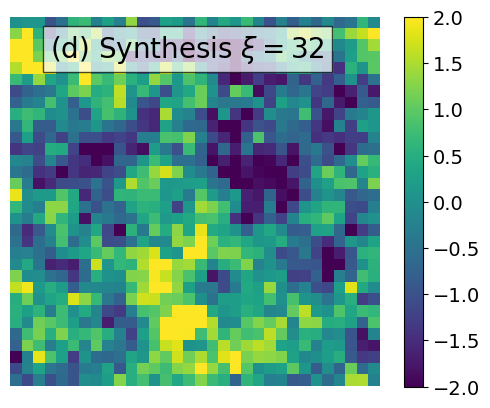}
\caption{
(a, c): Training samples of the Ornstein-Uhenbeck process for $\xi=4$ (a) and $\xi=32$ (c).
(b, d): Synthetized fields generated by WC-RG sampling for $\xi=4$ (b) and $\xi=32$ (d). 
}
\label{fig:OU_images}
\end{figure}

\begin{figure}
\includegraphics[width=0.75\linewidth]{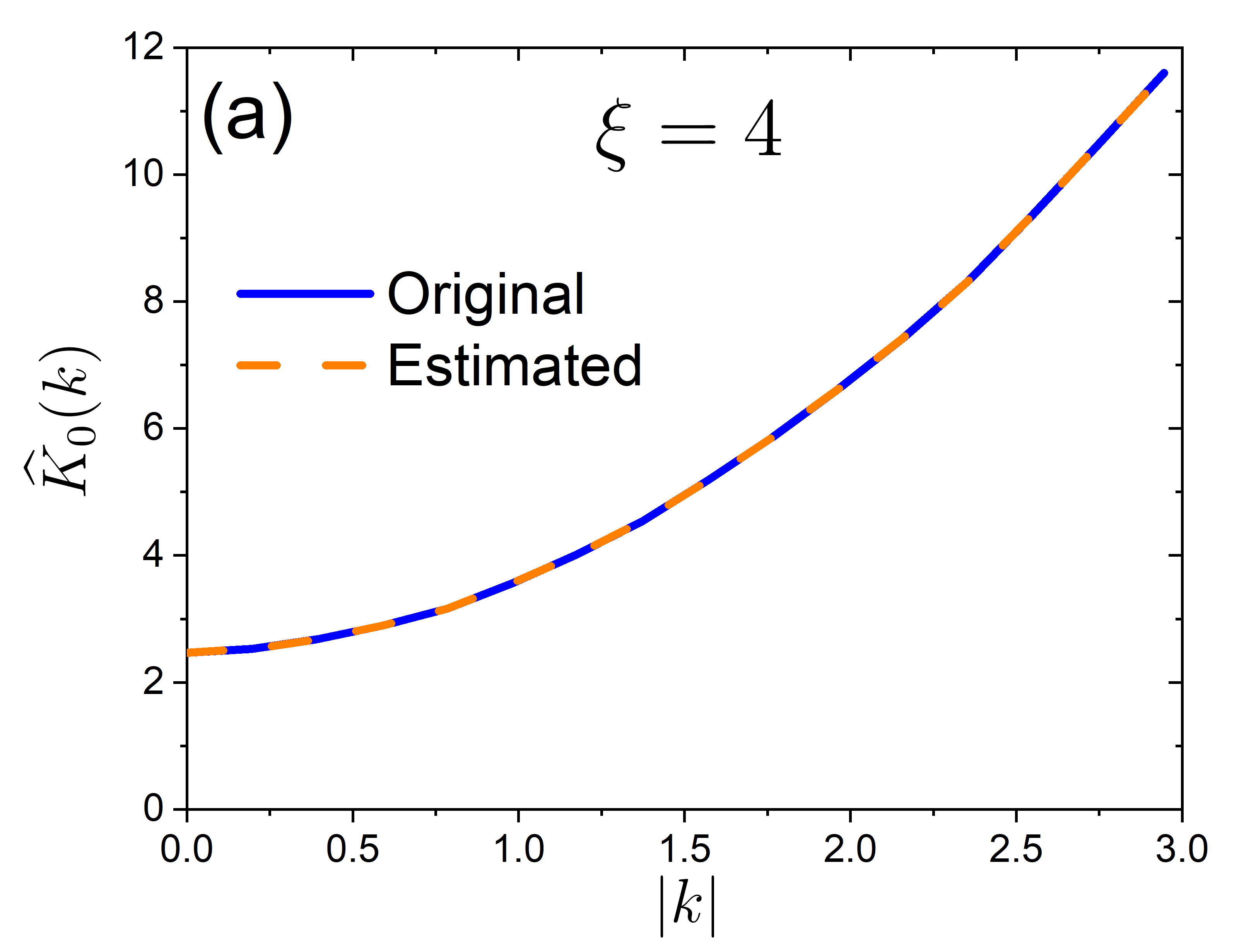}
\includegraphics[width=0.75\linewidth]{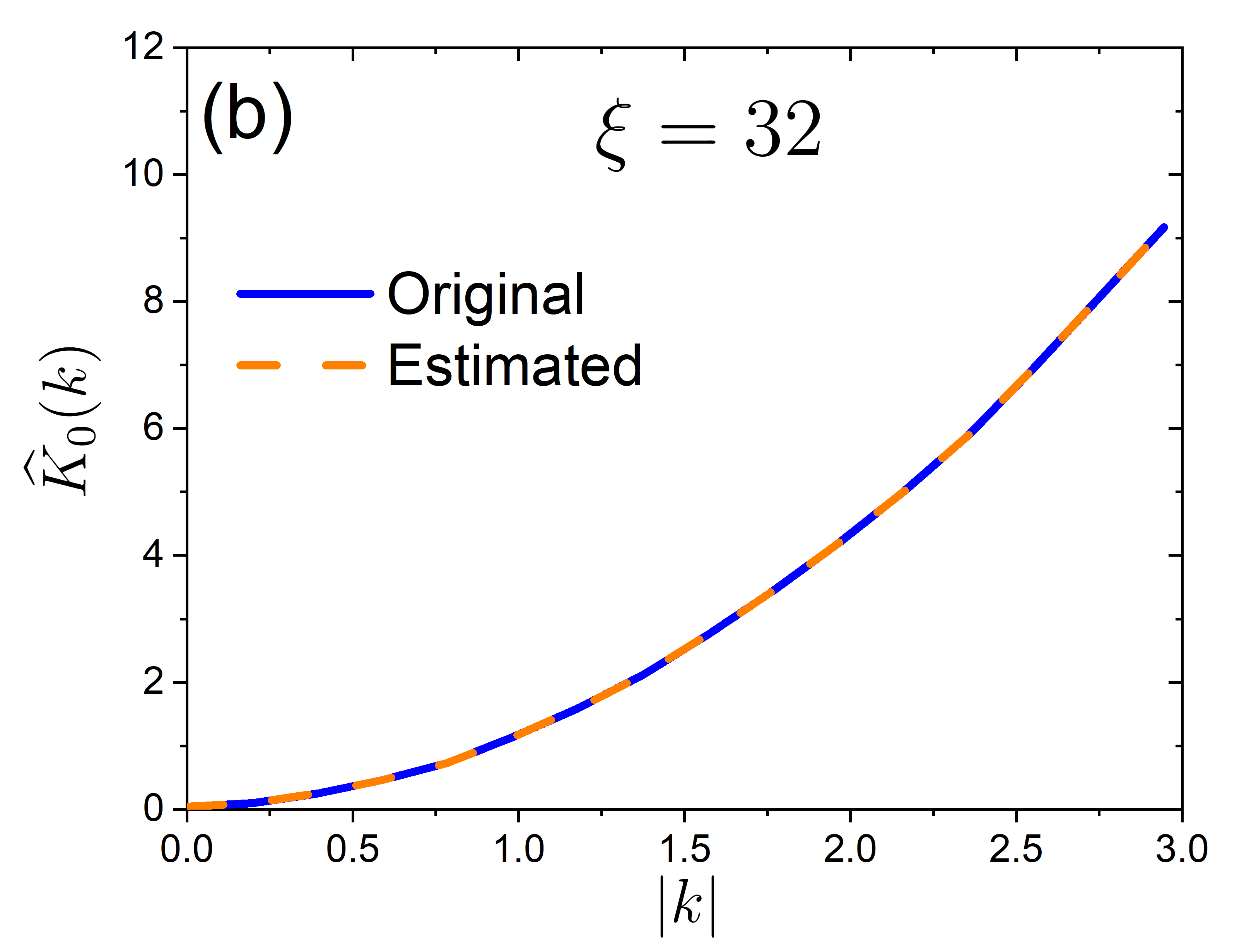}
\caption{Fourier eigenvalues $\widehat K_0(k)$ of the Ornstein-Uhlenbeck kernel $K_0$, radially averaged over constant Fourier wave-vector modulus $|k|$, for 
$\xi=4$ in (a) and $\xi=32$ in (b).
Original eigenvalues are shown with a solid curve and the estimated ones with a dashed curve.}
\label{fig:Gaussian-laplacian}
\end{figure}

We now present results obtained with WC-RG models and sampling of this Gaussian process.
Figures~\ref{fig:OU_images}(a, c) show typical samples of the Ornstein-Uhlenbeck process in Eq.~(\ref{geq}) of size $L = 32$, for $\xi = 4$ and
$\xi = 32$ (training data). 
These are short-ranged and long-ranged correlated fields, respectively.  
The values $\xi=L=32$ correspond to the critical point of the system.
Figures~\ref{fig:OU_images}(b, d) shows two 
samples synthesized by
the WC-RG coarse to fine sampling, for the same values of $\xi$. 
These visual textures can not be discriminated from the ones in Figs.~\ref{fig:OU_images}(a, c). Appendix \ref{app:gaussian_model} gives more details on these simulations. 

For a Gaussian process, we evaluate the accuracy of the WC-RG model by computing the precision of the estimated microscopic energy kernel $K_0$.
Figure~\ref{fig:Gaussian-laplacian} compares the (radially averaged) Fourier eigenvalues 
$\widehat K_0(k)$ of $K_0$ for the original and the estimated model by WC-RG, 
for $\xi = 4$ and $\xi = 32$. They perfectly superimpose because
the conditional probability models are exact at all scales, since the energy functions remain Gaussian. Appendix~\ref{app:gaussian_model} explains the implementation. It demonstrates that WC-RG calculations have no critical slowing down in presence of long range correlations $\xi = L$.

\subsection{The $\aphi^4$ Field-Theory}

The $\aphi^4$ field theory is the simplest model, which contains all the key ingredients 
of standard second-order phase transitions, such as large-scale collective behaviors, critical properties, long-range correlations, and self-similarity at the critical point. It has also played a central role in testing new techniques and ideas \cite{berges2002non}. 
Here, we follow the same strategy and apply the WC-RG method to the $\aphi^4$ field theory in two dimensions. 

The microscopic $\aphi^4$ model on a discrete lattice has a local potential introduced in Eq.~(\ref{nsodufsbis01}),
\begin{equation}
    \label{phi4eneransfd}
\HH_0 (\aphi _0) =\frac 1 2 \, \aphi _0^T K_0  \aphi _0, + C_0^T\,V(\aphi_0) .
\end{equation}
The quadratic kernel is $K_0 = -\beta\, \Delta$ where $\beta$ plays a role of inverse temperature.
and $\Delta$ is the Laplacian.
The local potential $C_0^T V(\varphi_0) = \sum_i C_0^T v(\varphi_0(i))$ has a
double well 
\begin{equation}
    \label{phi4eneransfd2}
C_0^T v (\varphi_0(i)) = \aphi^4_0(i) - (1+2\beta) \aphi^2_0(i) .
\end{equation}
Previous numerical work has shown the existence of a second-order 
phase transition in the thermodynamics limit at $\beta_c\simeq 0.67$~\cite{kaupuvzs2016corrections}: for $\beta < \beta_c$ the system is disordered, whereas for $\beta > \beta_c$ the system is in the ordered, or broken symmetry, phase.
\begin{figure}
\noindent
\includegraphics[width=0.48\linewidth]{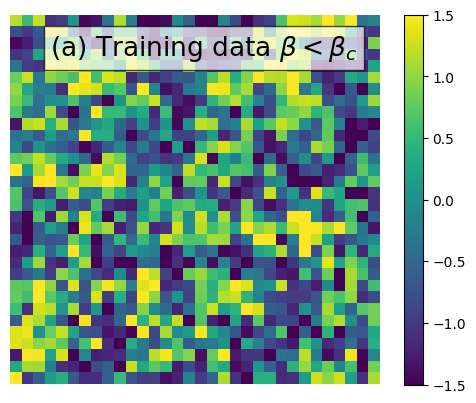}
\includegraphics[width=0.48\linewidth]{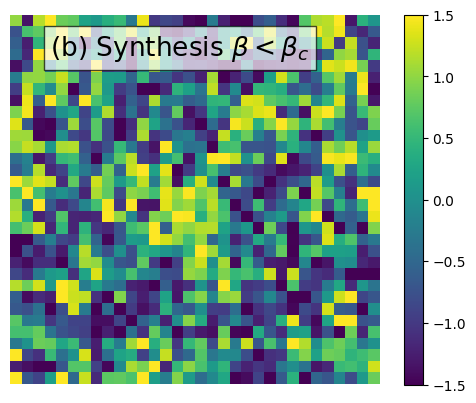}
\includegraphics[width=0.48\linewidth]{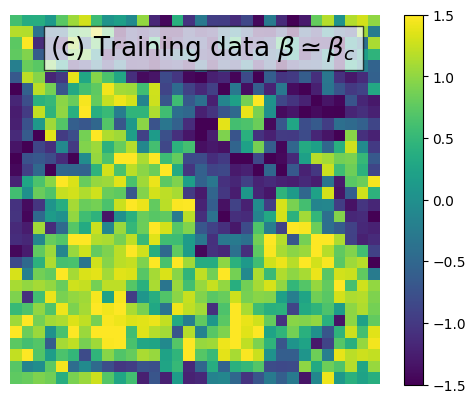}
\includegraphics[width=0.48\linewidth]{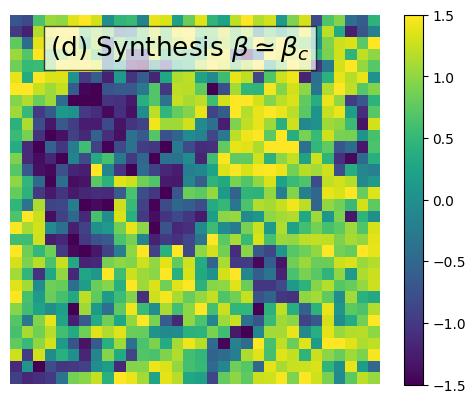}
\includegraphics[width=0.48\linewidth]{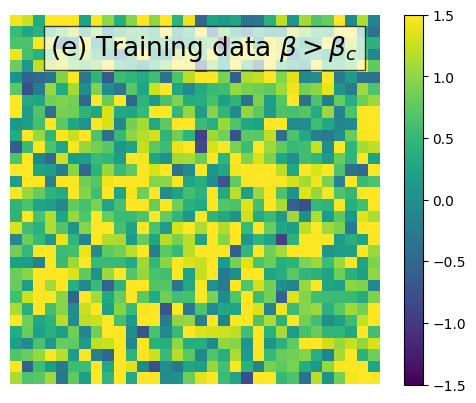}
\includegraphics[width=0.48\linewidth]{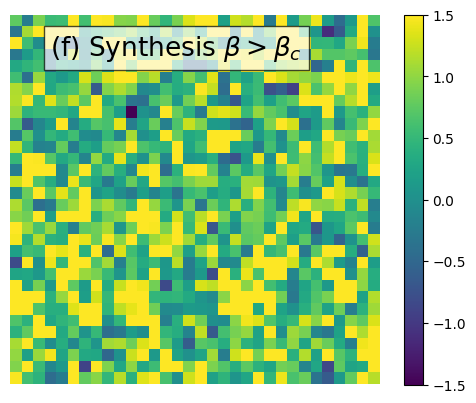}

\caption{Original training samples of  $\varphi^4$ and synthesized fields generated by WC-RG,  for $\beta=0.5<\beta_c$, $\beta=0.67 \simeq \beta_c$, and $\beta=0.76>\beta_c$ (from top to bottom).}
\label{fig:example_realizations_gen}
\end{figure}

\begin{figure}
\includegraphics[width=0.75\linewidth]{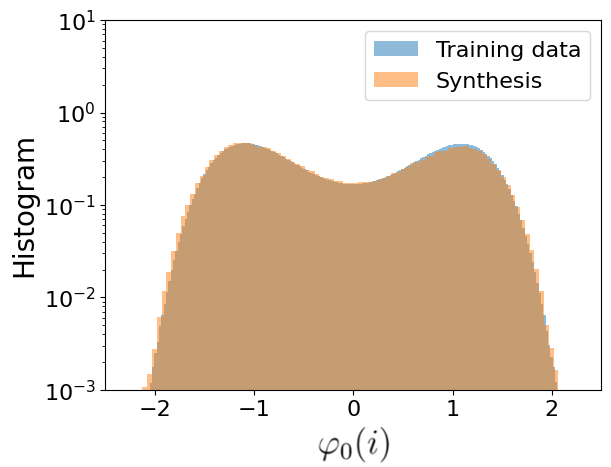}
\caption{Superimposed histograms of field values computed from 
$\varphi^4$ training data (blue) and WC-RG synthesized fields (orange), for $\beta \simeq \beta_c$.}
\label{fig:hist_phi4_critical}
\end{figure}

We consider a two-dimensional system with $L=32$.
We focus on four values of $\beta$: $\beta = 0.5, 0.6, 0.67$, and $0.76$, which covers the three different regimes of the model, the disordered ($\beta=0.5, 0.6$), critical ($\beta=0.67\simeq \beta_c$), and ordered ($\beta=0.76$) phases.
For each value of $\beta$, we generate $R = 10000$ samples of the field $\aphi_0$ which we shall use as \textit{training dataset} to perform the WC-RG. Typical configurations are shown in the left panels of Fig.~\ref{fig:example_realizations_gen}. See App.~\ref{app:phi_model} for more details. 

The right panels of Fig.~\ref{fig:example_realizations_gen} display
samples synthesized with a coarse to fine WC-RG, for $\beta=0.5, 0.67$ and $0.76$. They visually can not be discriminated from samples of the original process in the left panels,  even at $\beta \simeq \beta_c$.
Figure~\ref{fig:hist_phi4_critical} superimposes
the marginal histogram of field values $\varphi_0(i)$ obtained from the training dataset and the one generated by coarse to fine WC-RG sampling for $\beta \simeq \beta_c$. 
We also observe an excellent agreement.

\begin{figure}
\includegraphics[width=0.75\linewidth]{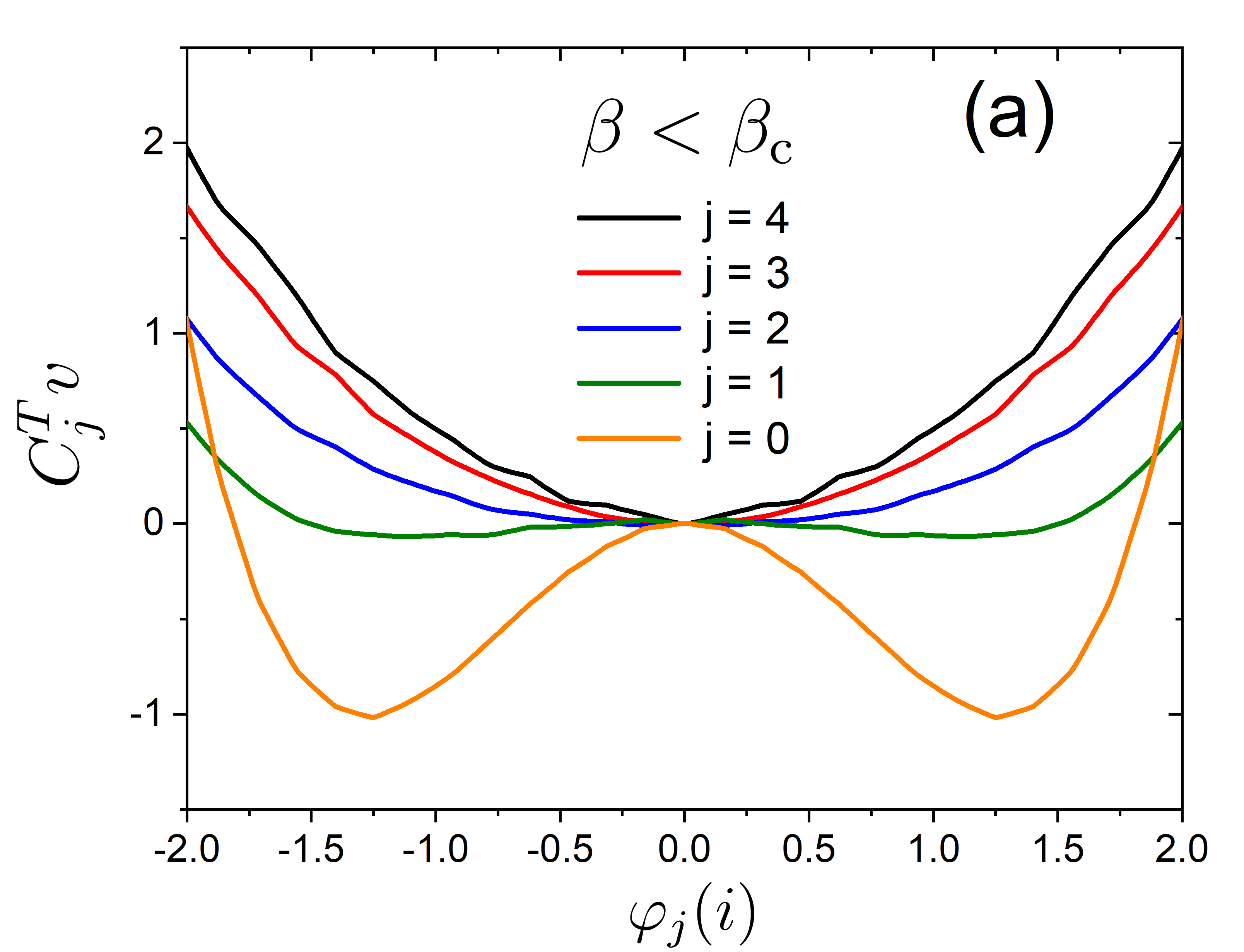}
\includegraphics[width=0.75\linewidth]{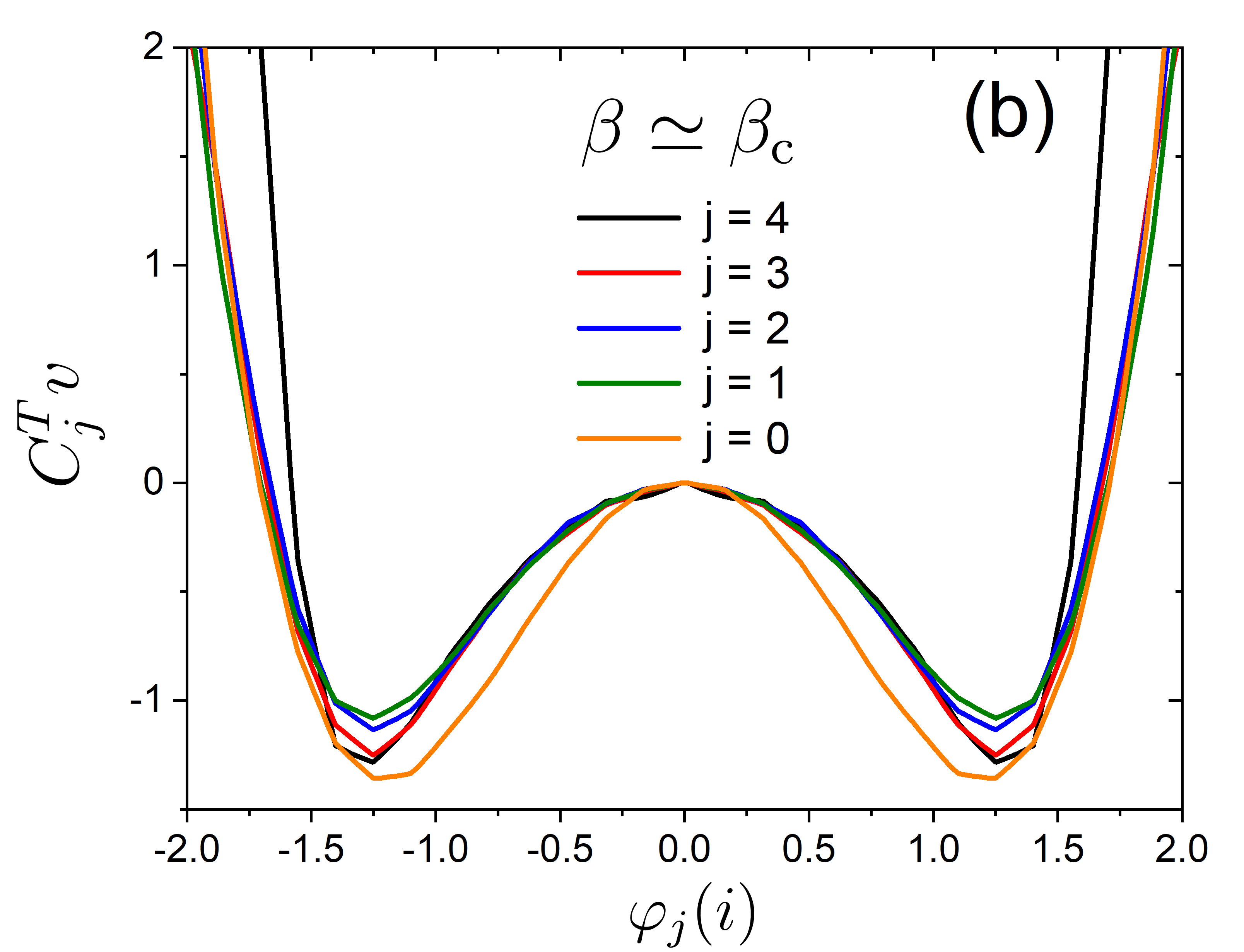}
\includegraphics[width=0.75\linewidth]{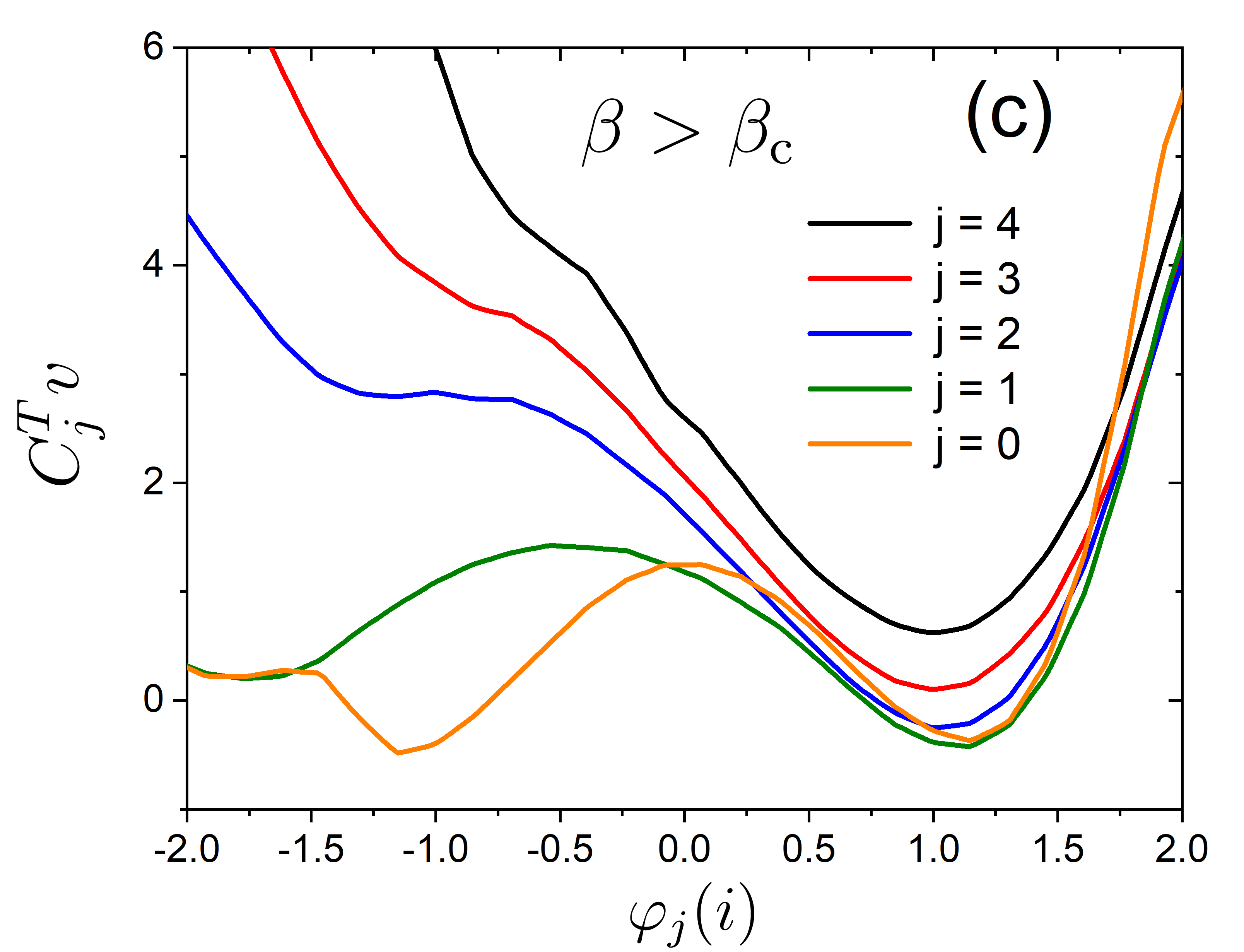}
\caption{WC-RG Flow of non-linear potentials $C_j^T v(\varphi_j(i))$ across scales $2^j$, 
for $\beta=0.5$ in (a), $\beta = 0.67 \simeq \beta_c$ in (b), and $\beta=0.76$ in (c).
}
\label{fig:self_similarity}
\end{figure}

Figure~\ref{fig:self_similarity}(a) shows the non-linear local potential $C_j^T v$ 
calculated at each scale $2^j$ for $\beta<\beta_c$. 
Recall that $C_j^T V (\aphi_j)=\sum_i C_j^T v(\aphi_j(i))$ in Eqs.~(\ref{nsodufsbis0}) and (\ref{ensdfiweef}). At large scales, the potential
has a single-well centered at zero because the disordered phase is nearly Gaussian at coarser scales. During the WC-RG flow from coarse to fine scales, $C_j^T v$ progressively acquires a double-well shape. This figure illustrates the {\it inverse RG flow} induced by WC-RG. It looks like a backward run of the forward Non-Perturbative RG flows \cite{berges2002non}. The analogous WC-RG flow for $\beta \simeq \beta_c$ is presented in Fig.~\ref{fig:self_similarity}(b). The potential $C_j^T v$ has a double well shape which remains stable for many intermediate scales. It is a manifestation of the critical fixed point of the WC-RG flow associated with a second-order phase transition. 
Finally,  Fig.~\ref{fig:self_similarity}(c) presents the result in the ordered phase. We start from a broken symmetry phase at a large scale, where the data are all in the same positively magnetized phase. Remarkably, the WC-RG recovers the symmetric potential at the finest scale.
In theory, the non-linear potential should have a slight asymmetry because an infinitesimal positive field has to be applied to obtain the positively magnetized phase at a large scale. Yet, this effect is very small and not detected numerically.
The ability of WC-RG to recover the microscopic (broken) symmetry is important in applications of this method to characterize properties of microscopic energies.

\begin{figure}
\includegraphics[width=0.75\linewidth]{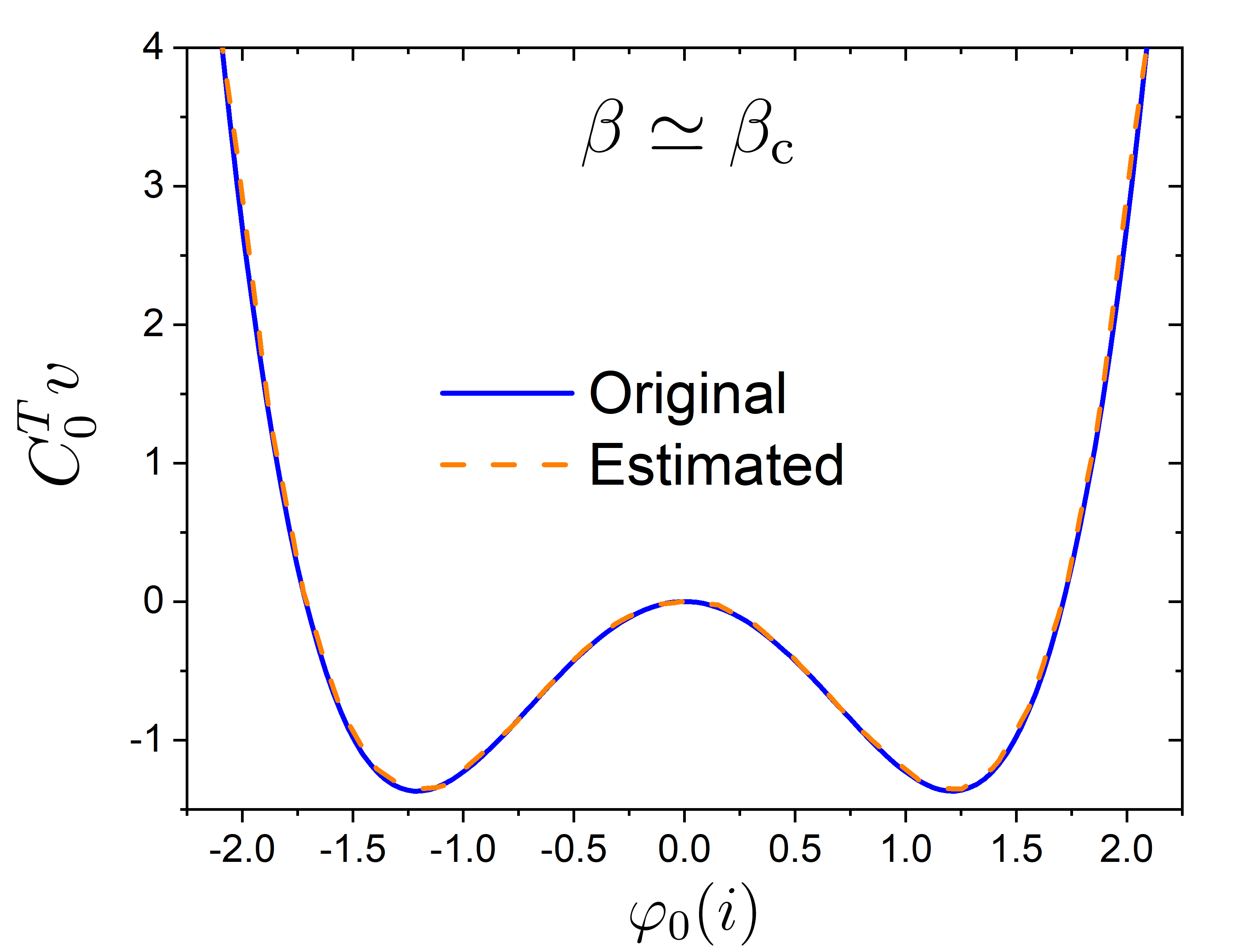}
\caption{Comparison for $\beta \simeq \beta_{\rm c}$ of the original non-linear local potential $C_0^T v(\varphi_0(i))$ at the microscopic scale (solid curve) and its 
WC-RG estimation (dashed curve). }
\label{fig:result_tc1}
\end{figure}

\begin{figure}
\includegraphics[width=0.45\linewidth]{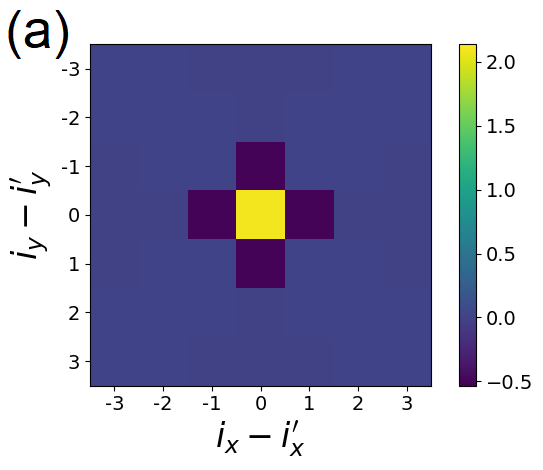}
\includegraphics[width=0.5\linewidth]{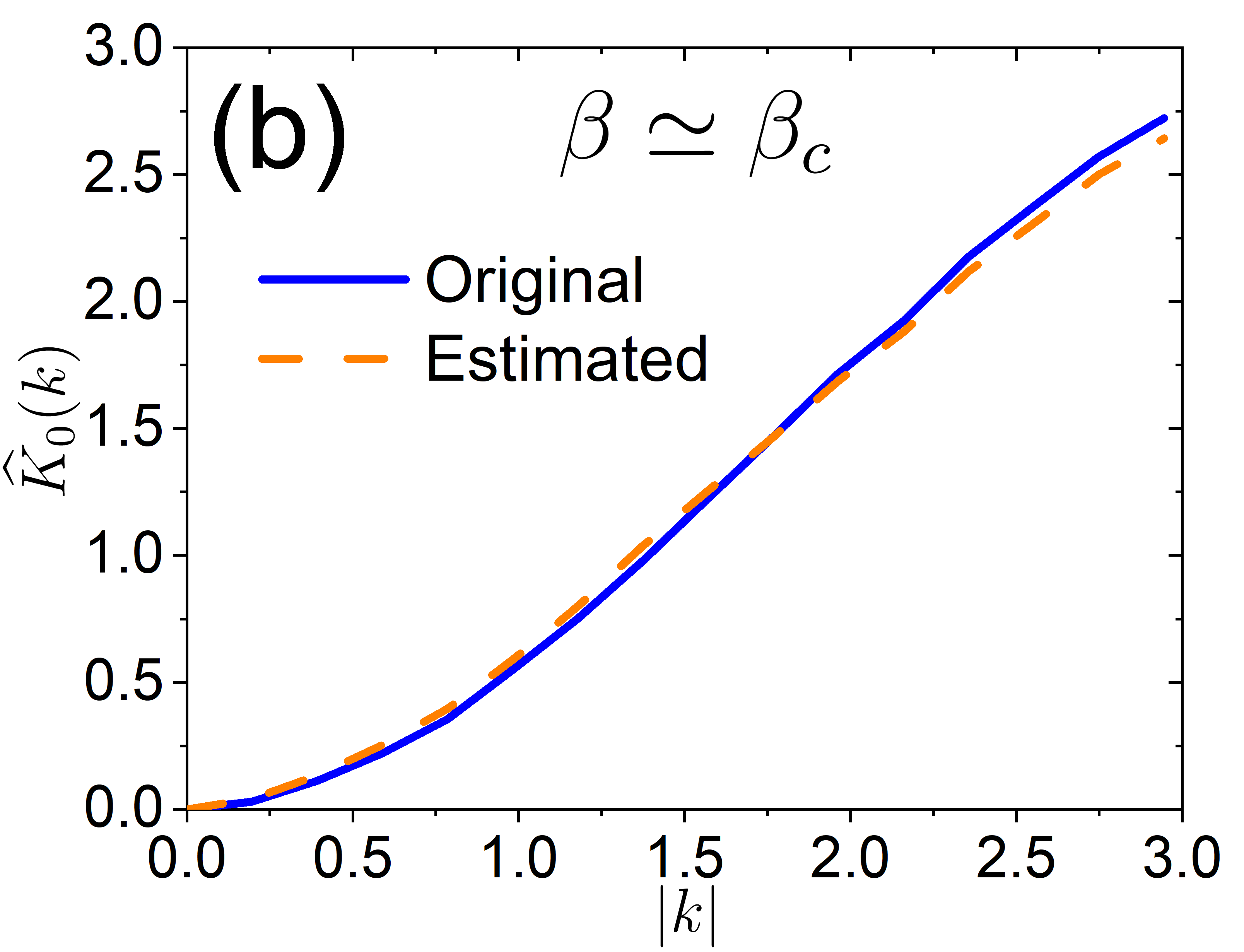}
\caption{(a): Estimated convolutional kernel $K_0(i-i')$ in the $d=2$ dimensional space for $\varphi^4$. (b): Comparison for $\beta \simeq \beta_c$ of the Fourier eigenvalues $\widehat K_0(k)$ of $K_0$ of the original model (solid curve) and the one estimated by the WC-RG (dashed curve).  }
\label{fig:result_tc2}
\end{figure}

The accuracy of the WC-RG model of $\varphi^4$
is evaluated by comparing the values of $K_0$ and $C_0$, which define the microscopic energy
function in Eq.~(\ref{phi4eneransfd}).
We focus on the phase transition $\beta \simeq \beta_c$, 
which is the most challenging case.
Figure~\ref{fig:result_tc1} shows that the estimated local
potential at the finest scale is nearly
equal to the estimated one. 
Figure~\ref{fig:result_tc2}(a) shows the convolution kernel of $K_0$ estimated 
by WC-RG. Figure~\ref{fig:result_tc2}(b) superimposes the eigenvalues of
the original Laplacian $K_0$ with its WC-RG estimation.
The excellent agreement demonstrates that the WC-RG provides a precise estimation of 
the $\varphi^4$ microscopic energy.

Since the $\varphi^4$ model is a priori known to be local,
to regress the free energies $\overline F_j$ we impose that
$\widetilde C_j=\overline{C}_{j+1}$ for all $1 \leq j \leq J-1$ and $\widetilde C_J=C_J$ in Eq.~(\ref{Ingssdcouprs23}). This is 
a usual assumption in RG treatments of the $\varphi^4$ model.
In our case, we can validate a posteriori this assumption by
computing the resulting linear regression error of the free-energy.
It is measured by the ratio between the mean-square 
regression error $\ell (\widetilde \K_j^{\star})$ in Eq.~(\ref{menasqaloss}) and the  average squared free-energy $\lb\bFF_j^2 (\aphi_j) \rb_{p_j}$. 
The relative error with this locality assumption
is of the order of $10^{-5}$ for all the $\beta$ that we considered. 
It confirms a posteriori that
a local potential provides an accurate model of $\varphi^4$ training examples.

\subsection{Cosmological Data}
\label{sec:Cosmological data}

\begin{figure*}
\includegraphics[width=0.15\linewidth]{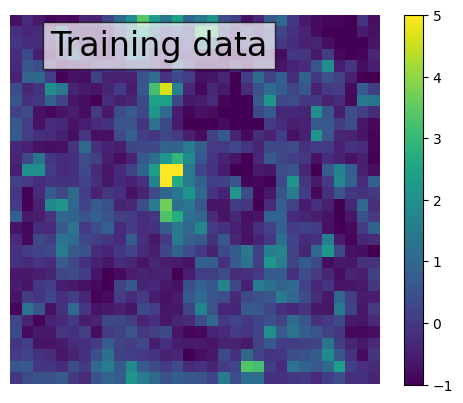}
\includegraphics[width=0.15\linewidth]{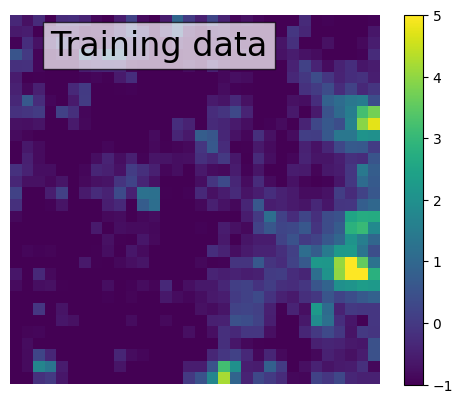}
\includegraphics[width=0.15\linewidth]{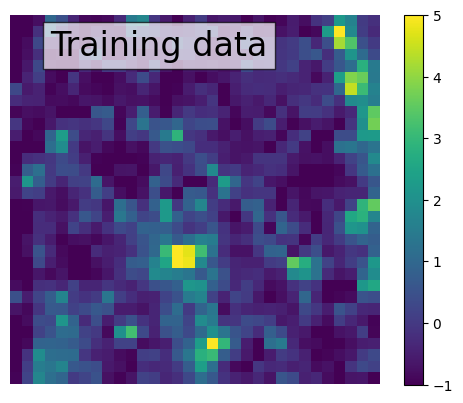}
\includegraphics[width=0.15\linewidth]{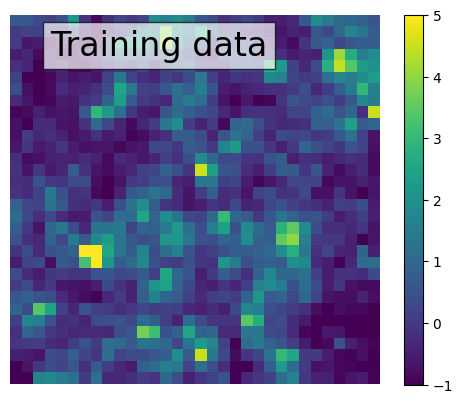}
\includegraphics[width=0.15\linewidth]{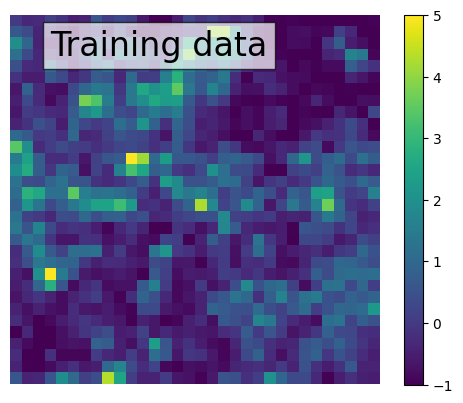}
\includegraphics[width=0.15\linewidth]{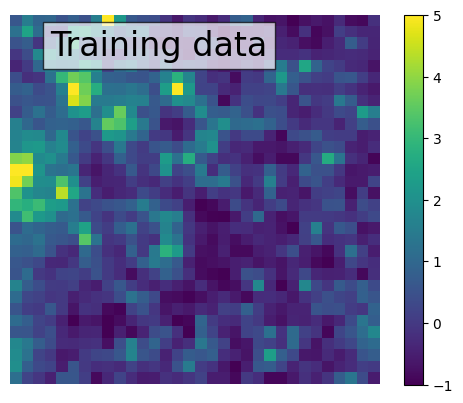}
\includegraphics[width=0.15\linewidth]{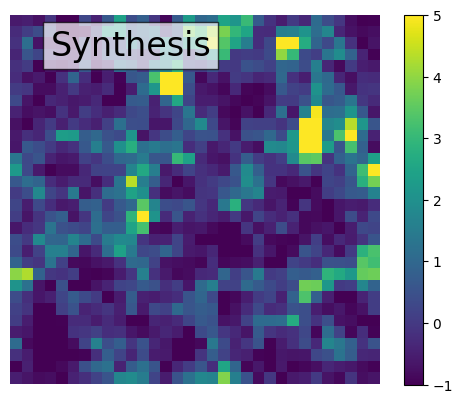}
\includegraphics[width=0.15\linewidth]{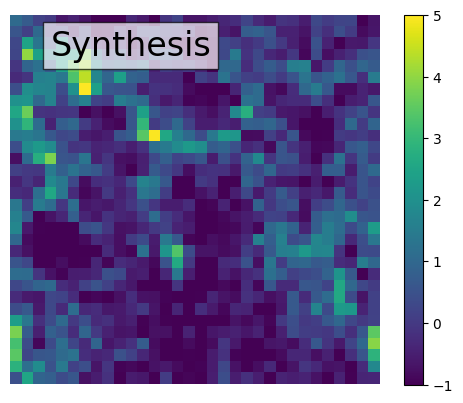}
\includegraphics[width=0.15\linewidth]{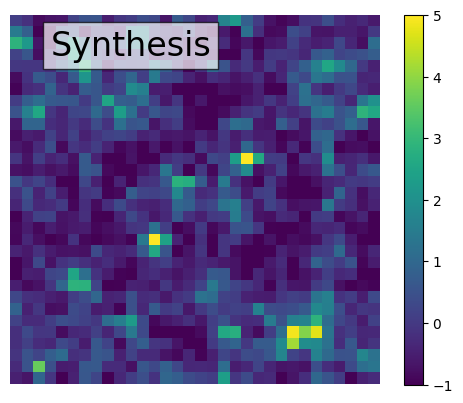}
\includegraphics[width=0.15\linewidth]{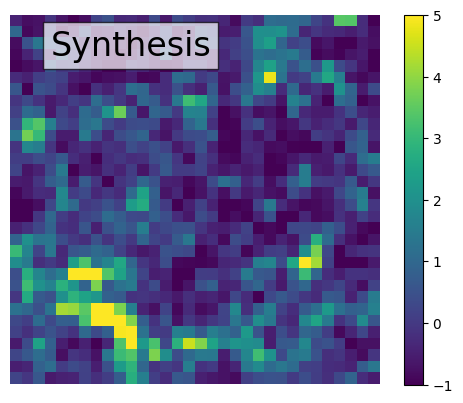}
\includegraphics[width=0.15\linewidth]{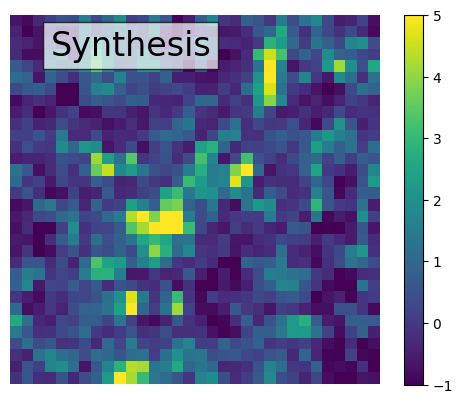}
\includegraphics[width=0.15\linewidth]{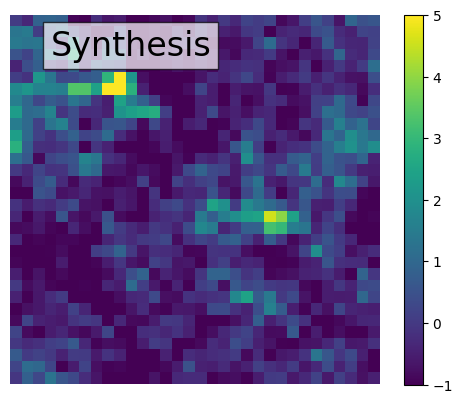}
\caption{Weak gravitational lensing maps in cosmology. Top: Training dataset from the Columbia lensing group~\cite{matilla2016dark,gupta2018non}. Bottom: Synthesized fields generated by the WC-RG sampling algorithm.}
\label{fig:Euclid_result_1}
\end{figure*}

\begin{figure}
\includegraphics[width=0.75\linewidth]{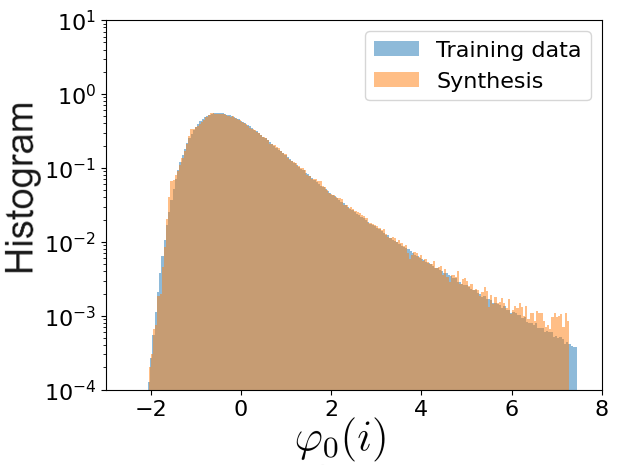}
\caption{Superposition of the normalized histogram of $\varphi_0(i)$ for the weak lensing training data (blue) and samples of multiscale WC-RG models (orange).}
\label{fig:Euclid_result_2}
\end{figure}

This section applies WC-RG estimation and sampling to model weak lensing images in Cosmology. Gravitational lensing deforms images of background objects such as galaxies near a foreground mass~\cite{bartelmann2001weak,kilbinger2015cosmology}. Galaxy clusters are the largest gravitationally bound structures in the Universe, with approximately 80\% of cluster content in the form of dark matter. It can cause strong and weak statistically coherent distortions of background sources on the order of 10\% (cluster weak lensing)~\cite{kilbinger2015cosmology}. A major scientific challenge is to study the weak gravitational lensing signature of large-scale structures in the Universe and to understand fundamental physics such as the nature of dark energy and the total mass of neutrinos. In particular, several groups have been trying to capture rich information that is beyond the traditional two-point statistics, using non-Gaussian statistics through high order moments ~\cite{fu2014cfhtlens,liu2015cosmological,kacprzak2016cosmology,martinet2018kids,shan2018kids}. 
More recently, iterated wavelet transforms called scattering transform have been used to generate weak lensing images conditioned by scattering moments \cite{Cheng_2021}. However, none of these techniques are able to define the explicit microscopic energy function which generates such fields.

In the following, we apply the WC-RG to study the statistics of weak gravitational lensing maps. 
We use a set of simulated convergence maps computed by the Columbia lensing group~\cite{matilla2016dark,gupta2018non} as a training dataset. It can be considered as test convergence maps for the next-generation space telescope, {\it Euclid}~\cite{laureijs2011euclid}. 
Details about the data and numerical implementation can be found in App.~\ref{App:cosmology}. 
The top row of Fig.~\ref{fig:Euclid_result_1} shows several examples of the convergence maps used as the training dataset. Compared to the $\varphi^4$ field theory images, one sees rare high amplitude local fluctuations (shown in yellow), reflecting a higher concentration of dark matter. This characterizes a more complex statistics 
typical of highly non-Gaussian processes (with a long tail, see below). 

On the bottom row of Fig.~\ref{fig:Euclid_result_1}, we show the 
synthesized images generated by WC-RG sampling using the estimated scale interaction energy functions introduced in Sec.~\ref{Coafinsec}. The original (top row) and synthesized (bottom row) define textured images which can not be discriminated visually. It shows that 
the WC-RG model is able to capture the high amplitude non-Gaussian fluctuations. 
A more quantitative analysis is performed by comparing the histograms of  $\varphi_0(i)$ for the training dataset and the fields generated by sampling the 
WC-RG models.  Figure~\ref{fig:Euclid_result_2} shows that both distribution are nearly
equal. In particular, the WC-RG model reproduces the 
long tail involving rare events, which has been hampering various approaches in statistical analysis.
It demonstrates that WC-RG is a powerful tool to tackle highly non-trivial scientific problems, including cosmology.

We now turn to the recovery of the energy function at the microscopic scale. 
For weak lensing, the energy function has a different meaning than for the $\varphi^4$ model in thermal equilibrium. Since cosmological data are obtained from an inherently {\it out-of-equilibrium} process, the energy function is not the Hamiltonian in the sense of a generator of the dynamics. However, it provides a compact and explicitely interpretable parametrisation of the high-dimensional probability distribution. It can be regarded as an "effective" Hamiltonian leading to a Boltzmann-like representation of the probability distribution. Looking for an effective Hamiltonian in a non-equilibrium problem is generally a challenge. It has been possible to obtain it only in a few cases~\cite{derrida2002large,bertini2015macroscopic}, mostly in simple $d=1$ systems. The estimation of the microscopic energy function of a cosmological system is particularly challenging because of the existence of long-range interactions resulting due to gravitation. 

\begin{figure}
\includegraphics[width=0.45\linewidth]{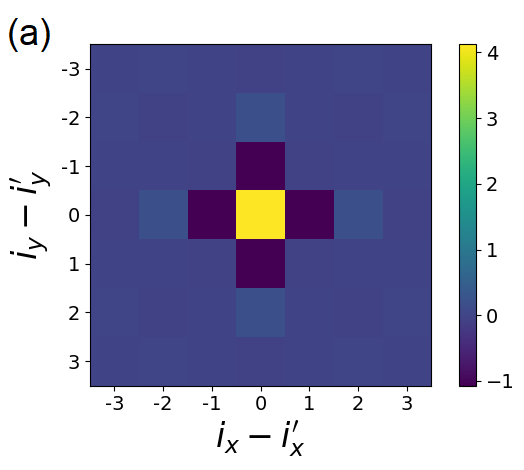}
\includegraphics[width=0.5\linewidth]{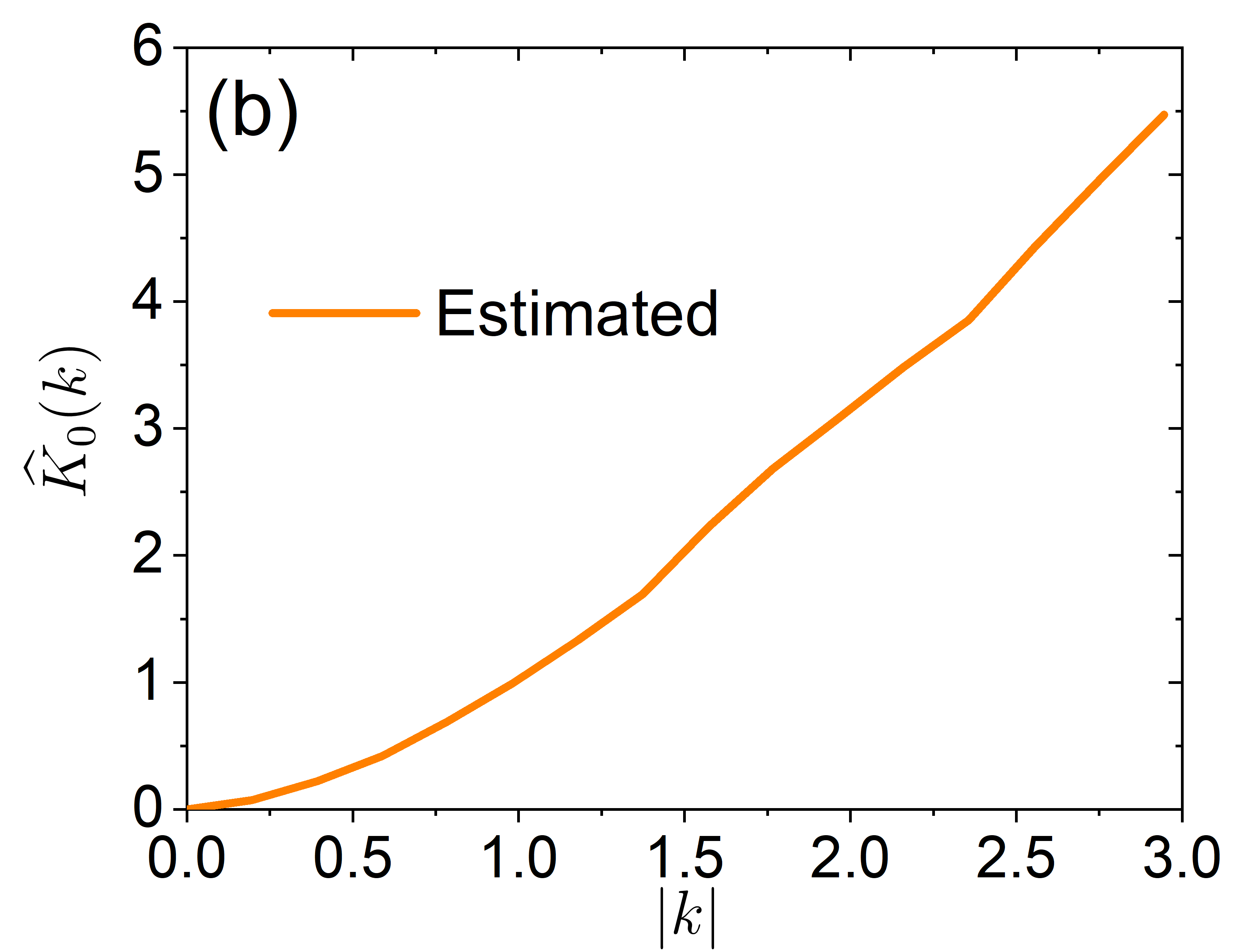}
\caption{(a) Estimated convolutional kernel $K_0 (i-i')$ in the $d=2$ dimension space
from the WC-RG model of weak gravitational lensing maps. (b) Fourier eigenvalues
$\widehat K_0 (k)$.}
\label{fig:result_eculid}
\end{figure}

A WC-RG multiscale microscopic energy model is decomposed in Eq.~(\ref{Ingssdcouprs223}) 
as a sum of local potential interactions at all scales,
\begin{equation}
\label{Ingssdcouprs2234}
\HH_0 (\aphi_0) =  \frac{1}{2}\aphi_0^T K_0 \aphi_0 +  \sum_{j=0}^J\, {C'_j}^T V (\aphi_j).
\end{equation}
We saw that $C'_0 = \overline C_1$ and $C'_j = \overline C_{j+1} - \widetilde C_j$ for $j \geq 1$, where
$\overline C_j^T V$ is potential of the interaction energy $\overline E_j$ 
and $\widetilde C_j^T V$ is the potential of the free energy $\overline F_j$.
For stationary fields, $K_0$ is a convolution operator. 
Figure~\ref{fig:result_eculid} displays the estimated convolution kernel $K_0(i-i')$ and its Fourier eigenvalues $\widehat K_0(k)$. It 
is the counterpart of Fig.~\ref{fig:result_tc2} for $\varphi^4$. It appears to be
close to a discrete Laplacian, which was not expected. 
We also find that the operators $K_j$ computed at all scales $2^j \geq 1$ remain close to 
a discrete Laplacian. 

\begin{figure}
\includegraphics[width=0.75\linewidth]{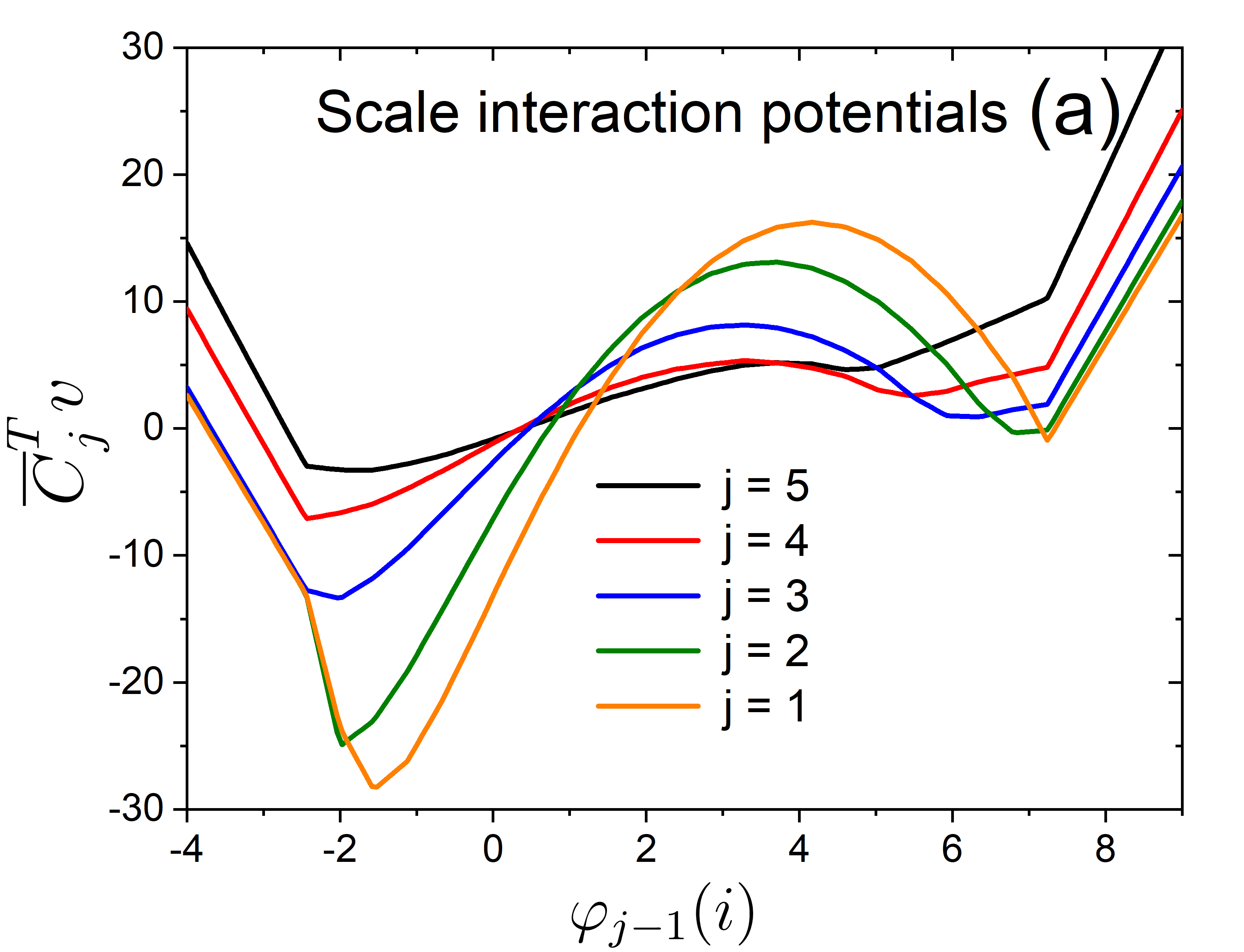}
\includegraphics[width=0.75\linewidth]{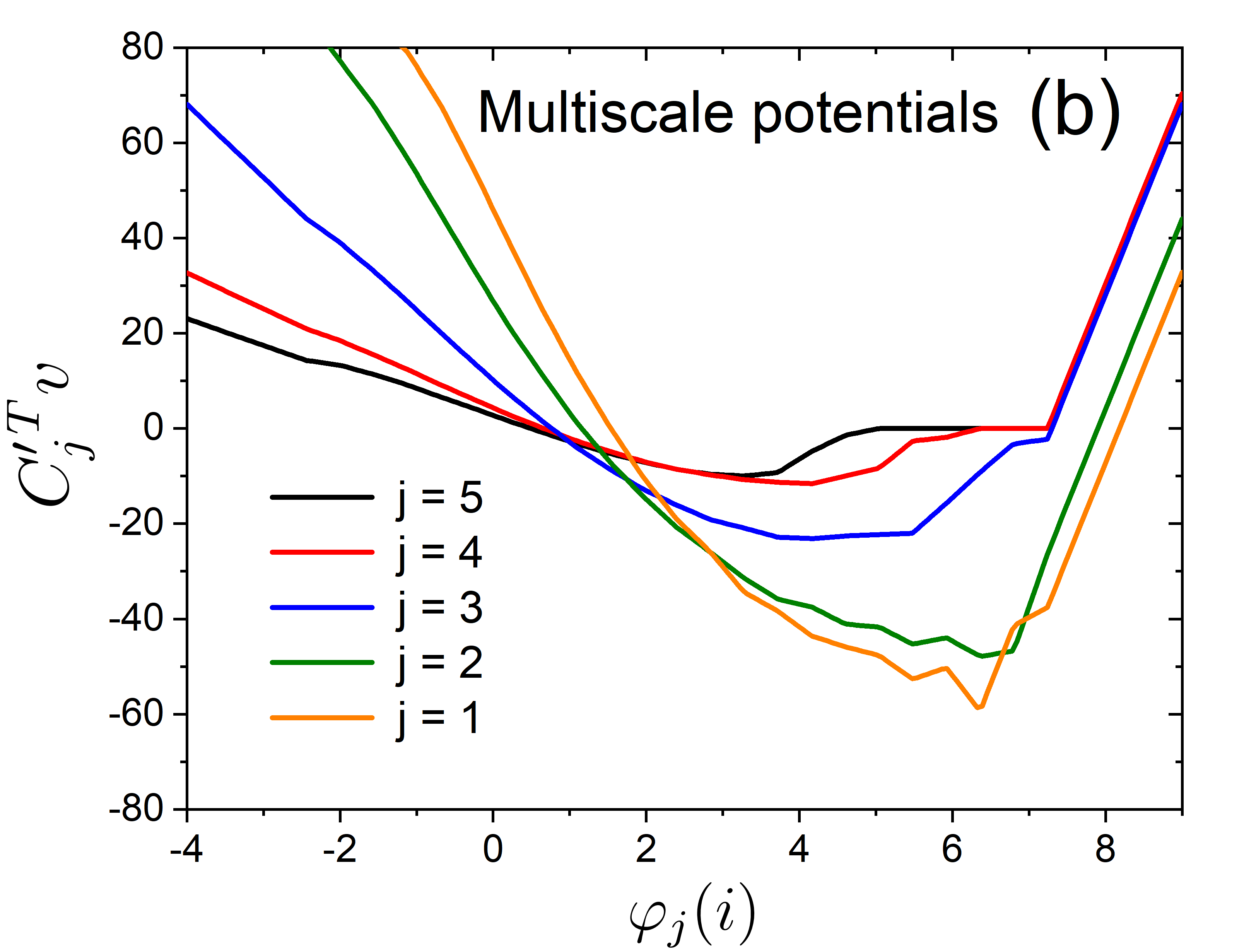}
\caption{
(a): Scale interaction potential $\overline C_j^T v((\aphi_{j-1}(i))$ across scales
$2^j$ estimated from weak-lensing maps. (b): Multiscale potentials ${C'_j}^T v(\aphi_j(i))$ of the microscopic energy function $E_0$ for $j \geq 1$.}
\label{fig:result_eculid_potential}
\end{figure}

Figure~\ref{fig:result_eculid_potential}(a) shows the evolution of the estimated interaction potential, $\overline C_j^T V(\aphi_{j-1}) = \sum_i \overline C_j^T v(\aphi_{j-1}(i))$
across scales. As opposed to $\varphi^4$, 
Fig.~\ref{fig:result_eculid_potential}(b) shows that the multiscale potential terms
$C'_j = \overline C_{j+1} - \widetilde C_j$ are non-zero at all scales $2^j$ for 
$1 \leq j \leq J$. They capture non-local interactions in the
microscopic scale energy in Eq.~(\ref{Ingssdcouprs2234}). 
The multiscale potentials ${C'_j}^T v(\aphi_j(i))$ in Fig.~\ref{fig:result_eculid_potential}(b) favors large values of the field at smaller scales, which increases the relevance of the secondary metastable minimum of $\overline C_j^T v(\aphi_{j-1}(i))$ in Fig.~\ref{fig:result_eculid_potential}(a). However, this effect diminishes at large scales thus avoiding the presence of large regions with high value of the field. These non-local, long-range interactions that favor high amplitude values locally but not globally, induce the long tail of the histogram in Fig.~\ref{fig:Euclid_result_2}. This tail corresponds to the bright yellow spots in the weak-lensing images of Fig.~\ref{fig:Euclid_result_1}. This interpretation echoes the aggregation of matters due to long-range gravitational forces.

\subsection{Stable and Unstable Representations of Probability Distributions}

Behind the fast convergence of WC-RG parameter estimation lies the fact that conditional probabilities of wavelet
fields provides
a stable parametrizations of large classes of probability distributions, even close to phase transitions. A WC-RG decomposes a distribution $p_0$ into 
 $p_0 = \alpha\, p_J \prod_{j=1}^J \overline p_j $, and it approximates each
conditional probability $\overline p_j (\dphi_j | \aphi_j)$ with a model
$\overline p_{\overline \theta_j} (\dphi_j | \aphi_j)$. 
Small errors on the conditional couplings
$\overline \theta_j$ do not 
strongly affect the properties of $p_0$. Indeed the conditional probability distribution $\overline p_j$ are not singular as the "fast degrees of freedom" $\dphi_j$ 
are well-behaved with short-range correlations.  
This is in stark contrast with standard energy-based models which directly approximate
$p_0$ by $p_{\theta_0}$. A small change of 
the coupling parameters in $\theta_0$ can lead to dramatic changes on $p_{\theta_0}$ near phase transitions.

\begin{figure}
\includegraphics[width=0.7\linewidth]{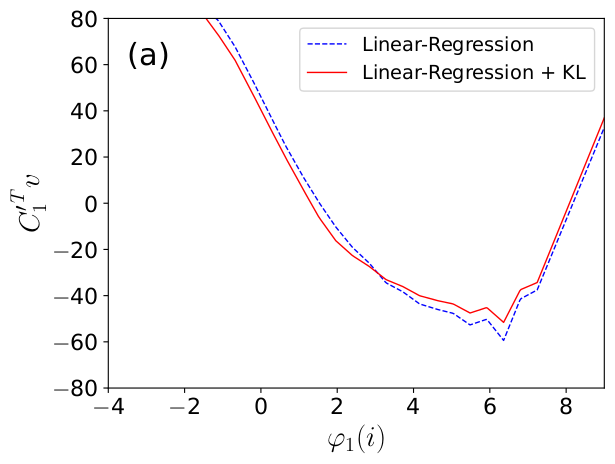}
\includegraphics[width=0.4\linewidth]{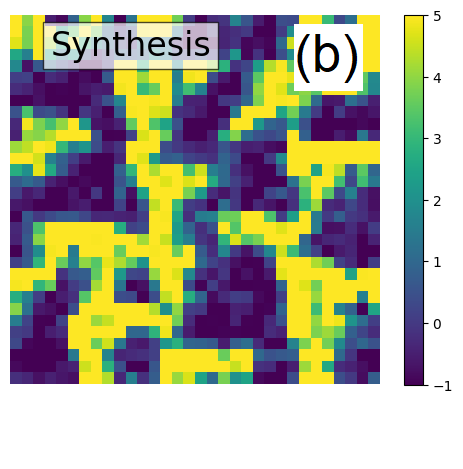}
\includegraphics[width=0.4\linewidth]{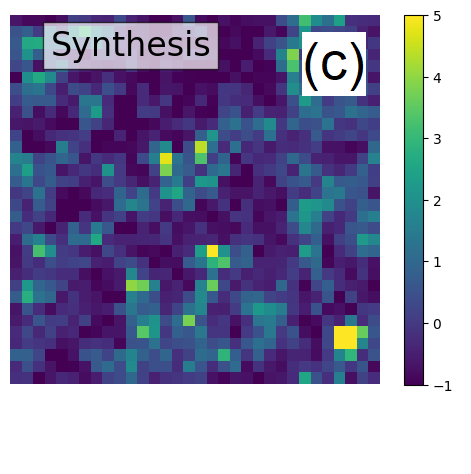}
\includegraphics[width=0.45\linewidth]{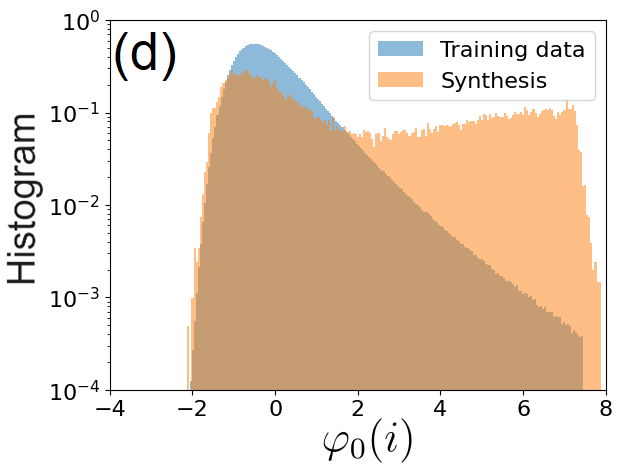}
\includegraphics[width=0.45\linewidth]{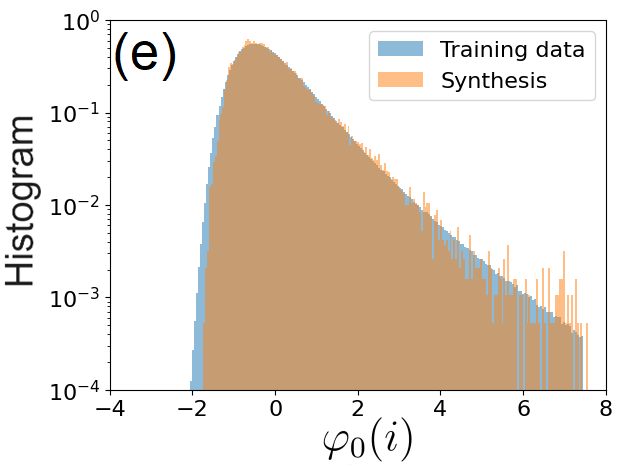}

\caption{
(a): Multiscale potential  ${C'_1}^T v(\aphi_1(i))$ at a fine scale $j=1$, used in
the microscopic energy $E_0$. The dashed curve potential is calculated
with a linear regression of the free energy $\overline F_1$. The full curve is obtained
with a direct KL-divergence minimization.
(b, c): Synthesized images by sampling using the
energy $E_0$. The images (b) and (c) correspond to the potentials ${C'_1}^T v$ shown in (a),
computed from $\overline F_1$ and by minimizing the KL divergence, respectively.
(d, e): Superimposed histograms of $\varphi_0(i)$ of
 synthesized images (in orange) and original weak lensing
 images (in blue). The large errors in the tail of histogram (d) correspond to high amplitude patches in image (b), whereas the histogram (e) of image (c) is well reproduced.}
 \label{fig:double_well_app}
\end{figure}

This is vividly illustrated over weak lensing maps. A model $p_{\theta_0}$ of $p_0$
is defined by the microscopic energy model in Eq.~(\ref{Ingssdcouprs2234}) defined by
$\theta_0 = (\frac 1 2 K_0\,,\, C'_0, C'_1, \cdots, C'_J)$, with
$C'_0 = \overline C_1$ and $C'_j = \overline C_{j+1} - \widetilde C_j$ for $j \geq 1$.
Figure~\ref {fig:double_well_app}(b) is a sample of $p_{\theta_0}$
calculated with an MCMC algorithm updating $\varphi_0$ directly. It has very different statistical properties than  original weak lensing images as in Fig. \ref{fig:Euclid_result_1}, which
clearly appears in the tail of the
superimposed histograms in Fig. \ref{fig:double_well_app}(d).
The excess tail corresponds to high amplitude clusters of sites organized in real space in an ant colony shape, see
Fig.~\ref{fig:double_well_app}(b). Their
typical width is 2 to 3 pixels, which indicates that the
multiscale potential parameter $C'_j$ has an error at $j=1$.
Such statistical errors do not appear with a coarse to fine wavelet 
sampling of the WC-RG model,
as shown in Fig.~\ref{fig:Euclid_result_2}. This model 
is parametrised by
the conditional coupling parameters $\overline \theta_j$, which include the
potential parameters $\overline C_j$ of $\overline E_j$.
The errors on $C'_j = \overline C_{j+1} - \widetilde C_j$ in $E_0$ is thus
produced by errors when estimating the potential parameters $\widetilde C_j$
of the free energies $\overline F_j$.
Yet, we are going to show that these errors are small and the
large statistical errors in Fig.~\ref{fig:double_well_app}(d) are due to instabilities of the coupling parametrisation $\theta_0$.

To demonstrate this property, we modify $\theta_0$ by only changing the
potential parameter $C'_j$ for $j=1$. We
initialise $C'_1$ with the value obtained from the free energy calculation. 
Its value is updated with a
gradient descent on the KL divergence $D_{\rm KL}(p_0 || p_{\theta_0})$,
with the algorithm of Sec. \ref{finegridestimat}. Figure \ref{fig:double_well_app}(a)
shows that this optimization produces a very small modification of ${C'_1}^T v$.
Yet, the images sampled from this slightly modified microscopic energy $E_0$ have a totally
different results in Fig. \ref{fig:double_well_app}(c), which now match the weak lensing maps. Indeed their histogram in
Fig. \ref{fig:double_well_app}(e) is superimposed over the histogram of weak lensing images.

 This numerical experiment shows that the coupling parameter $\theta_0$ of $E_0$ was estimated with a good accuracy through free energy regressions. However, this coupling parametrisation of $E_0$ is highly unstable. Such instabilities also appear at phase transitions.
WC-RG conditional probability representation circumvents this problem, by relying only
on the parameters $\overline \theta_j$ of wavelet conditional probabilities, which are stable. It thus leads to more reliable generative models of complex many-body problems.

\subsection{Absence of Critical Slowing Down for WC-RG}
\label{numtheory}

    

\label{finegrid}
\begin{figure}
\includegraphics[width=0.48\linewidth]{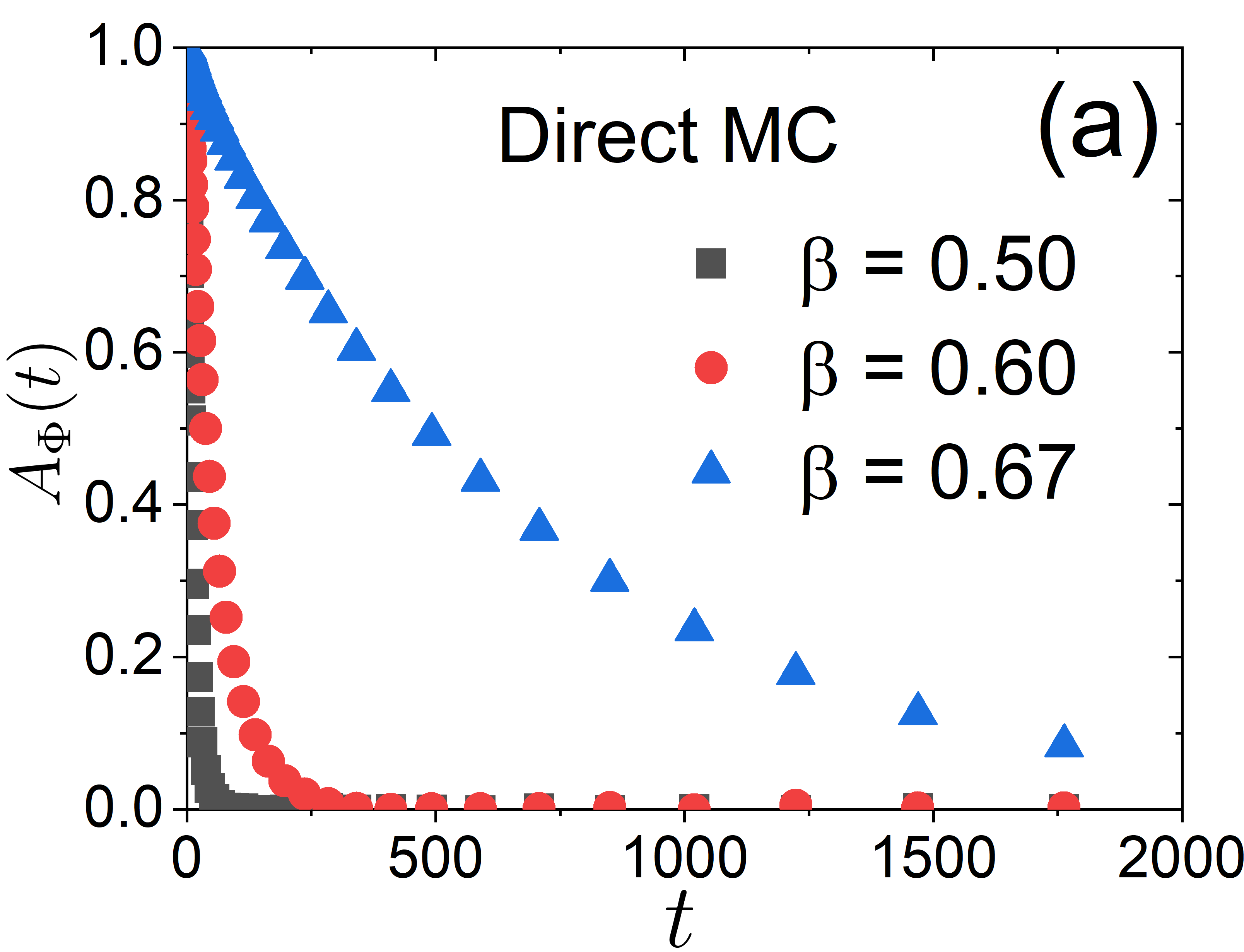}
\includegraphics[width=0.48\linewidth]{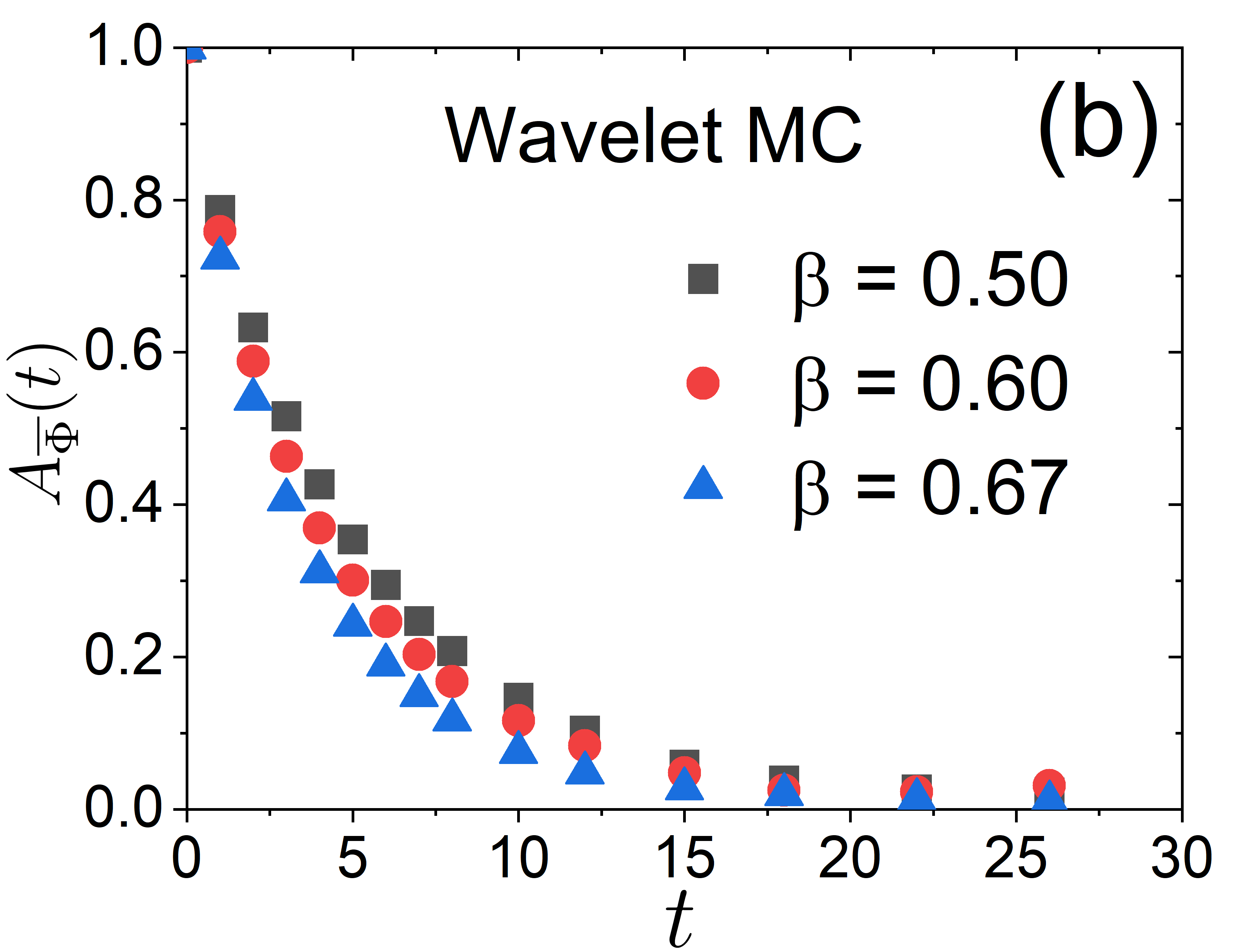}
\includegraphics[width=0.48\linewidth]{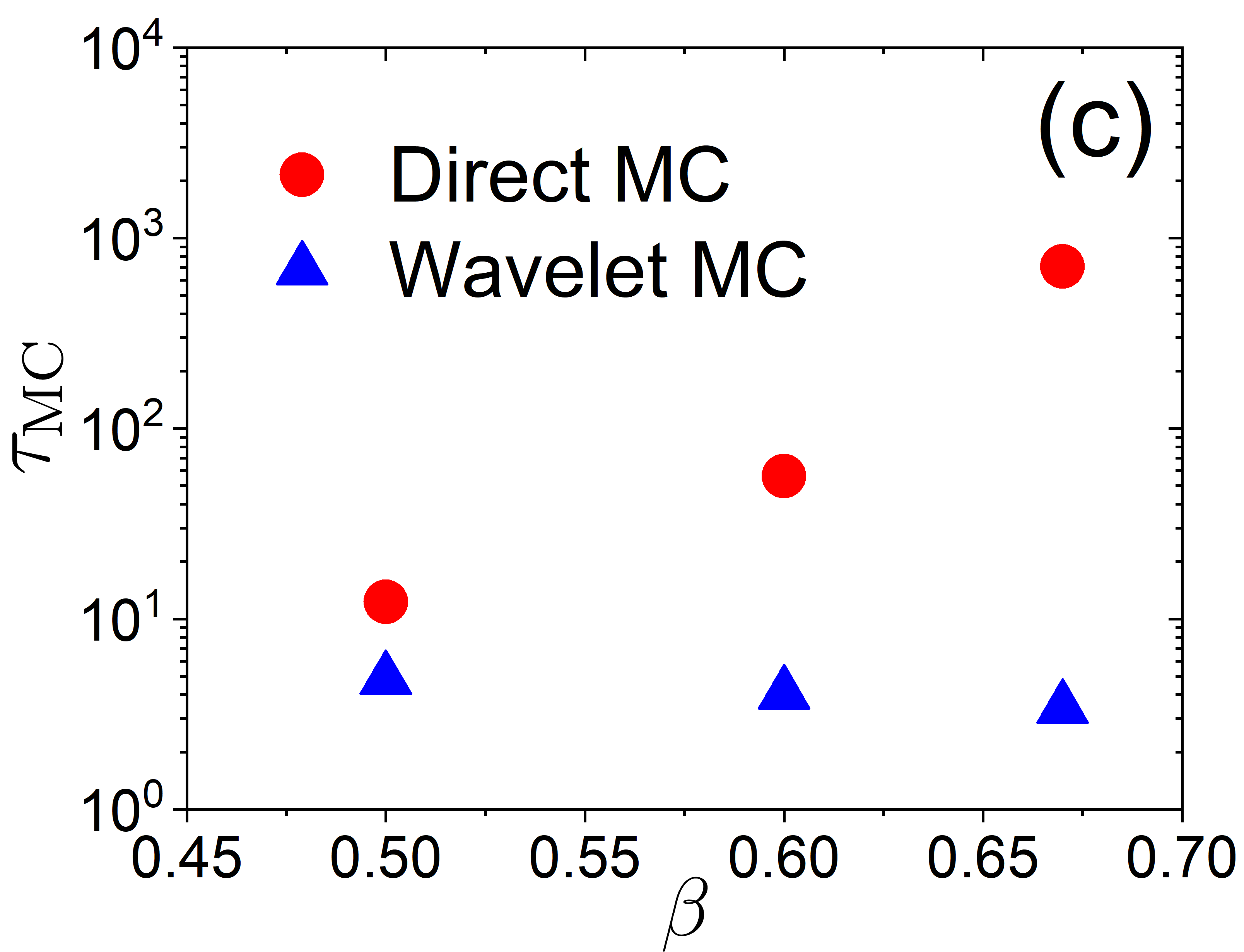}
\includegraphics[width=0.48\linewidth]{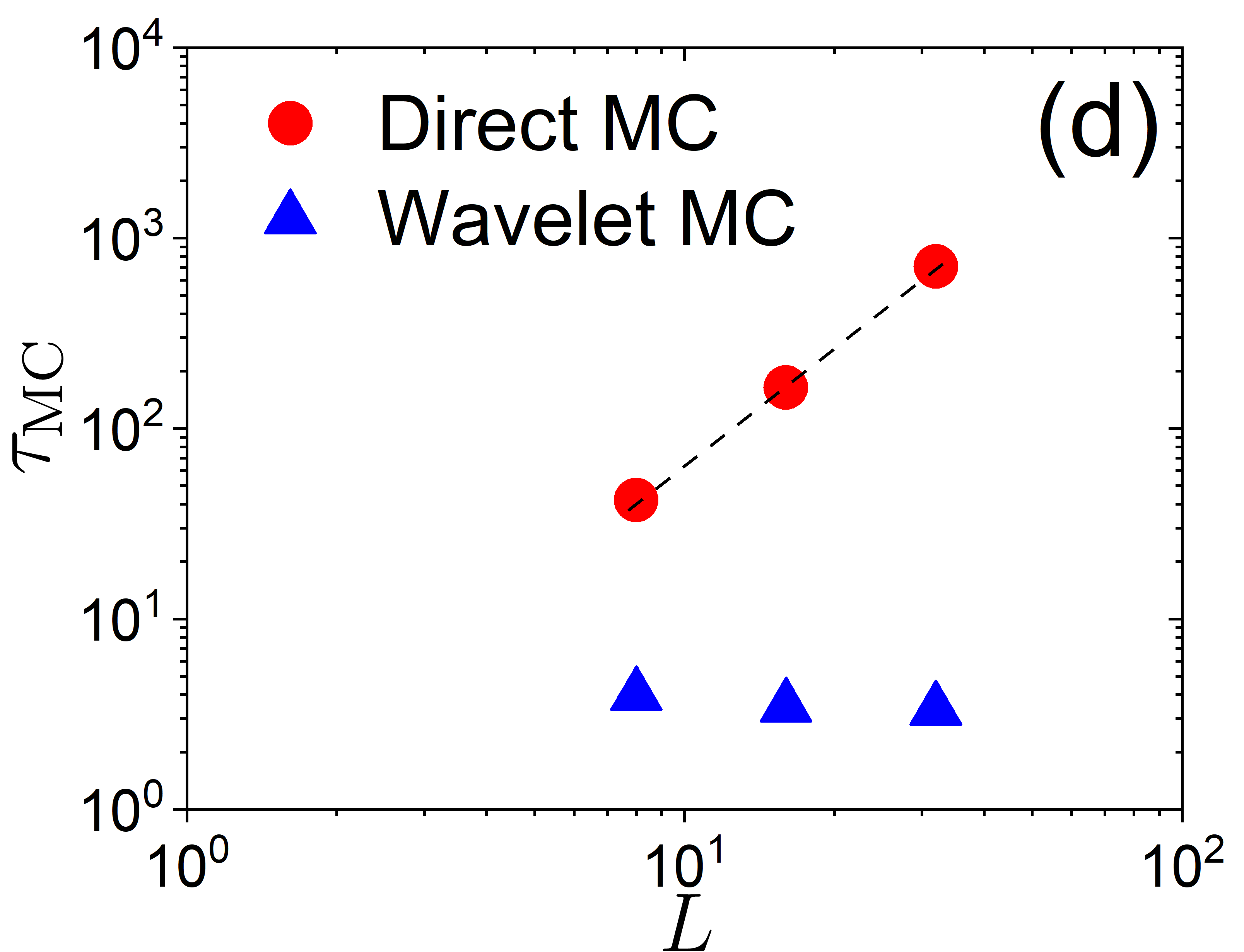}
\caption{
(a, b): Auto-correlation functions, for direct MC simulations 
updating $\varphi_0$ in (a), and for wavelet MC simulations updating $\dphi_1$ in (b). It is computed for different $\beta$ approaching $\beta_c$ for the $\varphi^4$ model.
(c, d): Decorrelation timescale $\tau_{\rm MC}$ for the direct and wavelet MC simulations.
It is computed as a function of $\beta$ with $L=32$ in (c), and as a function of $L$ with $\beta \simeq \beta_c$ in (d).
The dashed line in (d) corresponds to $\tau_{\rm MC} \sim L^z$ with $z=2.04$.
}
\label{fig:MC_phi4A}
\end{figure}

\label{finegrid}
\begin{figure}
\includegraphics[width=0.48\linewidth]{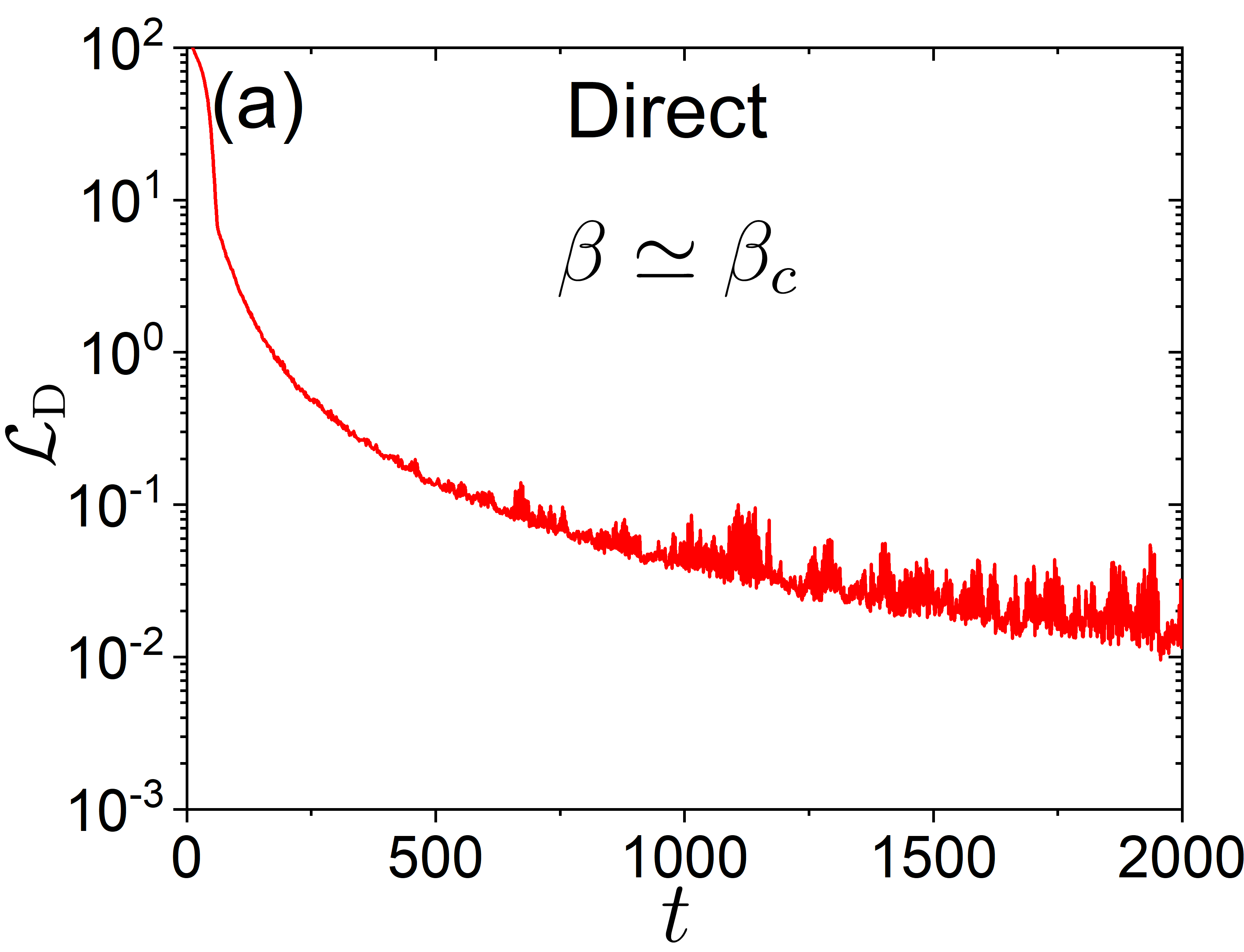}
\includegraphics[width=0.48\linewidth]{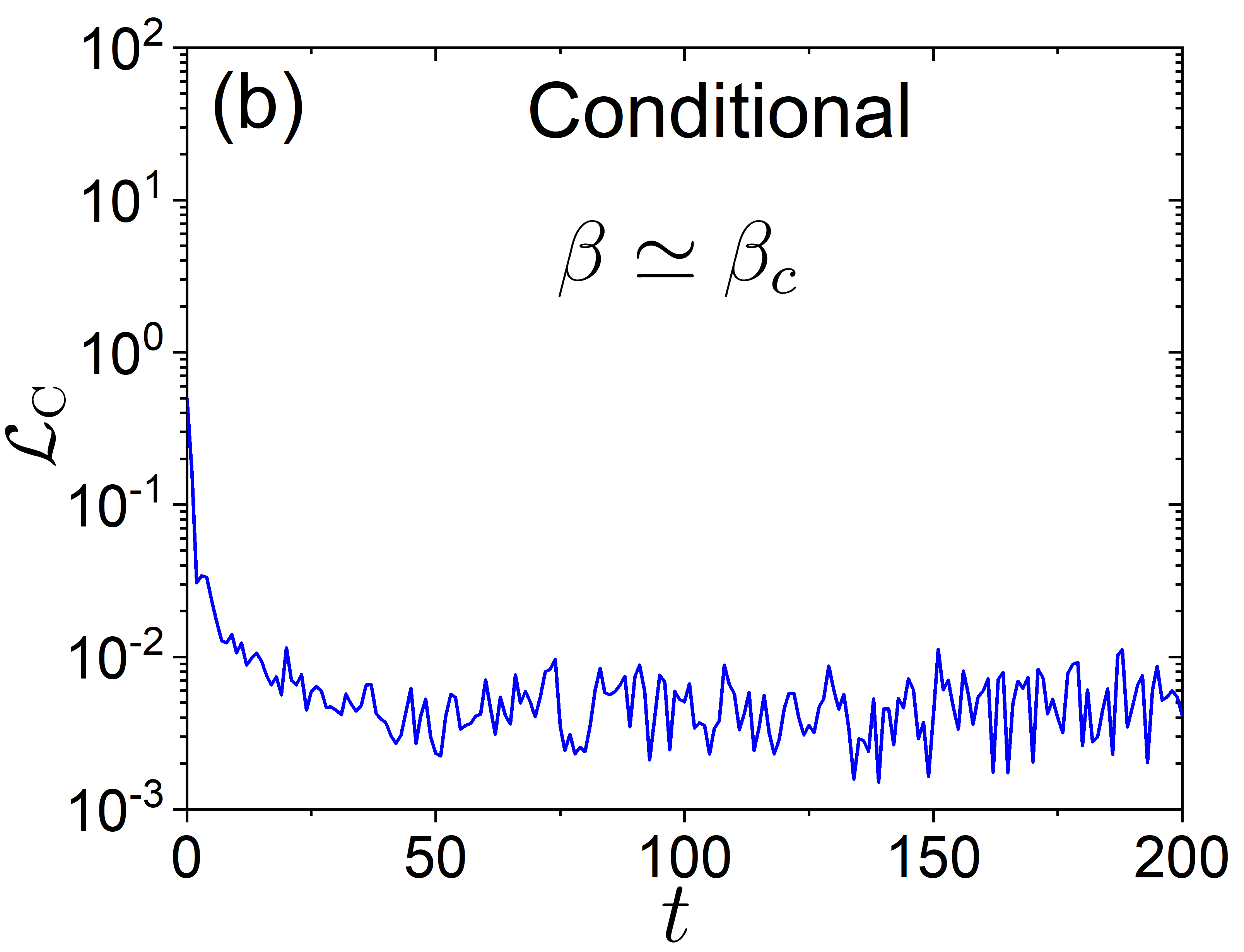}
\caption{
Moment errors as a function of the number of gradient-descent steps $t$.
(a): Error 
$\mathcal{L}_{\rm D}$ in Eq.~(\ref{LD}) for the direct algorithm.
(b): Error $\mathcal{L}_{\rm C}$ in Eq.~(\ref{LC}) for the wavelet conditional algorithm.
}
\label{fig:GD_phi4}
\end{figure}

This section illustrates the numerical stability of WC-RG calculations over the $\aphi^4$ model. In particular, we contrast WC-RG with standard direct coupling estimation approaches plagued by critical slowing down close to the critical point. As discussed in Sec.~\ref{precondHamilsec}, the critical slowing down appears both in the gradient descent dynamics and to estimate moments by MCMC.
Numerical details are given in App.~\ref{app:numerical_methods}. 
For Gaussian models, the absence of critical slowing down of WC-RG algorithms
is demonstrated in App.~\ref{app:gaussian_model}.

For the $\varphi^4$ model, we begin by studying the critical slowing down of the MCMC for direct coupling estimation when $\beta$ approaches $\beta_c$, and when $L$ is increased at $\beta \simeq \beta_c$. 
The MCMC simulations shown in
Fig. \ref{fig:MC_phi4A} are performed at the last gradient descent step when the coupling parameters have converged to an optimal value.
Let us denote $\aphi_0^{(t)}$ the finest scale field at a Monte-Carlo time step $t$, that is evolving with
the MCMC simulation. 
One MC time step corresponds to $L^2$ trial MC updates over an image of $L^2$ pixels.
The magnetization is written by $\Phi(t) =|L^{-2} \sum_i \aphi_0^{(t)}(i)|$. Its time auto-correlation function $A_\Phi(t)$ is given by
\begin{equation}
\label{eq:autocorrelation_Phi}
    A_\Phi(t) = 
\frac{    \left\langle \delta \Phi(t)\delta \Phi(0) \right\rangle_{\rm time}}{\left\langle \delta \Phi^2 \right\rangle_{\rm time}}
    ,
\end{equation}
where $\langle \cdots \rangle_{\rm time}$ denotes the time average under the stationary state and $\delta \Phi(t)=\Phi(t)-\langle \Phi \rangle_{\rm time}$.
Figure~\ref{fig:MC_phi4A}(a) plots $A_\Phi(t)$ for a direct Monte-Carlo update of 
$\aphi_0$, for different values of $\beta$ approaching $\beta_c$. The decorrelation timescale $\tau_{\rm MC}$ defined by $A_{\Phi}(\tau_{\rm MC})=1/e$ increases in Fig.~\ref{fig:MC_phi4A}(c). The critical slowing down also appears when $L$ increases at $\beta \simeq \beta_c$. As expected,  Fig.~\ref{fig:MC_phi4A}(d) shows that $\tau_{\rm MC}$ diverges as $\tau_{\rm MC} \sim \mathcal{O}(L^z)$ with $z\simeq 2$ at the critical point
\cite{zinn2002quantum}, whereas it remains finite for $\beta<\beta_c$. 

In WC-RG, the MCMC is performed at each scale on the wavelet fields $\dphi_j$, while $\aphi_j$ is fixed. We call it a {\it wavelet MC}. Figure~\ref{fig:MC_phi4A}(b) shows the
resulting auto-correlation function $A_{\overline \Phi}(t)$ of $\overline \Phi(t)$, given by
\begin{equation}
\overline \Phi(t)= \frac{1}{3 (L2^{-j})^{2}}\, \sum_{m, i} |\dphi_j^{(t)}(m,i)| .
\label{eq:auto_WCRG}
\end{equation}
It is displayed at the finest scale $j=1$, but coarser scales behave similarly and require fewer computations. The results show that, remarkably, $A_{\overline \Phi}(t)$ decays at the same rate for all $\beta$, which is very different from a direct MC updating 
of $\varphi_0$. The decorrelation time scale
$\tau_{\rm MC}$  is defined by $A_{\overline \Phi}(\tau_{\rm MC})=1/e$. Observe that it
remains nearly constant as a function of $\beta$ in Fig.~\ref{fig:MC_phi4A}(c), and as a function of $L$ in Fig.~\ref{fig:MC_phi4A}(d). It means that the wavelet MC is not affected at all by the critical slowing down.

We now study the convergence of the gradient descent,
which estimates coupling parameters by minimising a KL divergence. It also
involves a critical slowing down near the phase transition. 
The error is measured
by the squared norm of moments errors. It
measures the amplitudes of the parameter gradients in Eqs.~(\ref{gradse2}) and (\ref{gradse20}).
For the direct gradient descent algorithm of Sec. \ref{finegridestimat}, 
the moment error is 
\begin{equation}
\label{LD}
\mathcal{L}_{\rm D}(t)=||\left\langle \U_j  \right\rangle_{p_{\K_{j}^{(t)}}} - 
 \left\langle \U_j \right\rangle_{p_j}||^2,
 \end{equation}
 where $t$ is the gradient descent time step.
 It is computed at the finest scale $j = 0$.
 For the wavelet conditional gradient descent of Sec. \ref{Hessian-ga}, the moment error is
 \begin{equation}
 \label{LC}
 \mathcal{L}_{\rm C}(t)=||\lB \bU_j  \rB_{p_{\bK_j^{(t)}}} - 
 \lB \bU_j \rB_{p_{j-1}}||^2 .
 \end{equation}
 It is only shown at the finest scale $j=1$, where its decay is the slowest.
 Figure~\ref{fig:GD_phi4} compares the
 decay of $\mathcal{L}_{\rm D}(t)$ and $\mathcal{L}_{\rm C}(t)$, 
 near the phase transition $\beta \simeq \beta_c$. 
 For each method we compute numerically the maximum step size $\epsilon$ such that the gradient decent converges and leads to an accurate microscopic energy function $E_0$. 
Numerical results shows that $\epsilon=0.005$ for the direct algorithm, whereas
$\epsilon=0.5$ is $100$ times larger  for the conditional gradient descent.
As a result, $\mathcal{L}_{\rm C}$ in Fig.~\ref{fig:GD_phi4} (a) has
a much slower decay than $\mathcal{L}_{\rm C}$ in Fig.~\ref{fig:GD_phi4} (b).
It confirms that the WC-RG is also able to precondition the gradient descent Hessian in the $\varphi^4$ case and thus eliminate the critical slowing down.


The remarkable performance of WC-RG comes from the fact that RG can handle a singular theory close to a critical point, contrary to perturbation theory, which treats on the same footing all degrees of freedom. Close to a critical point, where the susceptibility to changes in the coupling parameters diverges, perturbation theory is bound to fail. On the contrary, since RG always works on the "fast short-scale degrees of freedom", it is not affected by the critical point, and one can perform approximation on the RG flow safely \cite{delamotte2012introduction}. In our context, the direct coupling estimation plays the same role as a perturbation theory. It  fails when approaching the critical point 
because it works directly on the degrees of freedom and coupling parameters plagued by a singular behavior. Instead, the WC-RG works at each scale on the wavelet fields corresponding to the shortest scales, which are non-critical and hence fast. This is the essential physical ingredient that makes WC-RG numerically stable and unaffected by the critical slowing down.

\section{Conclusions and Discussions}

The wavelet conditional renormalization group (WC-RG) provides a new representation of high-dimensional probability distributions, as a product of conditional probabilities. They are associated with scale interaction energy functions, which can be estimated from limited sets of data. It addresses a major open issue in physics and machine learning. 
This approach is tightly related to the renormalization group theory developed for second-order phase transitions. Fluctuations of the field, or "fast degrees of freedom", are represented at each scale in a wavelet orthogonal basis. 

The WC-RG differs from a standard forward RG in several aspects. It does not suppose that the microscopic energy is
known {\it a priori} but it provides an estimation of this energy from data.
A standard RG computes the flow of coupling parameters defined over the whole energy function. A WC-RG rather computes the flow of conditional coupling parameters,
which specify conditional probabilities of wavelet fields given coarse-grained fields.
We showed that estimations of conditional coupling parameters by maximum likelihood 
is well conditioned and avoids critical slowing down. A WC-RG generates new fields
by sampling wavelet conditional probability distributions, which also circumvents the critical slowing down at phase transitions. 
Explicit expression of the microscopic energy can be recovered from
WC-RG estimations, with the thermodynamic integration and a linear regression that are introduced. 
This was applied to Gaussian and $\varphi^4$ models in thermal equilibrium, and to weak gravitational lensing maps in non-equilibrium.
The study of the cosmological data is particulary challenging since the underlying distribution is unknown {\it a priori}. 

We introduced local potential models at each scale, which can
capture long-range interactions at the microscopic scale.
These multiscale Ansatz open the possibility to build models of complex physical fields including turbulences. Such models have indeed been studied in Ref.~\cite{Allys_2020,zhang2021maximum} by specifying dependencies across scales through correlations of phases and amplitudes of wavelet coefficients. It would also be interesting to study applications of WC-RG to disordered systems, for which it could provide a new efficient sampling method of low-temperature configurations.

Many publications have pointed out the similarities between RG transformations and deep network architectures~\cite{mehta2014exact,lin2017does,iso2018scale,roberts2022principles}, and  flow based modeling~\cite{li2018neural}.
The computational architecture of WC-RG bares some close resemblance with deep generative networks, particularly U-Nets. This paper suggests that it may be a conditional RG which is calculated 
in many deep neural networks. 
Similarly to deep networks, a WC-RG decomposes the field at multiple scales with a wavelet transforms which cascades convolutions and subsamplings. Local potentials are computed with piecewise linear rectifiers which connects wavelet coefficients at a given scale, stored in a network layer. The normalisation is achieved by a normalisation of variances, which corresponds to batch normalisations in deep neural networks. Network parameters correspond to coupling interactions in a wavelet conditional RG, which are learned with a stochastic gradient descent algorithm. Field generations are achieved with a coarse to fine algorithm similarly to U-Net architectures. However, as opposed to deep neural networks, the optimisation involves an MCMC Metropolis sampling algorithm. It is computationally slow but guarantees to reach the optimal solution. Replacing Metropolis random propositions by appropriate determinist propositions would fill the gap with deep neural networks. 

\begin{acknowledgements}
We acknowledge funding from the French government under management of Agence Nationale de la Recherche as part of the ``Investissements d’avenir'' program, reference ANR-19-P3IA-0001 (PRAIRIE 3IA Institute) and from the Simons Foundation collaboration ``Cracking the Glass Problem'' (No. 454935 to G. Biroli). 
We also thank Zoltan Haiman for sharing the convergence maps obtained by the Columbia Lensing group, and Antonin Brossollet and Simon Coste for careful reading of the manuscript. 
 \end{acknowledgements}

\appendix

\section{Wavelet Bases and Representation of Operators}
\label{app1}

The fast wavelet transform decomposes a field $\varphi_0$ by using orthogonal filters $G$ and $\wG$. 
Appendix \ref{wavefiltdesign} briefly reviews the
construction of such filters in any spatial dimension $d$.
A wavelet transform stabilizes renormalization group computations
because wavelet bases nearly diagonalize operators
involved in the calculation of large classes of energy functions.
Appendix \ref{app1.2} explains that the wavelet fields $\dphi_j$
computed by the fast wavelet transform of Sec.~\ref{RGwavebasis} are
decomposition coefficients in a 
wavelet orthonormal basis of ${\bf L^2}(\R^d)$. 
Theorem \ref{Specsn} is proved in App.~\ref{Theorem1Proof} by showing
that a singular homogeneous operator, such as a Laplacian in a Gaussian model, is diagonally
dominant in a wavelet orthonormal basis.
We used the PyWavelets software~\cite{lee2019pywavelets} to compute the fast wavelet transform.

\subsection{Wavelet filter design}
\label{wavefiltdesign}

We review the construction of wavelet filters $\LL$ and $\wG$ which
satisfy the orthogonality conditions in Eq.~(\ref{consdsf}).
We  begin in spatial dimension $d = 1$ and then introduce a separable extension of
such filters in any dimension $d$.

\subsubsection{One-dimensional filters}
In dimension $d=1$, the coarse-graining filter $\LL^{\scalebox{0.6}{(1)}}$ has a
Fourier transform concentrated at
low wave-numbers.
The superscript, $(1)$, denotes $d=1$.
The wavelet filter $\wG^{\scalebox{0.6}{(1)}}$ 
instead computes high wave-number variations. 
The orthogonality conditions in Eq.~(\ref{consdsf}) have been 
proved to be satisfied \cite{Mallat:89b}
if the Fourier series
$\widehat \LL^{\scalebox{0.6}{(1)}}(k)$ of $\LL^{\scalebox{0.6}{(1)}}(n)$ satisfies for all $k \in [0,2\pi]$,
\begin{equation}
  \label{quadrature0}
  |\widehat \LL^{\scalebox{0.6}{(1)}} (\om)|^2 + |\widehat \LL^{\scalebox{0.6}{(1)}} (\om + \pi)|^2 = 2~~\mbox{and}~~\widehat \LL^{\scalebox{0.6}{(1)}} (0) = \sqrt 2 ,
\end{equation}
and
\begin{equation}
\label{supnsdfius}
\widehat \LL^{\scalebox{0.6}{(1)}} (k) > 0~~\mbox{for}~~k \in [0,\pi/2] .
\end{equation}
The Fourier series $\widehat \wG^{\scalebox{0.6}{(1)}}(k)$ of the wavelet filter $\wG^{\scalebox{0.6}{(1)}}(n)$
satisfies
\begin{equation}
  \label{quadrature}
\widehat \wG^{\scalebox{0.6}{(1)}} (k) = e^{-i \om}\, \widehat \LL^{\scalebox{0.6}{(1)}*} (k + \pi),
\end{equation}
where $i = \sqrt{-1}$ and $*$ is the symbol for the complex conjugate. This implies that
\begin{equation}
  \label{quadrature}
\wG^{\scalebox{0.6}{(1)}} (n) = (-1)^{1-n} \LL^{\scalebox{0.6}{(1)}} (1-n).
\end{equation}

The Haar filter is a simple solution of Eq.~(\ref{quadrature0}) given by
\begin{equation}
    \label{Haarfilt}
\LL^{\scalebox{0.6}{(1)}}(n) = 2^{-1/2}~~\mbox{if}~~n=0,1~~\mbox{and}~~\LL^{\scalebox{0.6}{(1)}}(n) = 0~~\mbox{otherwise}.
\end{equation}
It computes the one-dimensional Kadanoff block averaging in Eq.~(\ref{blockav}).
The corresponding wavelet filter is
$\wG^{\scalebox{0.6}{(1)}}(0) = -2^{-1/2}$, $\wG^{\scalebox{0.6}{(1)}}(1) = 2^{-1/2}$ and $\wG^{\scalebox{0.6}{(1)}}(n) = 0$ if $n \neq 0, 1$.
It iteratively computes Haar wavelet fields with Eq.~(\ref{Harwnsdf}). 

The Shannon filter is another simple solution of Eq.~(\ref{quadrature0})
whose Fourier transform is supported in $k \in [-\pi/2,\pi/2]$, 
\begin{equation}
    \label{Shannonfilt}
\widehat \LL^{\scalebox{0.6}{(1)}} (\om) = 2^{1/2}~{\bf 1}_{|\om| \leq \pi/2},
\end{equation}
where ${\bf 1}$ is the characteristic function.
Next section shows that Haar and Shannon filters lead to Haar and Shannon wavelets, which are badly localized in the Fourier and real-space domains, respectively. Therefore, we shall rather use 
Daubechies filters defined in Ref.~\cite{daubechies1992ten}. 
They  define wavelets which are well
localized {\it both} in the real-space and Fourier domains, which is important to efficiently represent operators involved in the calculations of energy functions.

\subsubsection{Separable multidimensional filters}
In dimension $d$, wavelet filters which satisfy the orthogonality
conditions in Eq.~(\ref{consdsf}) can be defined 
as separable products of the one-dimensional filters $\LL^{\scalebox{0.6}{(1)}}$ and $\wG^{\scalebox{0.6}{(1)}}$. 
The $d$ dimensional low-pass filter $\LL^{\scalebox{0.6}{(d)}}$ is constructed as
\begin{equation}
    \label{low-passfi}
\LL^{\scalebox{0.6}{(d)}}(n_1,n_2,...,n_d) = \prod_{\ell = 1}^d \LL^{\scalebox{0.6}{(1)}}(n_l).
\end{equation}
Let us write $\wG_0^{\scalebox{0.6}{(1)}} = \LL^{\scalebox{0.6}{(1)}}$ and $\wG_1^{\scalebox{0.6}{(1)}} = \wG^{\scalebox{0.6}{(1)}}$.
In dimension $d$, there are $2^d-1$ wavelet channels with the associated $2^d-1$ filters $\wG_m^{\scalebox{0.6}{(d)}}$ for $1 \leq m \leq 2^d-1$. These are obtained with different separable products of $\wG_0^{\scalebox{0.6}{(1)}}$ and $\wG_1^{\scalebox{0.6}{(1)}}$.
By using a binary
digit $b_1, b_2, ..., b_d$ with $b_\ell \in \{0,1\}$, we define
\begin{equation}
    \label{high-passfi}
\wG_m^{\scalebox{0.6}{(d)}}(n_1,n_2,...,n_d) = \prod_{\ell = 1}^d \wG_{b_\ell}^{\scalebox{0.6}{(1)}} (n_\ell) .
\end{equation}
For example, in $d=2$, there are three channels ($m=1, 2, 3$):
\begin{eqnarray}
\wG_1^{\scalebox{0.6}{(2)}}(n_1, n_2) &=& \LL^{\scalebox{0.6}{(1)}} (n_1) \wG^{\scalebox{0.6}{(1)}} (n_2), \nonumber \\
\wG_2^{\scalebox{0.6}{(2)}}(n_1, n_2) &=& \wG^{\scalebox{0.6}{(1)}} (n_1) \LL^{\scalebox{0.6}{(1)}} (n_2), \nonumber  \\
\wG_3^{\scalebox{0.6}{(2)}}(n_1, n_2) &=& \wG^{\scalebox{0.6}{(1)}} (n_1) \wG^{\scalebox{0.6}{(1)}} (n_2).  \nonumber
\end{eqnarray}
One can verify that
these filters satisfy the orthogonality condition in Eq.~(\ref{consdsf})~\cite{stephane1999wavelet}.
In this paper, the filter $\LL^{\scalebox{0.6}{(d)}}$ and the vector of $2^d-1$ filters $\wG_m^{\scalebox{0.6}{(d)}}$ are often denoted simply as $G$ and $\overline G$, respectively.

\subsection{Wavelets Bases of ${\bf L^2}(\R^d)$ from filters}
\label{app1.2}

A fast wavelet transform computes wavelet fields $\dphi_j$ as a cascade of filtering and subsamplings with the wavelet filters $\LL$ and $\wG$. We explain
that these wavelet fields can be rewritten as decomposition coefficients
in a wavelet orthonormal basis of the Hilbert space ${\bf L^2}(\R^d)$ of square-integrable functions, $\int |f(x)|^2 dx < \infty$. These wavelet bases are obtained by dilations and translations
of wavelet functions, which result from the cascade of wavelet filters $\LL$ and $\wG$.

\subsubsection{Wavelet bases}
The fast wavelet transform computes $\aphi_j$ and $\dphi_j$ by iterating $j$
times on $\LL$ and $\wG$ from $\aphi_0$.
Discrete wavelets are the equivalent filters which
relate $\aphi_j$ and $\dphi_j$ to $\aphi_0$. 
We first set all normalization factors $\gamma_j = 1$ for simplicity.
Since $\aphi_j = \LL\aphi_{j-1}$, where $\LL$ is a convolution and
subsampling operator, we get
\begin{equation}
\label{coarsening}
\aphi_j = (\LL)^j\, \aphi_0~.
\end{equation}
Since $G$ is convolution and subsampling by $2$, one can verify that
$(G)^j$ computes the inner product with an equivalent
filter denoted by $\widetilde \psi^0_j$, which is translated at intervals $2^j$:
\[
\aphi_j (i) = \sum_{i'} \aphi_0 (i')\, \widetilde \psi^0_j (i' - 2^j i) .
\]
Similarly, since $\dphi_{j} = \wG\aphi_{j-1}$, we get
\begin{equation}
\label{coarsening2}
\dphi_j = \wG\,(\LL)^{j-1}\, \aphi_0~.
\end{equation}
We thus verify that
wavelet fields are computed as inner products with $2^d -1$ different 
wavelet filters $\widetilde \psi^m_j$ translated at intervals $2^j$,
\[
\dphi_j (m, i) = \sum_{i'} \aphi_0 (i')\, \widetilde \psi^m_j (i' - 2^j i) .
\]
Note that the superscript $m$ on $\widetilde \psi^m_j$ specifies the wavelet channels, and it is not an exponent. The same convention will be used below.

The multiresolution theory in Ref.~\cite{Mallat:89b} proves that
$2^{-jd/2}\, \widetilde \psi^0_j (2^{-j} i)$ 
converges to a scaling function $\psi^0 (x)$ with $x \in \R^d$,
whose Fourier transform is 
\begin{equation}
\label{scansdf}
    \widehat \psi^0(\om) = \prod_{p=1}^{\infty} \frac{\widehat \LL (2^{-p} \om)}{\sqrt 2}~.
\end{equation}
Moreover, each discrete wavelet $2^{-jd/2}\, \widetilde \psi^m_j (2^{-j} i)$ for
$1 \leq m \leq 2^d-1$ converges to a wavelet $\psi^m (x)$  with $x \in \R^d$,
whose Fourier transform is
\begin{equation}
\label{scansdf2}
    \widehat \psi^m (\om) = \frac{\widehat {\wG}_m (2^{-1} \om)}{\sqrt 2} \, \widehat \psi^0 (2^{-1} \om) .
\end{equation}
The function $\psi^0 (x)$ is an averaging filter called a {\it scaling function} which satisfies $\int \psi^0(x)\, dx = 1$,
whereas $\psi^m (x)$ for $1 \leq m \leq 2^d-1$ are called wavelet functions which
satisfy $\int \psi^m(x)\, dx = 0$.
Dilated and translated wavelets are written by
\begin{equation}
\psi^m_{j,i} (x) = 2^{- j d/2}\, \psi^m (2^{-j} x - i), \label{eq:wavelets_from_mother}   
\end{equation}
for $0 \leq m \leq 2^d-1$.
The main result  proved in Refs.~\cite{Mallat:89b,Meyer:92c}
is that the family of wavelets,
\begin{equation}
    \label{waveletbasis}
\Big\{ \psi^m_{j,i} (x) \Big\}_{1 \leq m \leq 2^d-1\,,\,j \in \Z\,,\,i \in \Z^d},
\end{equation}
is an orthonormal basis of ${\bf L^2}(\R^d)$.

In the renormalization group decomposition, the wavelet transform
is normalized by $\gamma_j$ in Eq.~(\ref{fastdec3}). One can prove \cite{Mallat:89b} that
at any scale $2^j$, the family of scaling functions $\{ \psi^0_{j,i}(x) \}_{i \in \Z^d}$ is 
also orthonormal within ${\bf L^2}(\R^d)$. For any microscopic field $\varphi_0$ on a discrete lattice,
one can verify that there exists $\varphi \in {\bf L^2}(\R^d)$ such that for any $j \geq 0$
the coarse-graining approximations $\varphi_j$ provide the renormalized decomposition coefficients of $\varphi$ in these bases of scaling functions:
\begin{equation}
\label{waveorthcoeefs0}
\varphi_j (i) = a^{-1}_j\, \int_{\R^d} \varphi(x)\,\psi^0_{j,i} (x)\,  dx ,
\end{equation} 
where $a_j = \prod_{\ell=1}^j \gamma_\ell$ is the renormalization factor.
If $\varphi_0$ is stationary over $\Z^d$ then $\varphi(x)$ can be defined as a
stationary process in $\R^d$.
Similarly,  for $1 \leq m \leq 2^d-1$, the wavelet fields are normalized
decomposition coefficients of $\varphi(x)$ in the wavelet orthonormal basis:
\begin{equation}
\label{waveorthcoeefs}
\dphi_j (m, i) = a^{-1}_j\, \int_{\R^d} \varphi(x)\,\psi^m_{j,i} (x)\,  dx .
\end{equation}
The coarse-grained fields $\aphi_j(i)$ and wavelet fields $\dphi_j(m, i)$ computed by a fast wavelet transform thus correspond to decomposition coefficients of a field $\varphi(x)$ over orthogonal
functions in ${\bf L^2}(\R^d)$. This result is important to understand the action of
operators over such fields when computing energy functions. 

\subsubsection{Choice of wavelet to represent energy functions}
We now explain how to choose wavelet filters for a WC-RG.
Energy functions involve differential operators such as Laplacians or gradients, but they also include pointwise non-linearities as in the $\varphi^4$ model. Differential operators are diagonal on a Fourier basis, whereas polynomial pointwise non-linearities are local in the spatial domain (real-space) but produce global interactions between Fourier coefficients. Both types of operators induce local interactions over wavelet coefficients if the wavelets $\psi^m(x)$ are sufficiently well localized in the spatial domain, and if their Fourier transform $\widehat \psi^m (k)$ are also well-localized along wave-vectors.

The wavelets $\psi^m(x)$ of a wavelet orthonormal basis
are entirely specified by the one-dimensional filter $\LL^{\scalebox{0.6}{(1)}}$ which
satisfies Eqs. ~(\ref{quadrature0}) and (\ref{supnsdfius}).
Indeed, we derive $\wG^{\scalebox{0.6}{(1)}}$ with
Eq.~(\ref{quadrature}), the separable $d$-dimensional filters $G^{\scalebox{0.6}{(d)}}$ and $\wG^{\scalebox{0.6}{(d)}}$ with
Eqs.~(\ref{low-passfi}) and (\ref{high-passfi}), and each $\widehat \psi^m$
with Eqs.~(\ref{scansdf}) and (\ref{scansdf2}). Therefore, the wavelet properties are adjusted
with an appropriate choice of filter $\LL^{\scalebox{0.6}{(1)}}$. 

If $\LL^{\scalebox{0.6}{(1)}}$ is a one-dimensional filter having
$s+1$ non-zero coefficients then one can verify that each $\psi^m(x)$ has a compact support of
width $s$ \cite{daubechies1992ten}.
Moreover if its Fourier transform satisfies
$\widehat \LL^{\scalebox{0.6}{(1)}} (k) = \sqrt{2} + \mathcal{O}(|k|^q)$ then one can verify that $\widehat \psi^m (k)$  for $1 \leq m \leq 2^d-1$ satisfies $|\widehat \psi^m (k)| = \mathcal{O}(|k|^q)$ at low wave-vectors $k$. The integer $q$ is called the number of vanishing moments of $\psi^m$
because this last properties implies that $\psi^m$ is orthogonal to any polynomial $P(x)$ of degree strictly less than $q$: $\int \psi^m (x)\, P(x) \,dx =0$.
For a fixed $q$, a Daubechies filter \cite{daubechies1992ten} is a filter $\LL^{\scalebox{0.6}{(1)}}$
satisfying Eqs.~(\ref{quadrature0}) and (\ref{supnsdfius}), having $q$ vanishing moments
and a support of minimum size $s = 2q-1$.

Haar wavelets are defined by the Haar filter in Eq.~(\ref{Haarfilt}).
In dimension $d=1$ (hence $m=1$ and we drop off the superscript), one can verify that it defines a Haar wavelet
$\psi (x) = {\bf 1}_{[0,1/2)} - {\bf 1}_{[1/2,1)}$.
It is discontinuous, with a 
compact support of size $s = 1$ and $q = 1$ vanishing moment.
The Kadanoff scheme corresponds to decomposition on a Haar wavelet basis.
Yet it is not sufficiently well localized in the Fourier domain to accurately
approximate singular differential operators with a nearly diagonal matrix.
Wilson instead performed approximate RG calculations \cite{wilson1971renormalization} using 
Shannon wavelets, obtained with
the Shannon filter in Eq.~(\ref{Shannonfilt}). In dimension $d = 1$, the
Fourier transform of $\psi$ is $\widehat \psi(k) = {\bf 1}_{[-2\pi,-\pi)} + {\bf 1}_{[\pi,2\pi)}$.
It is infinitely differentiable and has 
an infinite number of vanishing moments. It is therefore well localized in the
Fourier domain and provides a nearly diagonal approximation of differential operators. However, these wavelets have
infinite support with a slow spatial
decay in real space and are not absolutely integrable. Pointwise polynomial non-linearities
thus producie long-range interactions over Shannon wavelet coefficients, as in the Fourier case.

Haar and Shannon wavelets can be interpreted as Daubechies wavelets having respectively $q = 1$ and $q = \infty$ vanishing moments. To obtain accurate approximations of differential operators and pointwise non-linearities requires to choose $1 < q < \infty$.  In this paper, numerical calculations are performed with
a Daubechies wavelet having $q=4$ vanishing moments, that is called db4 wavelet (or D8 wavelet).
Figure \ref{fig20} shows the graph of a Haar, Shannon, and a Daubechies wavelet having
$4$ vanishing moments in $d=1$ dimension.

\subsection{Proof of Theorem \protect\ref{Specsn}}
\label{Theorem1Proof}

Theorem \ref{Specsn} considers a singular covariance operator
whose eigenvalues $\lambda_\aphi (k)$ have a power-law decay $|k|^{-\zeta}$
and shows that it is represented by
nearly diagonal matrices in a wavelet basis, for wavelets having
$q \geq \zeta / 2$ vanishing moments. Because the wavelet fields are normalized,
it proves that the 
eigenvalues $\lambda_{\dphi_j}$ of the covariance of $\dphi_j$
have a lower bound $A > 0$ and an upper bound $B < \infty$ which do
not depend upon $j$.

We first compute $A$ and then $B$.
We write $\lambda_{\aphi_{j-1}}$ the
eigenvalues of the covariance of $\aphi_{j-1}$ at the site $n$ and $n'$,
\[
C_{\aphi_{j-1}} (n,n') = \lb \aphi_{j-1} (n)\,\aphi_{j-1}(n') \rb_{p_{j-1}}. 
\]
Since $\dphi_j = \gamma_j^{-1} \wG \aphi_{j-1}$ and $\wG$ satisfies the unitary condition in Eq.~(\ref{consdsf}), 
we verify that the wavelet field $\dphi_j$ is obtained from $\aphi_{j-1}$
with an orthogonal projection weighted by $\gamma_j^{-1}$. As a result the covariance of $\dphi_j$ has eigenvalues
between the minimum and maximum eigenvalues of the covariance of $\aphi_{j-1}$ 
multiplied by $\gamma_j^{-2}$ and hence
\begin{equation}
    \label{infonsdf00}
 \inf \{ \lambda_{\dphi_{j}} \} \geq \gamma_j^{-2}\,\inf \{ \lambda_{\varphi_{j-1}} \} . 
\end{equation}
A lower bound of $ \lambda_{\dphi_{j}}$ is thus obtained
by computing a lower bound of $ \lambda_{\varphi_{j-1}}$.
We saw in Eq.~(\ref{waveorthcoeefs0}) that 
\[
\varphi_j (i) = a^{-1}_j\, \int_{\R^d} \varphi(x)\,\psi^0_{j,i} (x)\,  dx,
\]
where $a_j = \prod_{\ell=1}^j \gamma_\ell$,  and we get
\[
C_{\aphi_j}(n,n') = \frac{a_{j}^{-2}}{(2 \pi)^{d}}
 \int _{\R^d} \lambda_\varphi (\om)\,
\widehat \psi^{0}_{j,n} (k)\,\widehat \psi^{0}_{j,n'} (-k)\, dk .
\]
Since $\psi^{0}_{j,n} (x) = 2^{-dj/2} \psi^0(2^{-j} x - n)$ and
$\lambda_\varphi (\om) = c\, |\om|^{-\zeta}$, we get 
\begin{equation}
 C_{\aphi_j}(n,n') =    \frac{c\,a^{-2}_{j} 2^{j \zeta}}{(2 \pi)^{d}} 
\int_{\R^d} |\om|^{-\zeta}\, 
|\widehat \psi^{0}(k)|^2\,e^{i k (n - n')}\, dk .
\end{equation}

By rewriting this integral by a sum of integrals
over $[-\pi,\pi]^d$, this last integral can be rewritten as
\begin{equation}
    \label{covnasfoihsdfs}
C_{\aphi_j}(n,n')
= \frac{1}{(2 \pi)^{d}}
\int_{[-\pi, \pi]^d} \lambda_{\aphi_j}(k) \,
 e^{i k (n - n')}\, d k,
\end{equation}
where its eigenvalues for $k \in [-\pi,\pi]^d$ are
\begin{equation}
    \label{covnasfoihsdfs20}
\lambda_{\aphi_j}(k) = 
{c\,a^{-2}_{j} 2^{j \zeta}}
\sum_{\ell \in \Z^d} |\om + 2 \ell \pi|^{-\zeta}\, 
|\widehat \psi^{0}(k + 2 \ell \pi)|^2 .
\end{equation}
By selecting the first term $\ell = 0$ in the sum Eq.~(\ref{covnasfoihsdfs20}), we derive that for $k \in [-\pi,\pi]^d$,
\begin{equation}
\label{nsdufaasd}
\lambda_{\aphi_j}(k) \geq
{c\,a^{-2}_{j} 2^{j \zeta}\, \Gamma}\, \pi^{-\zeta}  ,
\end{equation}
where $\Gamma = \inf_{k \in [-\pi,\pi]^d} \{ |\widehat \psi^0(k)|^2 \}$.
One can prove that $\Gamma > 0$, because $\widehat \psi^0(k)$ is continuous,
and  Eq.~(\ref{supnsdfius}) with
Eq.~(\ref{scansdf}) guaranties that $\widehat \psi^0(k)$ does
not vanish on $[-\pi,\pi]^d$.

We derive a lower bound
of $\lambda_{\dphi_j}$ from Eqs.~(\ref{infonsdf00}) and (\ref{nsdufaasd}),
by inserting $a_{j} = \gamma_{j} a_{j-1}$,
\begin{equation}
\label{lowen8sdfa}
\inf \{ \lambda_{\dphi_j} \} \geq  c\,a^{-2}_{j}\, 2^{(j-1) \zeta} \, \Gamma\, \pi^{-\zeta}  .
\end{equation}

Let us now compute an upper bound of the eigenvalues
$\lambda_{\dphi_{j}}$ of the covariance of $\dphi_j$. 
The wavelet field $\dphi_j$ has zero average
$\lb \dphi_j (m, n) \rb_{p_{j-1}}  = 0$ and its covariance is thus
\[
C_{\dphi_j} (m, m', n, n') = \lb \dphi_j (m, n)\,\dphi_j(m', n') \rb_{p_{j-1}}  .
\]
Similarly to Eq.~(\ref{covnasfoihsdfs}), we verify that 
$C_{\dphi_j} (m,m',n,n') = C_{\dphi_j} (m,m',n-n')$
and hence that for fixed $m$, $m'$, it is diagonalized in a Fourier basis.
Let us write $\widetilde \lambda_{\dphi_j}(m,m',k)$ the eigenvalue at a wave-vector $k \in [-\pi,\pi]^d$. 
To compute the eigenvalues of $C_{\dphi_j} (m,m',n,n')$, for each $k$ we must also diagonalize the
matrix $\widetilde \lambda_{\dphi_j}(m,m',k)$ along $m,m'$. An upper bound of these eigenvalues 
is obtained by computing the trace of this matrix since all eigenvalues are positive
\begin{equation}
    \label{uppsdfn}
\sup \left\{ \lambda_{\dphi_j} \right\} \leq \sup_{k} \left\{ \sum_{m=1}^{2^d-1} \widetilde \lambda_{\dphi_j}(m,m,k) \right\}.
\end{equation}
Similarly to Eq.~(\ref{covnasfoihsdfs20}), we verify that 
for $k \in [-\pi,\pi]^d$,
\begin{equation}
\label{covnasfoihsdfs22}
 \widetilde \lambda_{\dphi_j}(m,m,k) =
{c\,a^{-2}_{j} 2^{j \zeta} }
\sum_{\ell \in \Z^d} |\om + 2 \ell \pi|^{-\zeta}\, 
|\widehat \psi^{m}(k + 2 \ell \pi)|^2.
\end{equation}
We then have an inequality,
\begin{eqnarray}
\nonumber
 \widetilde \lambda_{\dphi_j}(m,m,k)& \leq & 
{c\,a^{-2}_{j} 2^{j \zeta} }\Big(
|\om |^{-\zeta}\, 
|\widehat \psi^{m}(k)|^2\\
& & + \pi^{-\zeta}\, 
\sum_{\ell \in \Z^d - \{0\}}  
|\widehat \psi^{m}(k + 2 \ell \pi)|^2 \Big). \nonumber \\
\label{insdf978asd}
\end{eqnarray}
The first term is uniformly bounded 
for all $k \in [-\pi,\pi]^d$, because
the wavelets $\psi^m$ have $q \geq \zeta/2$ vanishing moments and hence $|\widehat \psi^m(k)| = \mathcal{O}(|k|^{q})$ for $k $ in the neighborhood of $0$. To control the
second term, observe that 
$\{\psi^m (x-i) \}_{i \in \Z^d}$ is an orthonormal 
family and hence 
\[
\int \psi^m (x-i)\,\psi^m(x)\,dx = \delta(i),
\]
where $\delta(i)$ is the Kronecker's delta.
Computing the Fourier transform along $i$ of this equality gives for all $\om \in [-\pi,\pi]^d$,
\begin{equation}
\label{nsdfuasdfsdd}
\sum_{\ell \in \Z^d} |\widehat \psi^{m}(\om + 2 \ell \pi)|^2 = 1 . 
\end{equation}
We derive from Eq.~(\ref{insdf978asd})
that there exists a finite $\Gamma'$ such that for all $k \in [-\pi,\pi]^d$,
\[
 \widetilde \lambda_{\dphi_j}(m,m,k) \leq
c\, a_{j}^{-2} 2^{j \zeta} \, \Gamma' .
\]
Inserting this inequality in Eq.~(\ref{uppsdfn}) proves that
\begin{equation}
\label{covnasfoihsdfs12}
\sup \{ \lambda_{\dphi_j} \} \leq (2^d-1)\,
c\, a_{j}^{-2} 2^{j \zeta} \, \Gamma' .
\end{equation}

To finish the proof, we 
relate $a_j$ to $2^{j \zeta}$.
The normalization in Eq.~(\ref{normalisation})
implies that 
\begin{equation}
    \label{normalisation6}
  \sum_{m=1}^{2^d-1}   \frac{1}{(2 \pi)^{d}}
\int_{[-\pi, \pi]^d} \widetilde \lambda_{\dphi_j}(m,m,k)\, dk =  \sum_{m=1}^{2^d-1} \langle |\dphi_j(m, n)|^2 \rangle
    = 1  .
\end{equation}
The equality in Eq.~(\ref{covnasfoihsdfs22})  gives
\begin{equation}
    \label{aauppsdfn4}
     a_{j}^{2} \, 2^{-j \zeta} =  c\,\rho
\end{equation}
with
\[
\rho =  \frac{1}{(2 \pi)^{d}}
\sum_{m=1}^{2^d-1} \int_{\R^d} |k|^{-\zeta}\, 
|\widehat \psi^{m}(k)|^2 \, dk.
\]
The constant $\rho$ is finite and strictly positive
because of the vanishing moment condition which imposes that 
$|\widehat \psi^{m}(k)|^2 = {\cal O}(|k|^{2q})$ with $q \geq \zeta/2$
and because 
each wavelet is normalized and hence
\[
\|\psi^m \|^2 = \frac{1}{(2 \pi)^{d}}
\int_{\R^d} 
|\widehat \psi^{m}(k)|^2 \, dk = 1 .
\]
Inserting this in Eq.~(\ref{covnasfoihsdfs12}), we prove with Eq.~(\ref{uppsdfn}) that
\begin{equation}
    \label{uppsdfn4}
\sup \{ \lambda_{\dphi_j} \} \leq (2^d - 1)\,\rho^{-1}\, \Gamma' = B ,
\end{equation}
which finishes the proof of the upper-bound. 
Inserting Eq.~(\ref{aauppsdfn4}) in the lower bound Eq.~(\ref{lowen8sdfa}) gives
\begin{equation}
\label{lowen8sdfa22}
\inf \{ \lambda_{\dphi_j} \} \geq \rho^{-1} \Gamma  (2\pi)^{-\zeta} = A > 0 ,
\end{equation}
which finishes the theorem proof.

\section{Scale Interaction Energies}

\subsection{Scale Interaction Potential and Coupling Across Scales}
\label{interintsce}

This appendix proves the expression Eq.~(\ref{Ingssdcouprs}) of the interaction potential 
$\bHH_j = \bK_j^T \bU_j$.

Equations (\ref{noiusdfaa2}) and (\ref{noiusdfaa}) imply that 
\[
\bK_j^T \bU_j (\aphi_{j-1}) = \K_{j-1}^T  \U_{j-1} (\aphi_{j-1}) - \K_{j-1}^T \U_{j-1} (P \aphi_{j-1}),
\]
where $P = \LL^T \LL$ and
\[
\K_{j-1}^T \U_{j-1} (\aphi_{j-1}) =  \frac 1 2 \aphi_{j-1}^T K_{j-1} \aphi_{j-1} + C_{j-1}^T V(\aphi_{j-1}) .
\]
It results that
\begin{equation}
    \label{Decomnsiudfha}
\bK_{j}^T \bU_j (\aphi_{j-1}) =
 A + C_{j-1}^T \Big(V(\aphi_{j-1} ) - 
V(P \aphi_{j-1}) \Big),
\end{equation}
with
\[
A = \frac{1}{2} \aphi_{j-1}^T K_{j-1} \aphi_{j-1} - \frac{1}{2}
 \aphi_{j-1}^T \LL^T \LL K_{j-1} \LL^T \LL \aphi_{j-1}.
 \]
With Eqs.~(\ref{fastdec3}) and (\ref{fastrec3}), we get
 \begin{align}
 \nonumber
 A = &\frac 1 2 \aphi_{j-1}^T K_{j-1} \aphi_{j-1} - \frac {\gamma_j^2} 2
 \aphi_j^T \LL K_{j-1} \LL^T \aphi_{j} \\
 \label{ensudfadfsdf}
 = & \frac {\gamma_j^2} 2
 \Big( 2 \, \dphi_j^T  \wG K_{j-1} \LL^T \aphi_{j} +  
 \dphi_j^T \wG K_{j-1} \wG^T \dphi_{j} \Big).
 \end{align}
We note that $\aphi_{j}^T  \LL K_{j-1} \wG^T \dphi_j= \dphi_j^T  \wG K_{j-1} \LL^T \aphi_{j}$ because $K_{j-1}$ is symmetric, $K_{j-1}^T=K_{j-1}$.
By iterating on Eq.~(\ref{fastrec3}) we can decompose $\aphi_{j}$ into wavelet fields at all scales,
\begin{eqnarray}
\aphi_{j} &=& \sum_{\ell=j+1}^J a_{j+1,\ell} \left(G^T \right)^{\ell-j-1} \overline G^T \dphi_\ell \nonumber \\ 
&\qquad& \qquad +  \ a_{j+1,J} \nonumber \, 
\left(G^T\right)^{J-j}  \dphi_{J+1} ,
\end{eqnarray}
with $a_{j+1,\ell} = \prod_{\ell' = j+1}^{\ell} \gamma_{\ell'}$ and $\dphi_{J+1} = \aphi_J$. Inserting this
equation in Eq.~(\ref{ensudfadfsdf}) gives
\begin{equation}
\label{DecompAnA}
A = \frac{1}{2} \sum_{\ell=j}^{J+1} \dphi_{j}^T \overline K_{j,\ell} \dphi_{\ell}
\end{equation}
with 
\begin{eqnarray}
\overline K_{j,j} &=& \gamma_j^2 \wG K_{j-1} \wG^T~~, \nonumber \\ 
\overline K_{j,J+1} &=& 2 \gamma_j^2 a_{j+1,J}\, \wG K_{j-1}  (G^T)^{J-j+1},
\end{eqnarray}
and for $j+1 \leq \ell \leq J$,
\[
\overline K_{j,\ell} = 2 \gamma_j^2 \, a_{j+1,\ell}\, \wG K_{j-1}  \left(G^T\right)^{\ell-j} \overline G^T.
\]
Inserting Eq.~(\ref{DecompAnA}) in Eq.~(\ref{Decomnsiudfha}) and prove Eq.~(\ref{Ingssdcouprs}) with $\overline C_j = C_{j-1}$.

 \subsection{Scale Interaction Free energy Calculation}
 \label{app:freeenergy}
 This section computes a regression of the scale interaction free-energy in Eq.~(\ref{freeEne2}) with a model
 $\widetilde \theta_j^T \widetilde U_j (\aphi_j)$ with a thermodynamic integration. 
 We saw in Eq.~(\ref{thermoint0}) that the mean-square error is minimized for
 \[
\widetilde \K_j^{\star} = \langle \wU_j \, \wU_j^T \rangle_{p_j}^{-1} \, \langle \bFF_j \, \wU_j \rangle_{p_j} ,
\]
 which requires to estimate 
 $\langle \overline F_j ^2 \rangle_{p_j}$ and $\langle \overline F_j \, \widetilde U_j \rangle_{p_j}$.
 We estimate these two moments with an empirical average of
 $ \overline F_j ^2 (\aphi_{j,r})$ and $\overline F_j (\aphi_{j,r})\, \widetilde U_j (\aphi_{j,r})$
 over $R$ samples $\aphi_{j,r}$ of $p_j$, computed from $R$ samples $\aphi_{0,r}$ of $p_0$.
 
 Each $\overline F_j (\aphi_{j,r})$ is calculated with a thermodynamic integration. 
 As explained in Eq.~(\ref{interuqnadfimso}), we introduce
 a parameterized energy model which isolates 
 a quadratic term in $\dphi_j$ from the potential term multiplied by $\la \in [0,1]$:
   \begin{equation}
    \label{interuqnadfimso20}
\bK_{j}^T \bU_{j,\lambda} (\aphi_{j-1}) =  Q_{\aphi_j} (\dphi_j) + \lambda \,\overline C_j^T \,V (\aphi_{j-1}) ,
\end{equation}
so that for $\lambda = 0$ we get a Gaussian energy and for $\lambda = 1$ we recover the full scale interaction
energy in Eq.~(\ref{Decomnsiudfha}). It results from Eq.~(\ref{ensudfadfsdf})  that $Q_{\aphi_j}$ can be written by
\begin{equation}
    \label{interuqnadfimso22}
Q_{\aphi_j} (\dphi_j) = \frac{\gamma_j^2} 2 (
2 \dphi_j^T K_j^{\rm hl} \aphi_j + \dphi_j^T K_j^{\rm hh} \dphi_j ) - \overline C_j^T \,V (P \aphi_{j-1}),
\end{equation}
with
\[
K_j^{\rm hl} =  
  \wG K_{j-1} \LL^T ~~\mbox{and}~~
  K_j^{\rm hh} = \wG K_{j-1} \wG^T .
\]
The scale interaction energy model in Eq.~(\ref{interuqnadfimso20}) defines a conditional probability parameterized
by $\la$,
\[
\overline p_{\bK_{j}, \la}(\dphi_j | \aphi_j) = \frac{e^{-\bK_{j}^T \bU_{j,\la} (\aphi_{j-1})}}{\overline Z_{j,\la}(\aphi_j)}\,,
\]
where $\overline Z_{j,\la}(\aphi_j)$ is the normalization factor.
The thermodynamic integration method computes the free-energy by integrating a derivative in $\lambda$,
which yields the following equation:
\begin{eqnarray}
  \label{Rndsoihsdfad}
  \bFF_j (\aphi_j) &=& \bFF_{j, \la=1} (\aphi_j) \nonumber \\
  &=& \bFF_{j, \la=0}(\aphi_j)
 +
\int_0^1  \Big\langle \overline C_j^T \,
           V (\aphi_{j-1}) \Big\rangle_{\overline p_{\overline \theta_{j}, \la}}
             d \lambda , \nonumber \\
\end{eqnarray}
where the free energy for $\la = 0$ is
\[
\bFF_{j,\la=0} (\aphi_j) = -\log \int e^{-Q_{\aphi_j} (\dphi_j)}\, d \dphi_j .
\]
By inserting Eq.~(\ref{interuqnadfimso22}), 
a Gaussian integral calculation gives
\begin{equation}
    \label{Gausnsdfoibsdfa}
\bFF_{j,\la=0} (\aphi_j) = \frac{1}{2} \aphi_j^T M_j \aphi_j - \overline C_j ^T  V(P \aphi_{j-1})  + \tilde c_j
\end{equation}
with
\[
M_j =  \gamma_j^2 (K_j^{\rm hl})^T (K_j^{\rm hh})^{-1} K_j^{\rm hl}  .
 \]
The constant $\tilde c_j$ is not calculated because we compute the free energy up to an additive constant.
As previously explained, the calculation of $\bFF_j(\aphi_j)$ is stable because it is performed over
the wavelet field $\dphi_j$ whose covariance is not singular. As a result, we can approximate the integral 
over $\lambda$ by a Riemann sum with few terms. 

Numerically, for each known realization $\aphi_{j,r}$ of $p_j$, 
we compute $\bFF_j (\aphi_{j,r})$ with Eq.~(\ref{Rndsoihsdfad}). 
We first evaluate $\bFF_{j,\la=0} (\aphi_{j,r})$
with Eq.~(\ref{Gausnsdfoibsdfa}). We then
approximate the integral over $\la$ in Eq.~(\ref{Rndsoihsdfad}) with
a Riemann sum of
$\langle \overline C_j^T \,
           V (\aphi_{j-1}) \rangle_{\overline p_{\overline \theta_{j}, \la}}$
with about $10$ values of $\la$ in $[0,1]$. For 
each fixed $\lambda$, each expected value is estimated by
computing a chain of samples $\dphi^{(t)}_{j,\la}$
of $\overline p_{\bK_{j}, \la}(\dphi_j | \aphi_{j,r})$, with wavelet MC updates at each
time $t$.
We then define
\[
\varphi^{(t)}_{j-1,r,\la} = 
 \gamma_j \LL^T \aphi_{j,r} + \gamma_j \G^T \dphi^{(t)}_{j,\la}~,
 \]
and for each $\aphi_{j,r}$ we perform an empirical average 
of all $\overline C_j^T \, V (\aphi^{(t)}_{j-1,r,\la})$ along the time variable $t$ of the MCMC chain.
Applying Eq.~(\ref{Rndsoihsdfad}) provides an estimation of
$\bFF_j (\aphi_{j,r})$ for each $\aphi_{j,r}$, from which we derive
$\langle \overline F_j ^2 \rangle_{p_j}$ and $\langle \overline F_j \, \widetilde U_j \rangle_{p_j}$ with
empirical averages along $r$.

\section{Methods for Numerical Experiments}
\label{app:numerical_methods}

This appendix contains numerical details on the three type of fields that are studied: Gaussians, $\varphi^4$, and weak gravitational lensing maps. For each system, we explain the numerical algorithms and give additional data supporting our conclusions. 

\subsection{Gaussian Field Theory}
\label{app:gaussian_model}

\noindent
{\bf Model}

The Gaussian model is a particular case of a local energy model. The
potential and coupling parameters in Eqs.~(\ref{nsodufs700}) and (\ref{nsodufs70}) 
are defined by setting the non-linear potential to zero:
$U_j =(  \varphi_j \varphi_j^T, 0)$ and
$\theta_{j} = (\frac 1 2 \,K_{j} , 0)$.
In this paper, we consider the Ornstein-Uhlenbeck Gaussian stochastic process introduced in Sec.~\ref{Gaussiansec}.
We define the model in the Fourier space on a lattice, given by 
\begin{equation}
    \widehat K_0(k) = \beta\left( |k|^2 + (2\pi/\xi)^2 \right),
\end{equation}
where $\xi$ is the correlation length that uniquely characterizes the process.
The overall factor $\beta$ is determined such that $\langle |\aphi_{0}(i)|^2 \rangle_{p_0}=1$.
We study systems with the size $L=8$, $16$, $32$, and 
$64$, varying $\xi=2, 4, 8, 16, 32$, and $64$.

\vspace{0.5cm}
\noindent
{\bf Direct coupling estimation}

The gradient decent dynamics of the direct coupling estimation 
in Eq.~(\ref{gradse2}) is given by
\begin{equation}
  {\K}_j^{(t+1)} - {\K}_j^{(t)} = \epsilon \,\Big(
 \langle \U_j  \rangle_{p_{\K_j^{(t)}}}- 
 \langle \U_j \rangle_{p_j} \Big).
 \label{eq:direct_coupling_app}
\end{equation}
For each time step $t$, $\theta_j^{(t)} = (\frac 1 2 \,K_j^{(t)}, 0)$ defines a Gaussian field whose covariance is the inverse of $K_j^{(t)}$
\begin{equation}
\langle \U_j \rangle_ {p_{\K_{j}^{(t)}}} = \lB  \varphi_j \varphi_j^T  \rB_{p_{\K_{j}^{(t)}}} = \left(K_{j}^{(t)}\right)^{-1}.
\end{equation}
We thus compute $\langle \U_j \rangle_ {p_{\K_{j}^{(t)}}}$ by inverting $ K_{j}^{(t)}$ instead of using
an MCMC algorithm.
For the initialization at $t=0$, we set $K_{j}^{(t=0)}=\sigma_j^{-2}Id$, where $\sigma_j^2 = 1$ is the normalised variance of $\aphi_j$.

\begin{figure}
\includegraphics[width=0.48\linewidth]{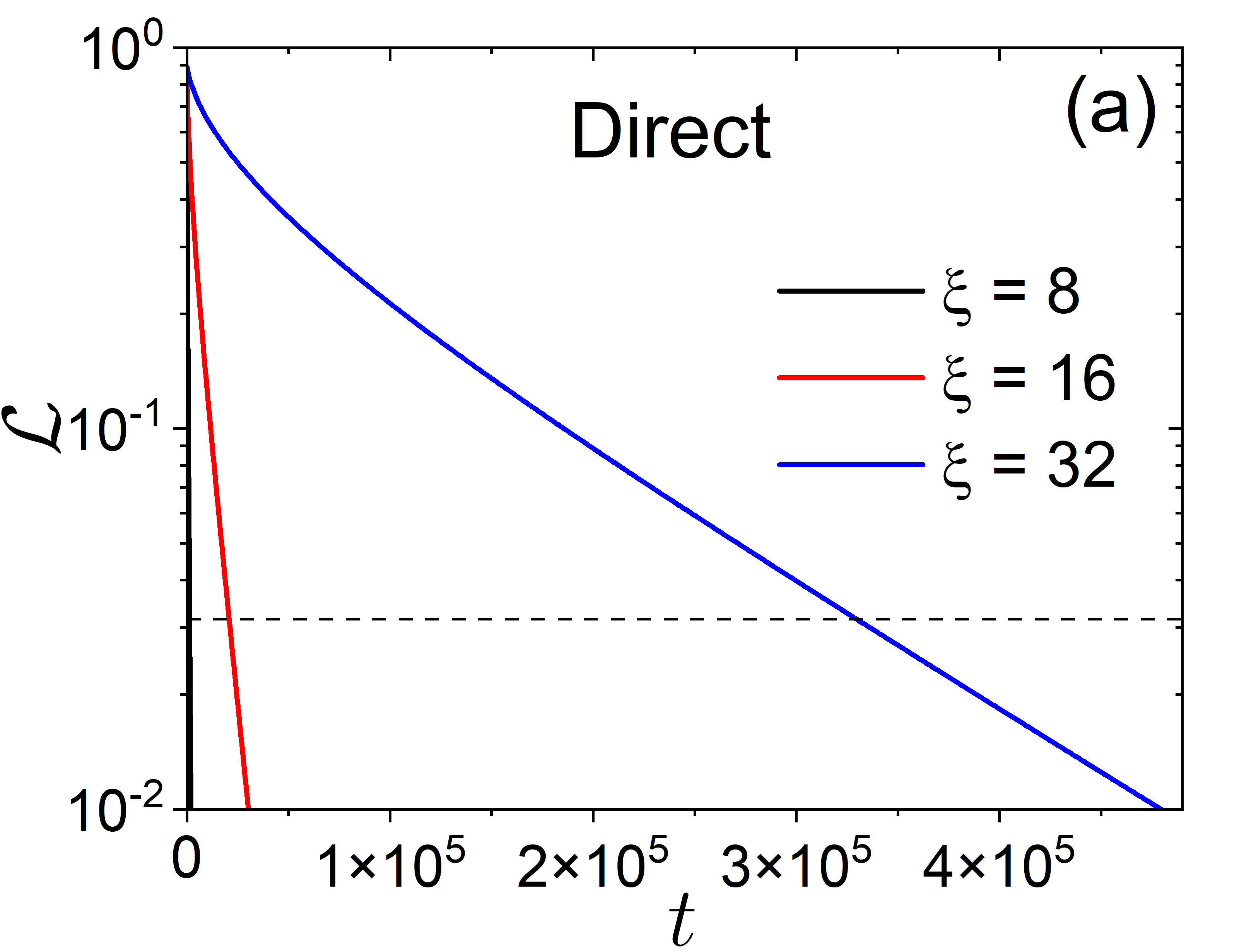}
\includegraphics[width=0.48\linewidth]{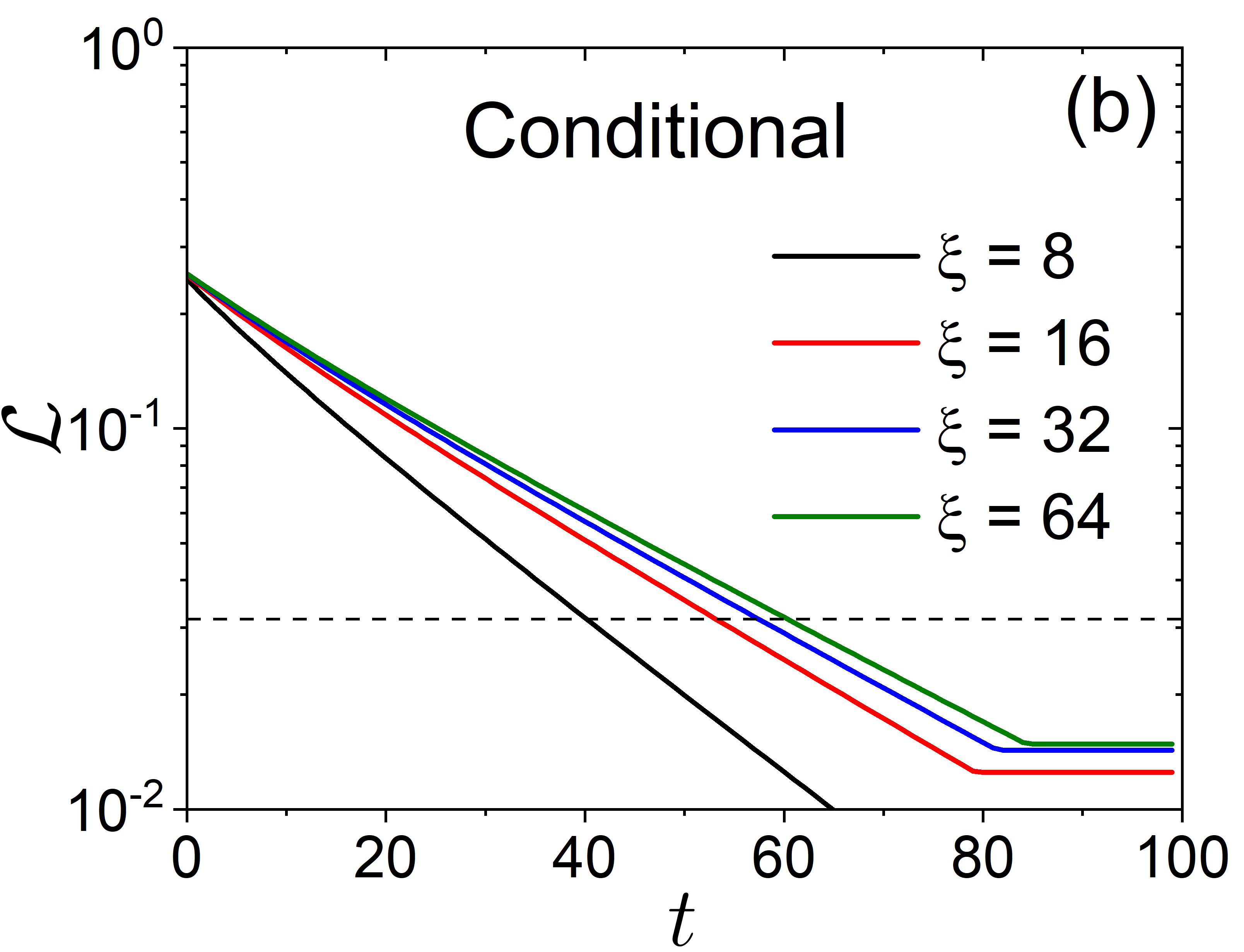}
\includegraphics[width=0.48\linewidth]{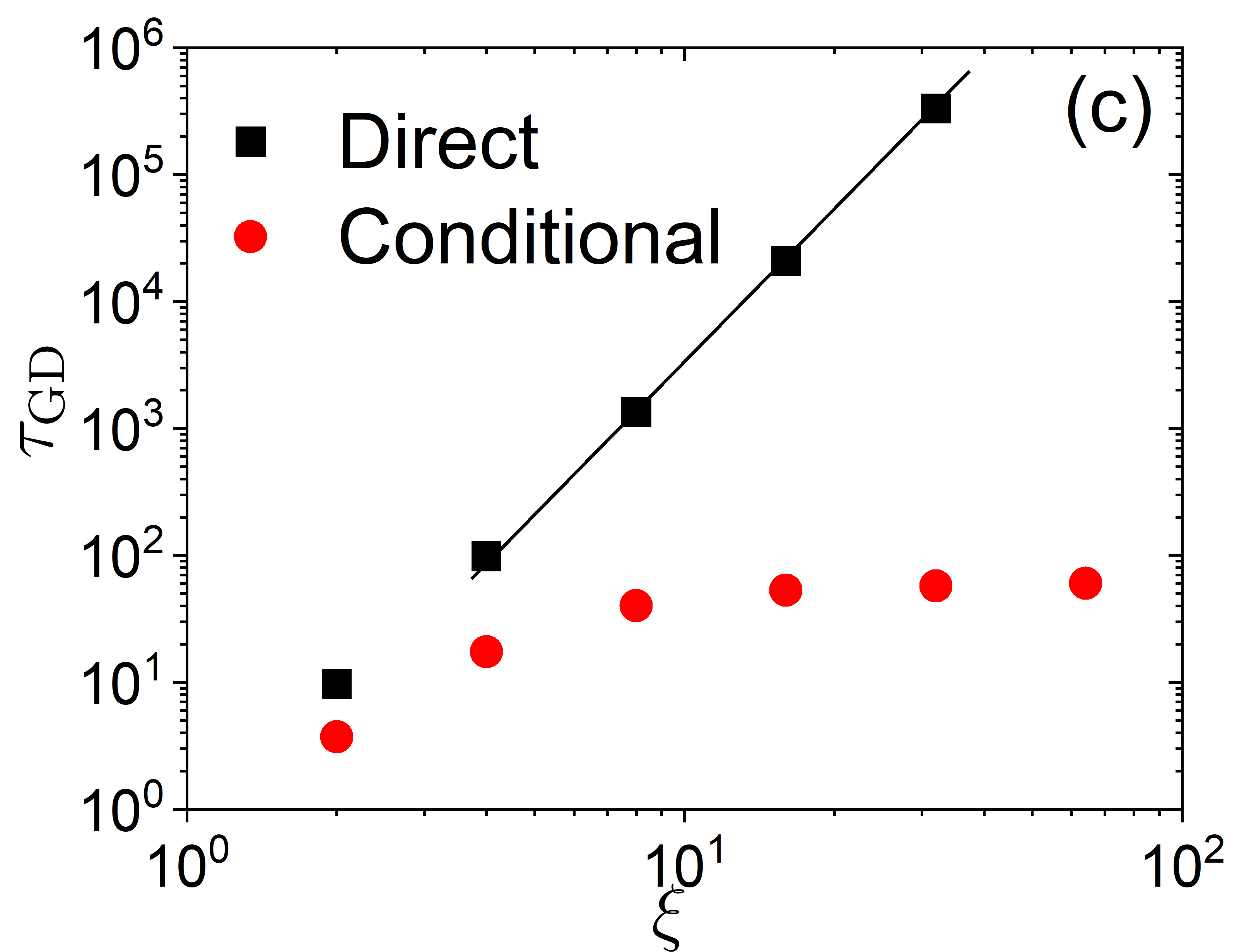}
\includegraphics[width=0.48\linewidth]{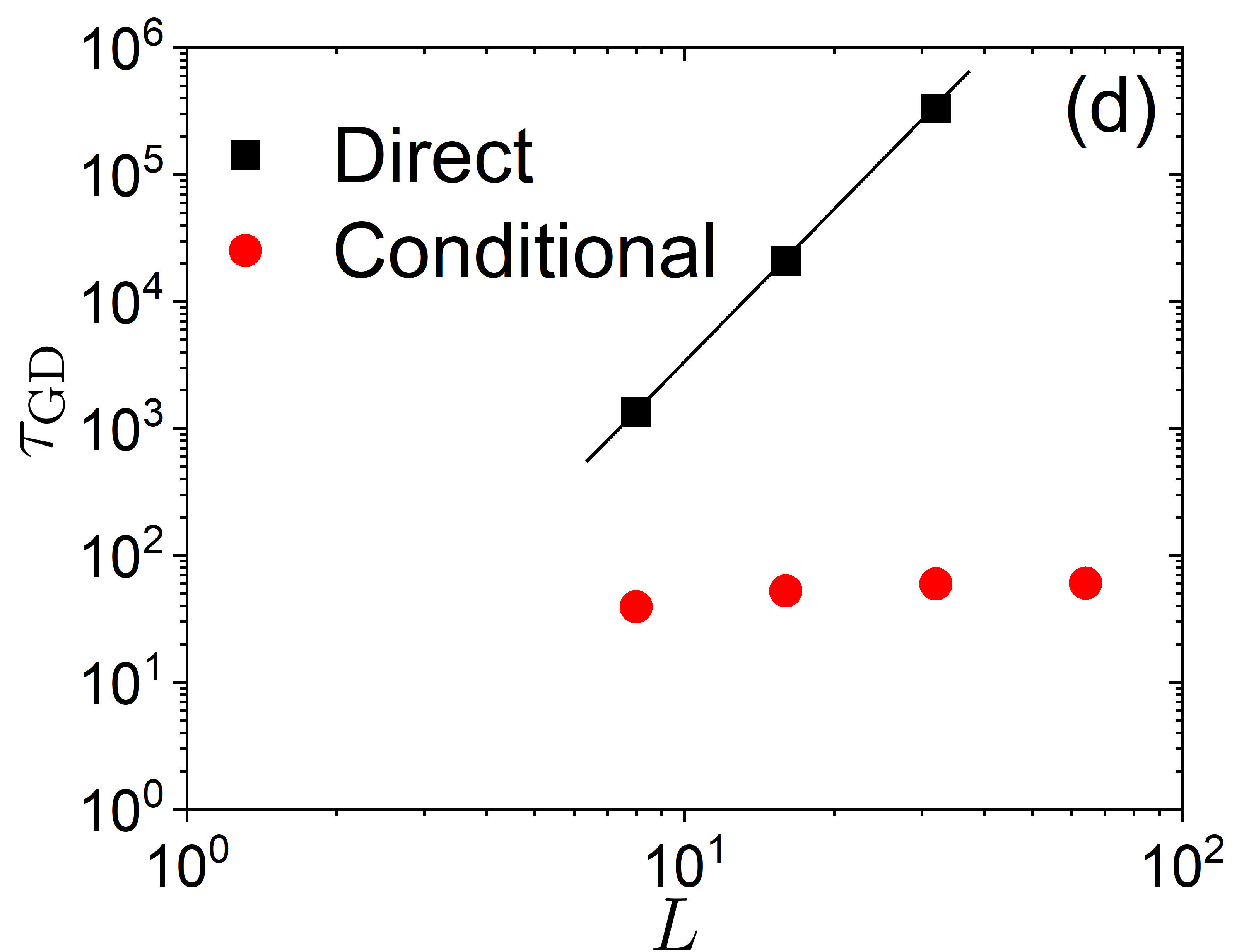}
\caption{
(a, b): Decay of the relative error $\mathcal{L}(t)$ for the direct coupling estimation with $L=32$ in (a), and for the conditional coupling estimation with $L=64$ in (b), for several $\xi$ up to critical point $\xi=L$.
The decay timescale $\tau_{\rm GD}$ is defined by $\mathcal{L}(\tau_{\rm GD})=10^{-1.5}$
(horizontal dashed line).
(c, d): Evolution of $\tau_{\rm GD}$ from the direct and conditional coupling estimations as a function of $\xi$ for $L$ fixed in (c), and at the critical point $L=\xi$ in (d).
The solid straight lines correspond to $\tau_{\rm GD} \sim \xi^4$ in (c) and $\tau_{\rm GD} \sim L^4$ in (d).
}
\label{fig:SD}
\end{figure}

Figure~\ref{fig:SD}(a) shows decay of a normalized error, $\mathcal{L}(t)=\|K_0^{(t)}-K_0\|_s/\|K_0\|_s$, 
where $||\cdot||_s$ is the spectral norm of a matrix.
For $L = 32$, the decay becomes progressively slower when $\xi$ increases. 
The timescale $\tau_{\rm GD}$ of the gradient decent dynamics is defined by
$\mathcal{L}(\tau_{\rm GD})=10^{-1.5} \simeq 0.03$.
Figure~\ref{fig:SD}(c) plots $\tau_{\rm GD}$ versus $\xi$. 
The decay rate is controlled by the maximum step size $\epsilon_{\rm max}$ of $\epsilon$ in Eq.~(\ref{eq:direct_coupling_app}), where $\epsilon_{\rm max}$ is the inverse of the maximum eigenvalue $\lambda_{\rm max}$ of the Hessian matrix $H_{\theta_0}$ in Eq.~(\ref{Hessian1}).
For a Gaussian model, $H_{\theta_0}$ contains only the $4^{\rm th}$-order moments of $\varphi_0$ and they can be written by the product of the second-order moments (Wick's theorem). It
thus indicates that $\tau_{\rm GD}$ varies like
\begin{equation}
    \tau_{\rm GD} \sim \epsilon_{\rm max}^{-1} = \lambda_{\rm max} \sim \left( \widehat K_0^{-1}(k=0) \right)^2 \sim \xi^4.
\end{equation}
The numerical results in Fig.~\ref{fig:SD}(c) follows this asymptotic growth. 
The gradient descent dynamics of the direct coupling estimation has a critical slowing down when the spatial correlation increases. 
Figure~\ref{fig:SD}(d) shows the critical scaling by plotting $\tau_{\rm GD}$ as a function of $L$, at the critical point $\xi=L$.
This scaling is worse than the mixing timescale of a direct MC, namely, $\tau_{\rm MC} \sim \mathcal{O}(\xi^z)$ when $\xi \approx L$ with $z = 2$.

\vspace{0.5cm}
\noindent
{\bf Conditional coupling estimation}

The gradient decent dynamics for the conditional coupling parameters $\bK_{j}$ is computed by Eq.~(\ref{gradse20}),
\begin{equation}
\label{eq:conditional_coupling_app}
\bK_{j}^{(t+1)} - \bK_{j}^{(t)} = \epsilon \,\Big(
\lB \bU_j  \rB_{p_{\bK_{j}^{(t)}}} - 
\lB \bU_j \rB_{p_{j-1}} \Big),
\end{equation}
where $\bU_j = \left( \dphi_j \dphi_j^T, \dphi_{j+1} \dphi_j^T,0 \right)$.
Since $p_{\bK_j^{(t)}}(\aphi_{j-1}) =  \overline p_{\bK_j^{(t)}}(\dphi_j | \aphi_j) \,p_j(\aphi_j)$ is also Gaussian, the
second order moments $\lB \bU_j  \rB_{p_{\bK_{j}^{(t)}}}$ can be also computed by the matrix inversion
without doing an MCMC simulation.
In the WC-RG framework, the conditional coupling estimation is performed scale per scale,  from $j=J$ down to $j=1$.
Figure~\ref{fig:SD} (b) shows the gradient descent dynamics at $j=1$ with $L=64$. It corresponds to the last scale of WC-RG and hence to the highest computational cost among all $j$.
We plot the relative error $\mathcal{L}(t)=\|K_0^{(t)}-K_0\|_s/\|K_0\|_s$ as in Fig.~\ref{fig:SD} (a) in order to compare both results.
For each time step $t$, $K_0^{(t)}$ is recovered with Eq.~(\ref{eq:reconstruction_gaussian}).
As expected, $\mathcal{L}(t)$ has a much faster decay for the conditional coupling estimation than for the direct one.
Note that $\mathcal{L}(t)$ saturates at a small value when $t$ is large. This small error is introduced by the elimination of terms for $\ell >j+1$ in Eq.~(\ref{interaposnit2}).
The decay timescale $\tau_{\rm GD}$ is also defined by
$\mathcal{L}(\tau_{\rm GD})=10^{-1.5}$.
Figure~\ref{fig:SD} (c) plots $\tau_{\rm GD}$ versus $\xi$ (with $L=64$), and 
Fig.~\ref{fig:SD} (d) plots $\tau_{\rm GD}$ versus $L$ (with $L=\xi$). As expected from Fig.~\ref{fig:SD} (b), the decay timescale $\tau_{\rm GD}$ of the conditional coupling estimation is much smaller, and becomes nearly constant when $\xi$ increases for a fixed $L$ and $L$ increases at the critical point. It shows that the conditional coupling estimation is not affected by the critical slowing down. 

\vspace{0.5cm}
\noindent
{\bf Monte-Carlo simulations}

Although it is not needed, we sample Gaussian models with 
MCMC simulation to test the evolution of the mixing time.
We have two MC simulation schemes. The first one is a {\it direct MC} using a direct
update of $\aphi_j$ to estimate $\langle \U_j  \rangle_{p_{\K_j^{(t)}}}$ in the direct coupling estimation in Eq.~(\ref{eq:direct_coupling_app}).
The second one is the {\it wavelet MC} which updates $\dphi_j$ given $\aphi_j$. It is used
to estimate $\lB \bU_j  \rB_{p_{\bK_j^{(t)}}}$ in the conditional coupling estimation in Eq.~(\ref{eq:conditional_coupling_app}), and to compute samples $\dphi_j$ in the WC-RG sampling algorithm of Sec.~\ref{Coafinsec}. 

\begin{figure}
\includegraphics[width=0.48\linewidth]{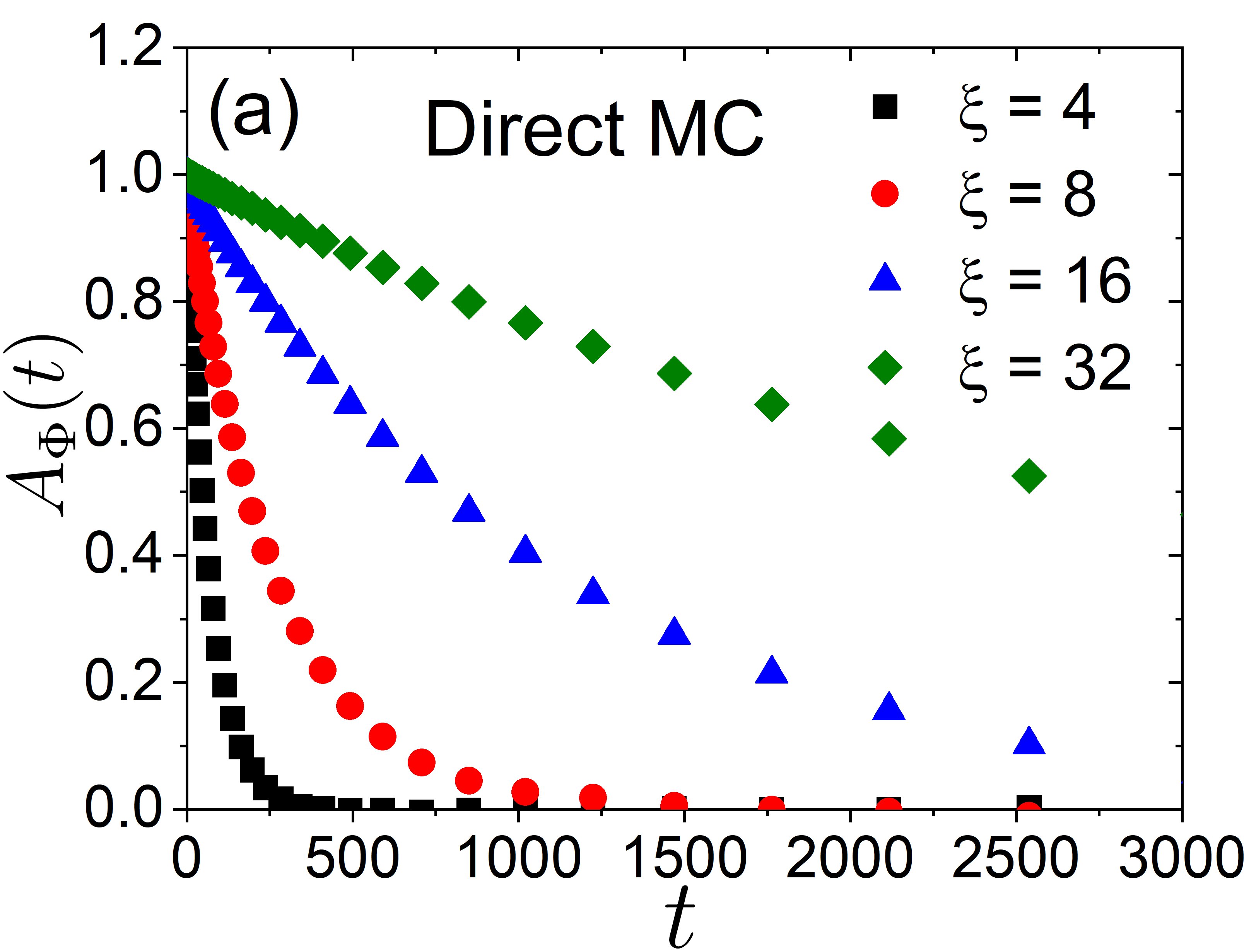}
\includegraphics[width=0.48\linewidth]{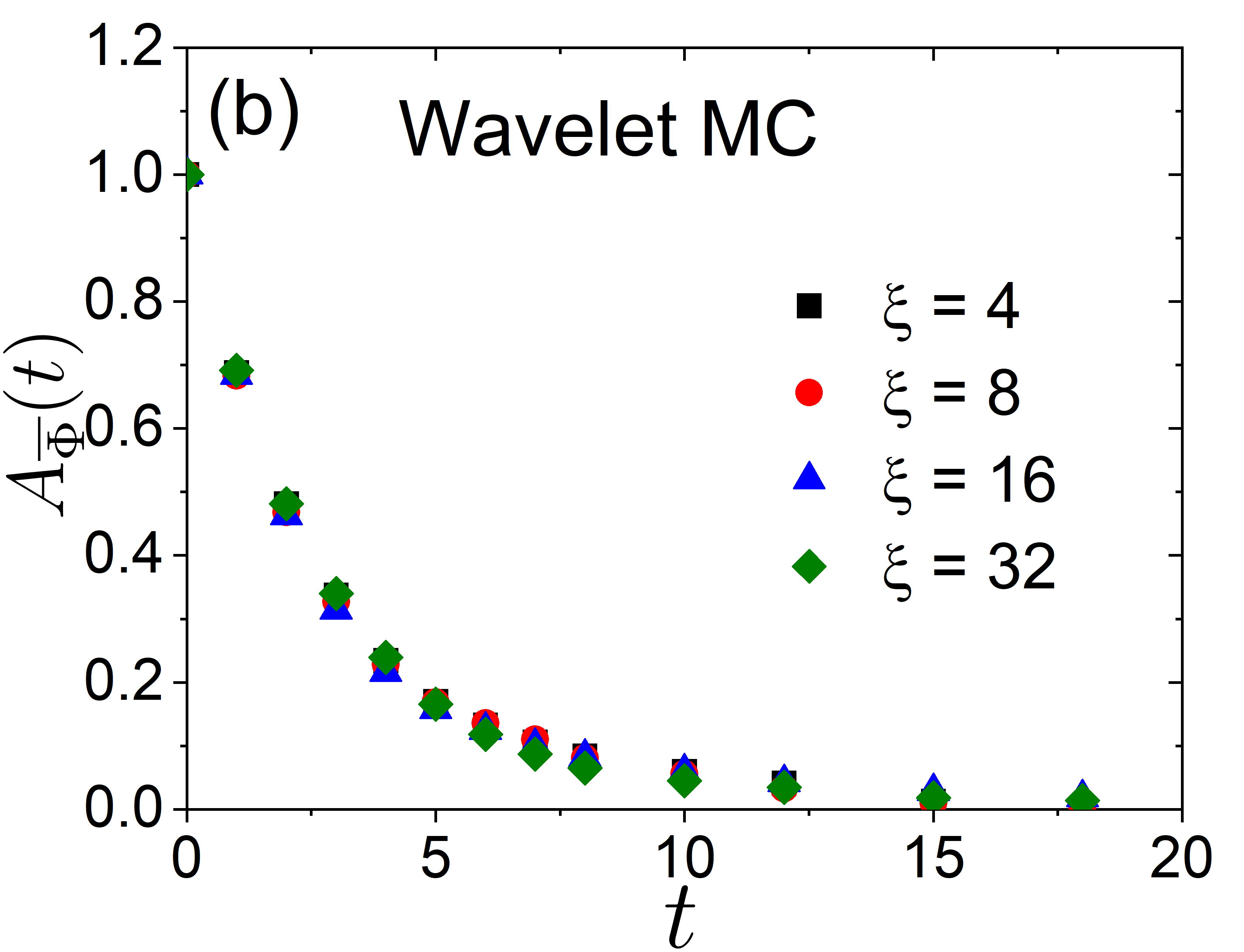}
\includegraphics[width=0.48\linewidth]{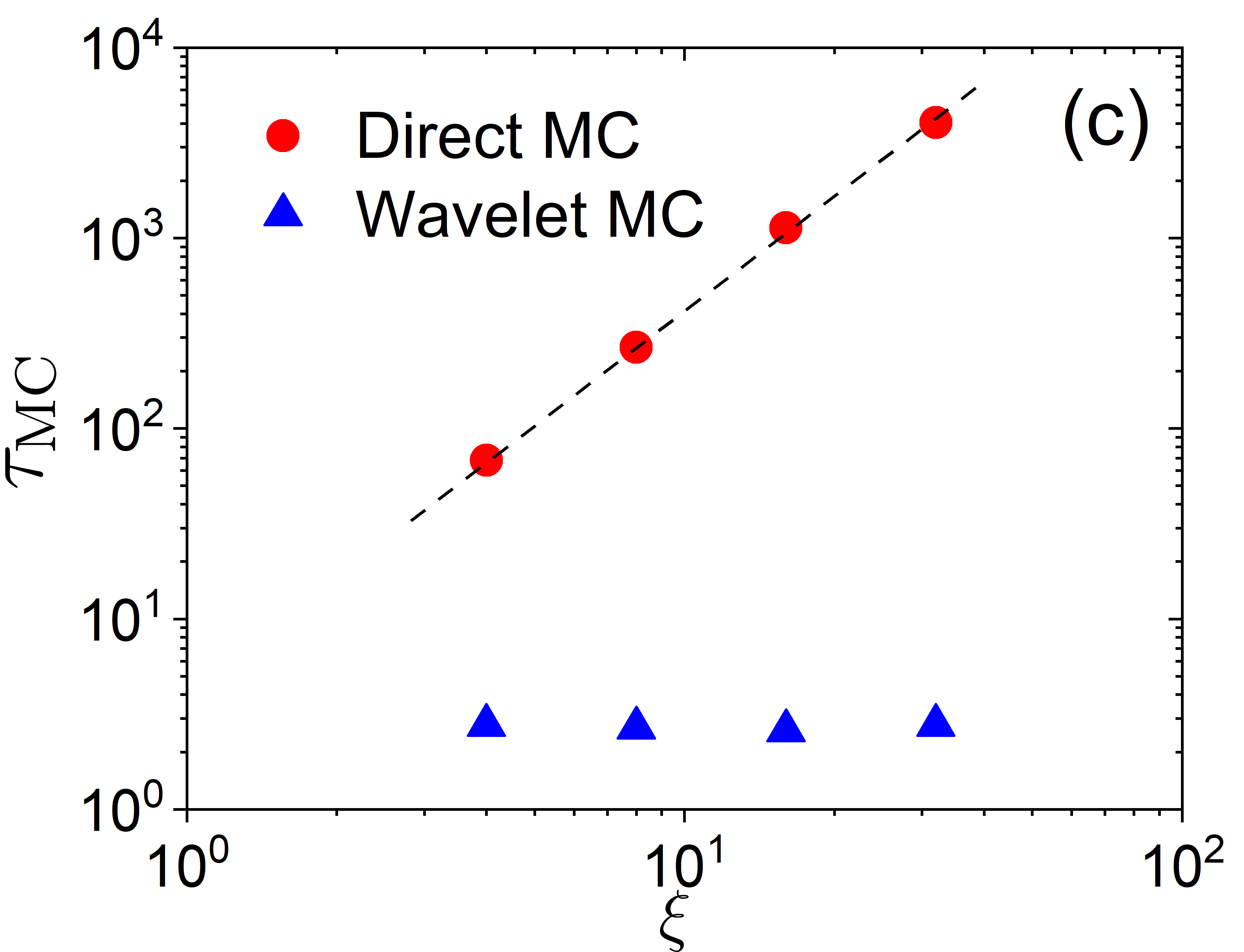}
\includegraphics[width=0.48\linewidth]{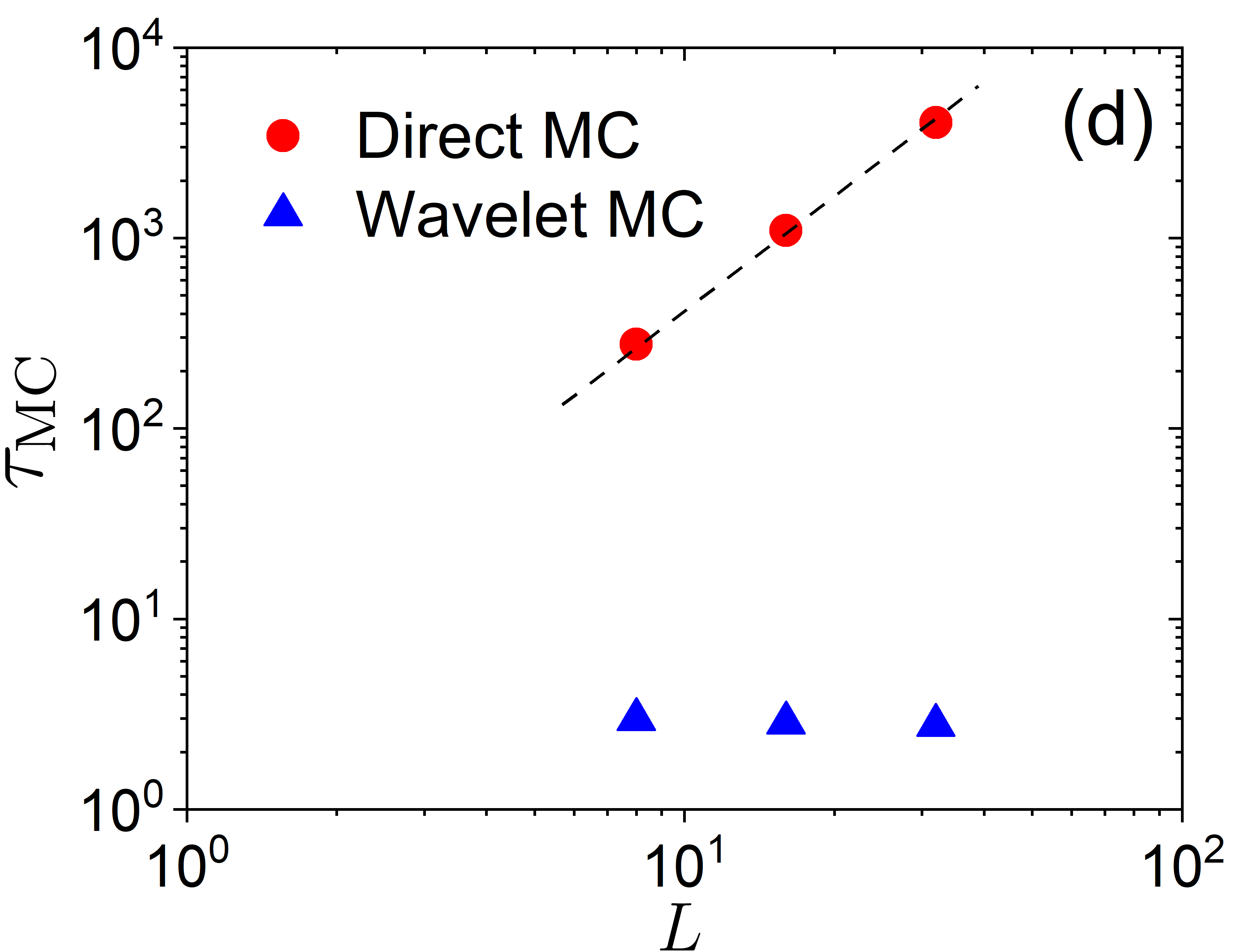}
\caption{MCMC simulations at the stationary state for the Gaussian model. (a, b): Auto-correlation  function for direct MC simulations in (a) and for the wavelet MC simulations in (b), for several $\xi$ with $L=32$.
(c, d): Timescale $\tau_{\rm MC}$ at which the auto-correlation is equal to $e^{-1}$.
In (c), $\xi$ varies whereas $L=32$. In (d), $\tau_{\rm MC}$ is computed at the critical point $\xi=L$, for different $L$. The dashed straight lines in (c) and (d) correspond to $\tau_{\rm MC} \sim \xi^2$ and $\tau_{\rm MC} \sim L^2$, respectively.}  
\label{fig:MC_gauss}
\end{figure}

For the direct MC simulation, we make the following proposal for a site $i$ chosen randomly: $\varphi_j(i) \to \varphi_j(i) + \delta\, (r-1/2)$, where $r$ is a random number uniformly distributed in $(0, 1]$.
We set $\delta=6.0$ to minimize 
the mixing time $\tau_{\rm MC}$ (defined below), at the critical point. The proposed update is accepted or rejected with the standard Metropolis-Hastings rule.
In order to estimate the decorrelation timescale at $j=0$, we compute the time auto-correlation function $A_\Phi(t)$ with Eq.~(\ref{eq:autocorrelation_Phi}).
A unit MC time step for the direct MC corresponds to $L^2$ MC trials (one per pixel on average).  MC simulations are computed up to $2 \times 10^6$ MC steps.
The last $10^6$ MC steps are used to compute the auto-correlation function $A_\Phi(t)$, by discarding the first $10^6$ MC steps. 
We compute ten independent samples to measure $A_\Phi(t)$.
Figure~\ref{fig:MC_gauss} (a) shows $A_\Phi(t)$ for several $\xi$ for $L=32$. The decay of $A_\Phi(t)$ becomes significantly slower when increasing $\xi$.
To evaluate the decay rate, we define a decorrelation (or mixing) timescale $\tau_{\rm MC}$ at which $A_\Phi(\tau_{\rm MC})=1/e$.
Figure~\ref{fig:MC_gauss} (c) shows $\tau_{\rm MC}$ versus $\xi$ for $L=32$, which verifies that $\tau_{\rm MC} \sim \xi^2$.
Figure~\ref{fig:MC_gauss} (d) shows a critical slowing down with $\tau_{\rm MC} \sim L^2$ when $L=\xi$ increases. In both plots, we find that $\tau_{\rm MC} \sim \xi^z$ and $\tau_{\rm MC} \sim L^z$ with $z=2$. This is justified by an argument given in Sec.~\ref{sec:Langevin} for the Ornstein-Uhlenbeck process, which shows that 
$\tau_{\rm MC} \sim \widehat K_0^{-1}(k=0) \sim \xi^2$

\vspace{0.5cm}
\noindent
{\bf Wavelet Monte-Carlo}

A wavelet MC updates only for the wavelet fields $\dphi_j$ given $\aphi_j$ fixed.
For each trial update, a channel $m$ and a site $i$ are chosen randomly to propose $\dphi_j(m, i) \to \dphi_j(m, i) + \delta\, (r-1/2)$.
We apply the standard Metropolis-Hastings rule by evaluating the scale interaction energy function $\overline E_j$. 
We set $\delta=12.0$ which minimizes the mixing time $\tau_{\rm MC}$ (defined below), at the critical point. 
We compute the auto-correlation function $A_{\overline \Phi}(t)$ with Eq.~(\ref{eq:auto_WCRG}) to estimate the decorrelation timescale. For the wavelet MC case, $3(L/2^j)^2$ MC trials correspond to one MC time step (one per pixel on average).
MC simulations are computed over $2 \times 10^3$ MC steps.
The last $10^3$ MC steps are used to compute $A_{\overline \Phi}(t)$, and 
we average over ten independent samples. 
The decorrelation timescale $\tau_{\rm MC}$ is also defined by 
$A_{\overline \Phi}(\tau_{\rm MC})=1/e$.
Figure~\ref{fig:MC_gauss} (b) plots $A_{\overline \Phi}(t)$ which decays much more quickly than with a direct MC in Fig.~\ref{fig:MC_gauss} (a). 
Figure~\ref{fig:MC_gauss}(c) shows that
$\tau_{\rm MC}$ nearly does not depend on $\xi$. Figure~\ref{fig:MC_gauss}(d) also shows that it does not depend on 
$L$ at the critical point.
This is totally different from direct MC simulations.
These results confirm that the dynamics of the wavelet fields do not suffer from any critical slowing down.

\subsection{Two dimensional $\varphi ^4$ field theory}
\label{app:phi_model}

\noindent
{\bf Model}

\begin{figure}
\includegraphics[width=0.75\linewidth]{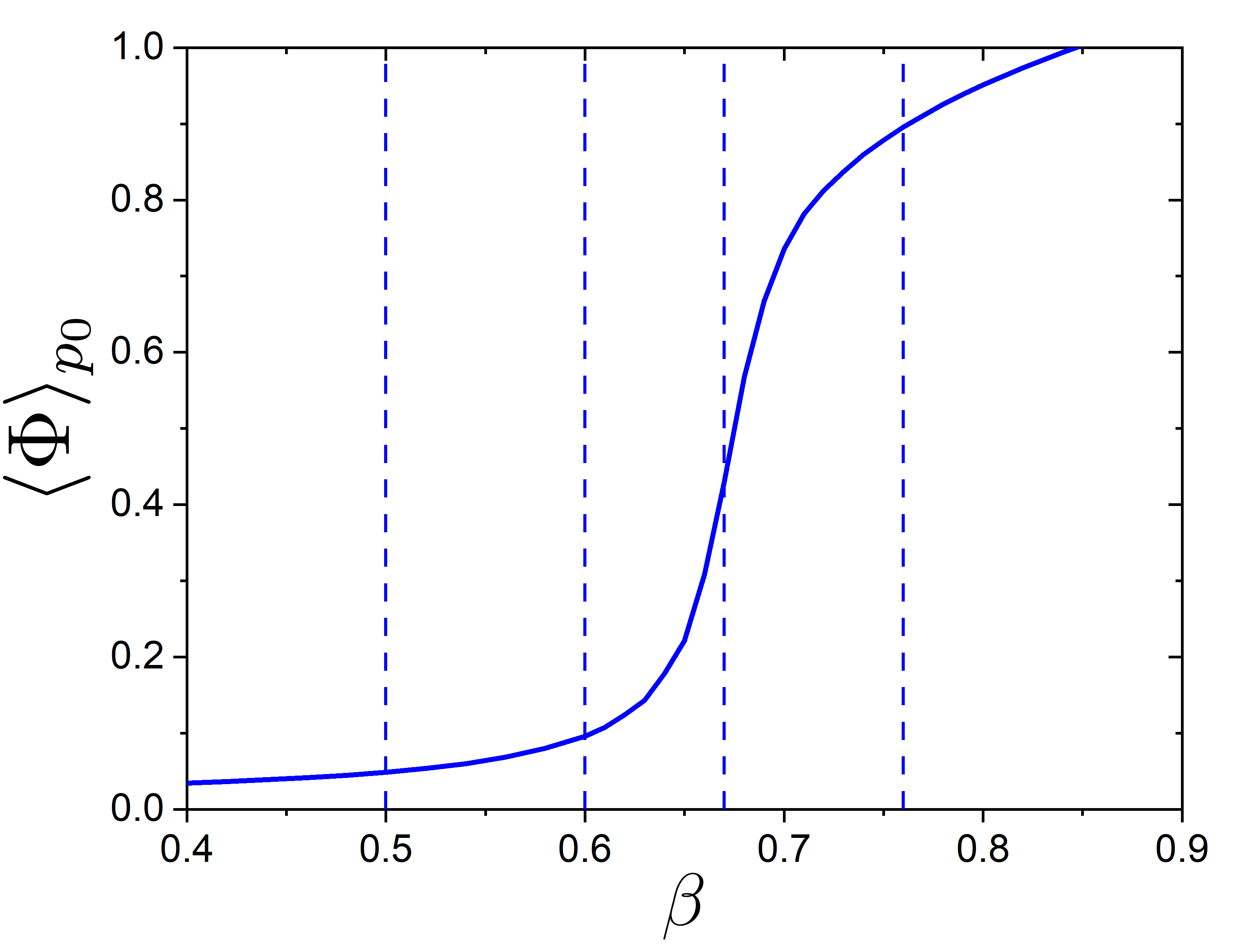}
\caption{Mean value of the magnetization of the $\aphi^4$ model computed over the training data set.
The four state points studied in this paper are shown by dashed lines: $\beta=0.50$, $0.60$, $0.67$, and $0.76$. 
}
\label{fig:example_realizations}
\end{figure}

The second model we consider is a lattice version of the $\varphi^4$ field theory introduced in Sec.~\ref{sec:intro_phi4}, which has been studied widely~\cite{milchev1986finite,hasenbusch1999monte,troster2005free,kaupuvzs2016corrections,zhong2018critical}.
We consider two-dimensional systems of linear size $L=32$.
Let $\Phi = L^{-2} |\sum_i \aphi_0(i)|$.
Figure~\ref{fig:example_realizations} shows that 
the mean magnetization $\langle \Phi \rangle_{p_0}$ increases with
 $\beta$. The effective critical value of $\beta$ for $L=32$ is found to be $
\beta_c\simeq 0.67$ by finding the peak of the  susceptibility. Because of finite size effects~\cite{binder1992finite}, the magnetization has a smooth crossover around $\beta_c$. 
We focus on four values: $\beta = 0.5$, $\beta = 0.6$, $\beta = 0.67 \simeq \beta_{\rm c}$ and $\beta = 0.76$, which are shown by vertical dashed line in Fig.~\ref{fig:example_realizations}. It covers disordered, critical, and ordered phases.
For each $\beta$, we generate $R = 10000$ statistically independent samples of the field $\aphi_0$, which we use as \textit{training dataset} to compute the WC-RG.

\vspace{0.5cm}
\noindent
{\bf Monte-Carlo simulations}

Monte-Carlo simulations for the $\varphi^4$ model are calculated as in the Gaussian model described above, except for the presence of a non-linear potential term in $E_j$ and $\overline E_j$.
To generate the training dataset, a direct MC is computed at the finest scale
$j=0$ with periodic boundary conditions.
It updates $\varphi_0(i)$ using MC trials with $\delta =3.0$~\cite{hasenbusch1999monte,zhong2018critical}.
For the direct and conditional coupling estimations in Eqs.~(\ref{gradse2}) and (\ref{gradse20}),
we estimate moments with a direct MC and a wavelet MC, respectively. 
The value of $\delta$ evolves dynamically during simulation so that the acceptance ratio remains nearly $50\%$.
To measure the auto-correlation functions $A_{\Phi}(t)$ and $A_{\overline \Phi}(t)$ in Fig.~\ref{fig:MC_phi4A}, we fix $\delta=3.0$ and $1.0$, respectively.

\vspace{0.5cm}
\noindent
{\bf Coupling parameter estimations}

In a direct coupling estimation, the gradient descent dynamics minimizes the KL divergence with Eq.~(\ref{gradse2}).  Prior information on symmetries reduces the number of coupling parameters that needs to be estimated. 
Due to the translational invariance, $K_j$ is defined by an $L/2^j \times L/2^j$ periodic kernel. 
We impose that $K_j$ is symmetric to transposition and reflection, which reduces the number of matrix elements to estimate.
The non-linear potential is defined by $36$ hat functions when
the system is in the ordered broken-symmetry phase with $\beta=0.76$.
When $\beta=0.5, 0.6$, and $0.67$, which correspond to disordered symmetric and critical phases, we use $17$ hat functions. 
For the WC-RG approach, the gradient descent dynamics in Eq.~(\ref{gradse2}) is 
computed scale per scale, from the coarsest scale, $j=5$ ($L/2^j=1$), down to $j=1$ ($L/2^j=16$). It is defined by 15 scale interaction matrices $\overline K_{j,\ell}$, and each of them is a circulant matrix having transposition and reflection symmetries. We further impose symmetries between different scale interaction matrices, originating from rotational symmetries of two-dimensional wavelet transform.
For the non-linear potential, we use the same set of the hat functions as the direct coupling estimation at all scales $2^j$.

\vspace{0.5cm}
\noindent
{\bf Thermodynamic Integration and Regression}

To estimate the energy function $E_j$ at each scale, we compute
a thermodynamic integration and a linear regression described in Sec.~\ref{app:freeenergy}.
The thermodynamic integration~\cite{frenkel2001understanding} starts with the Gaussian model, where the integration can be performed analytically. 
As explain in Sec.~\ref{app:freeenergy}, We multiply second term in Eq.~(\ref{Ingssdcouprs}) by $\lambda$ and estimate statistical averages of these non-Gaussian terms 
for each $\lambda$. We approximate the integral by a Riemann sum computed over
$10$ values of $\lambda$, uniformly distributed in $[0,1]$.
Figures~\ref{fig:result_tc1} and \ref{fig:result_tc2} 
compare the estimated energy function $E_0(\aphi_0)$ and the original one for $\beta \simeq \beta_c$.

\subsection{Cosmological data}
\label{App:cosmology}

We use a set of simulated convergence maps computed by the Columbia lensing group~\cite{matilla2016dark,gupta2018non}. These convergence maps are calculated by ray-tracing N-body simulations. The sample dataset is available at
\verb+ http://columbialensing.org/+
(Section "Dark Matter"). They simulate convergence maps of the next-generation space telescope {\it Euclid}. 
It is a mission of the European Space Agency whose goal is to map the geometry of the Universe and better understand dark matter and dark energy.
The cosmology is defined by the matter density parameter, $\Omega_{\rm m}=0.26$, and the root-mean-square matter fluctuation, $\sigma_8=0.8$.
We sub-sampled the $1024\times 1024$ maps by a factor of $4$ with a local averaging, and we extracted $32\times32$ patches from the sub-sampled dataset. We also introduced a cutoff of
the maximum amplitude of peaks, as shown in Fig.~\ref{fig:Euclid_result_2}. The resulting dataset was then normalized to have a mean $0$ and variance $1$.
We use $R =78126$ images of size $32\times32$ as a training dataset to compute
the WC-RG, but $R \simeq 3000$ is enough to get nearly the same results.

The gradient descent of WC-RG is computed with the same setting as the $\varphi^4$ model for $\beta=0.76$, which also has an asymmetric distribution. Namely, we use the same size of scale interaction matrices $\overline K_{\ell,j}$ for the three wavelet channels, and 
we decompose the nonlinear potential over hat functions. The number of hat functions is $23$. We also include ReLU functions at the edges, with coefficients set to $-10$ and $10$ at the left and right edges, respectively. These ReLU functions confine the field values over an appropriate interval.
This is needed for weak lensing data, which have a long-tail distribution that is
nearly Laplacian.

\providecommand{\noopsort}[1]{}\providecommand{\singleletter}[1]{#1}%
%


\providecommand{\noopsort}[1]{}\providecommand{\singleletter}[1]{#1}%
\begin{thebibliography}{84}%
\makeatletter
\providecommand \@ifxundefined [1]{%
 \@ifx{#1\undefined}
}%
\providecommand \@ifnum [1]{%
 \ifnum #1\expandafter \@firstoftwo
 \else \expandafter \@secondoftwo
 \fi
}%
\providecommand \@ifx [1]{%
 \ifx #1\expandafter \@firstoftwo
 \else \expandafter \@secondoftwo
 \fi
}%
\providecommand \natexlab [1]{#1}%
\providecommand \enquote  [1]{``#1''}%
\providecommand \bibnamefont  [1]{#1}%
\providecommand \bibfnamefont [1]{#1}%
\providecommand \citenamefont [1]{#1}%
\providecommand \href@noop [0]{\@secondoftwo}%
\providecommand \href [0]{\begingroup \@sanitize@url \@href}%
\providecommand \@href[1]{\@@startlink{#1}\@@href}%
\providecommand \@@href[1]{\endgroup#1\@@endlink}%
\providecommand \@sanitize@url [0]{\catcode `\\12\catcode `\$12\catcode
  `\&12\catcode `\#12\catcode `\^12\catcode `\_12\catcode `\%12\relax}%
\providecommand \@@startlink[1]{}%
\providecommand \@@endlink[0]{}%
\providecommand \url  [0]{\begingroup\@sanitize@url \@url }%
\providecommand \@url [1]{\endgroup\@href {#1}{\urlprefix }}%
\providecommand \urlprefix  [0]{URL }%
\providecommand \Eprint [0]{\href }%
\providecommand \doibase [0]{https://doi.org/}%
\providecommand \selectlanguage [0]{\@gobble}%
\providecommand \bibinfo  [0]{\@secondoftwo}%
\providecommand \bibfield  [0]{\@secondoftwo}%
\providecommand \translation [1]{[#1]}%
\providecommand \BibitemOpen [0]{}%
\providecommand \bibitemStop [0]{}%
\providecommand \bibitemNoStop [0]{.\EOS\space}%
\providecommand \EOS [0]{\spacefactor3000\relax}%
\providecommand \BibitemShut  [1]{\csname bibitem#1\endcsname}%
\let\auto@bib@innerbib\@empty
\bibitem [{\citenamefont {Perraudin}\ \emph {et~al.}(2019)\citenamefont
  {Perraudin}, \citenamefont {Srivastava}, \citenamefont {Lucchi},
  \citenamefont {Kacprzak}, \citenamefont {Hofmann},\ and\ \citenamefont
  {R{\'e}fr{\'e}gier}}]{perraudin2019cosmological}%
  \BibitemOpen
  \bibfield  {author} {\bibinfo {author} {\bibfnamefont {N.}~\bibnamefont
  {Perraudin}}, \bibinfo {author} {\bibfnamefont {A.}~\bibnamefont
  {Srivastava}}, \bibinfo {author} {\bibfnamefont {A.}~\bibnamefont {Lucchi}},
  \bibinfo {author} {\bibfnamefont {T.}~\bibnamefont {Kacprzak}}, \bibinfo
  {author} {\bibfnamefont {T.}~\bibnamefont {Hofmann}},\ and\ \bibinfo {author}
  {\bibfnamefont {A.}~\bibnamefont {R{\'e}fr{\'e}gier}},\ }\bibfield  {title}
  {\bibinfo {title} {Cosmological n-body simulations: a challenge for scalable
  generative models},\ }\href@noop {} {\bibfield  {journal} {\bibinfo
  {journal} {Computational Astrophysics and Cosmology}\ }\textbf {\bibinfo
  {volume} {6}},\ \bibinfo {pages} {1} (\bibinfo {year} {2019})}\BibitemShut
  {NoStop}%
\bibitem [{\citenamefont {Carrasquilla}\ and\ \citenamefont
  {Melko}(2017)}]{carrasquilla2017machine}%
  \BibitemOpen
  \bibfield  {author} {\bibinfo {author} {\bibfnamefont {J.}~\bibnamefont
  {Carrasquilla}}\ and\ \bibinfo {author} {\bibfnamefont {R.~G.}\ \bibnamefont
  {Melko}},\ }\bibfield  {title} {\bibinfo {title} {Machine learning phases of
  matter},\ }\href@noop {} {\bibfield  {journal} {\bibinfo  {journal} {Nature
  Physics}\ }\textbf {\bibinfo {volume} {13}},\ \bibinfo {pages} {431}
  (\bibinfo {year} {2017})}\BibitemShut {NoStop}%
\bibitem [{\citenamefont {No{\'e}}\ \emph {et~al.}(2019)\citenamefont
  {No{\'e}}, \citenamefont {Olsson}, \citenamefont {K{\"o}hler},\ and\
  \citenamefont {Wu}}]{noe2019boltzmann}%
  \BibitemOpen
  \bibfield  {author} {\bibinfo {author} {\bibfnamefont {F.}~\bibnamefont
  {No{\'e}}}, \bibinfo {author} {\bibfnamefont {S.}~\bibnamefont {Olsson}},
  \bibinfo {author} {\bibfnamefont {J.}~\bibnamefont {K{\"o}hler}},\ and\
  \bibinfo {author} {\bibfnamefont {H.}~\bibnamefont {Wu}},\ }\bibfield
  {title} {\bibinfo {title} {Boltzmann generators: Sampling equilibrium states
  of many-body systems with deep learning},\ }\href@noop {} {\bibfield
  {journal} {\bibinfo  {journal} {Science}\ }\textbf {\bibinfo {volume}
  {365}},\ \bibinfo {pages} {eaaw1147} (\bibinfo {year} {2019})}\BibitemShut
  {NoStop}%
\bibitem [{\citenamefont {Gabri{\'e}}\ \emph {et~al.}(2021)\citenamefont
  {Gabri{\'e}}, \citenamefont {Rotskoff},\ and\ \citenamefont
  {Vanden-Eijnden}}]{gabrie2021efficient}%
  \BibitemOpen
  \bibfield  {author} {\bibinfo {author} {\bibfnamefont {M.}~\bibnamefont
  {Gabri{\'e}}}, \bibinfo {author} {\bibfnamefont {G.~M.}\ \bibnamefont
  {Rotskoff}},\ and\ \bibinfo {author} {\bibfnamefont {E.}~\bibnamefont
  {Vanden-Eijnden}},\ }\bibfield  {title} {\bibinfo {title} {Efficient bayesian
  sampling using normalizing flows to assist markov chain monte carlo
  methods},\ }\href@noop {} {\bibfield  {journal} {\bibinfo  {journal} {arXiv
  preprint arXiv:2107.08001}\ } (\bibinfo {year} {2021})}\BibitemShut {NoStop}%
\bibitem [{\citenamefont {Schneidman}\ \emph {et~al.}(2006)\citenamefont
  {Schneidman}, \citenamefont {Berry}, \citenamefont {Segev},\ and\
  \citenamefont {Bialek}}]{schneidman2006weak}%
  \BibitemOpen
  \bibfield  {author} {\bibinfo {author} {\bibfnamefont {E.}~\bibnamefont
  {Schneidman}}, \bibinfo {author} {\bibfnamefont {M.~J.}\ \bibnamefont
  {Berry}}, \bibinfo {author} {\bibfnamefont {R.}~\bibnamefont {Segev}},\ and\
  \bibinfo {author} {\bibfnamefont {W.}~\bibnamefont {Bialek}},\ }\bibfield
  {title} {\bibinfo {title} {Weak pairwise correlations imply strongly
  correlated network states in a neural population},\ }\href@noop {} {\bibfield
   {journal} {\bibinfo  {journal} {Nature}\ }\textbf {\bibinfo {volume}
  {440}},\ \bibinfo {pages} {1007} (\bibinfo {year} {2006})}\BibitemShut
  {NoStop}%
\bibitem [{\citenamefont {Cocco}\ \emph {et~al.}(2018)\citenamefont {Cocco},
  \citenamefont {Feinauer}, \citenamefont {Figliuzzi}, \citenamefont
  {Monasson},\ and\ \citenamefont {Weigt}}]{cocco2018inverse}%
  \BibitemOpen
  \bibfield  {author} {\bibinfo {author} {\bibfnamefont {S.}~\bibnamefont
  {Cocco}}, \bibinfo {author} {\bibfnamefont {C.}~\bibnamefont {Feinauer}},
  \bibinfo {author} {\bibfnamefont {M.}~\bibnamefont {Figliuzzi}}, \bibinfo
  {author} {\bibfnamefont {R.}~\bibnamefont {Monasson}},\ and\ \bibinfo
  {author} {\bibfnamefont {M.}~\bibnamefont {Weigt}},\ }\bibfield  {title}
  {\bibinfo {title} {Inverse statistical physics of protein sequences: a key
  issues review},\ }\href@noop {} {\bibfield  {journal} {\bibinfo  {journal}
  {Reports on Progress in Physics}\ }\textbf {\bibinfo {volume} {81}},\
  \bibinfo {pages} {032601} (\bibinfo {year} {2018})}\BibitemShut {NoStop}%
\bibitem [{\citenamefont {Geman}\ and\ \citenamefont
  {Geman}(1984)}]{geman1984stochastic}%
  \BibitemOpen
  \bibfield  {author} {\bibinfo {author} {\bibfnamefont {S.}~\bibnamefont
  {Geman}}\ and\ \bibinfo {author} {\bibfnamefont {D.}~\bibnamefont {Geman}},\
  }\bibfield  {title} {\bibinfo {title} {Stochastic relaxation, gibbs
  distributions, and the bayesian restoration of images},\ }\href@noop {}
  {\bibfield  {journal} {\bibinfo  {journal} {IEEE Transactions on pattern
  analysis and machine intelligence}\ ,\ \bibinfo {pages} {721}} (\bibinfo
  {year} {1984})}\BibitemShut {NoStop}%
\bibitem [{\citenamefont {Portilla}\ and\ \citenamefont
  {Simoncelli}(2000)}]{portilla2000parametric}%
  \BibitemOpen
  \bibfield  {author} {\bibinfo {author} {\bibfnamefont {J.}~\bibnamefont
  {Portilla}}\ and\ \bibinfo {author} {\bibfnamefont {E.~P.}\ \bibnamefont
  {Simoncelli}},\ }\bibfield  {title} {\bibinfo {title} {A parametric texture
  model based on joint statistics of complex wavelet coefficients},\
  }\href@noop {} {\bibfield  {journal} {\bibinfo  {journal} {International
  journal of computer vision}\ }\textbf {\bibinfo {volume} {40}},\ \bibinfo
  {pages} {49} (\bibinfo {year} {2000})}\BibitemShut {NoStop}%
\bibitem [{\citenamefont {Zhu}\ \emph {et~al.}(1997)\citenamefont {Zhu},
  \citenamefont {Wu},\ and\ \citenamefont {Mumford}}]{zhu1997minimax}%
  \BibitemOpen
  \bibfield  {author} {\bibinfo {author} {\bibfnamefont {S.~C.}\ \bibnamefont
  {Zhu}}, \bibinfo {author} {\bibfnamefont {Y.~N.}\ \bibnamefont {Wu}},\ and\
  \bibinfo {author} {\bibfnamefont {D.}~\bibnamefont {Mumford}},\ }\bibfield
  {title} {\bibinfo {title} {Minimax entropy principle and its application to
  texture modeling},\ }\href@noop {} {\bibfield  {journal} {\bibinfo  {journal}
  {Neural computation}\ }\textbf {\bibinfo {volume} {9}},\ \bibinfo {pages}
  {1627} (\bibinfo {year} {1997})}\BibitemShut {NoStop}%
\bibitem [{\citenamefont {Kingma}\ and\ \citenamefont {Welling}(2014)}]{vae}%
  \BibitemOpen
  \bibfield  {author} {\bibinfo {author} {\bibfnamefont {D.}~\bibnamefont
  {Kingma}}\ and\ \bibinfo {author} {\bibfnamefont {M.}~\bibnamefont
  {Welling}},\ }\bibfield  {title} {\bibinfo {title} {Auto-encoding variational
  bayes},\ }in\ \href@noop {} {\emph {\bibinfo {booktitle} {Proc. of ICLR}}}\
  (\bibinfo {year} {2014})\BibitemShut {NoStop}%
\bibitem [{\citenamefont {Radford}\ \emph {et~al.}(2016)\citenamefont
  {Radford}, \citenamefont {Metz},\ and\ \citenamefont {Chintala}}]{Chintala}%
  \BibitemOpen
  \bibfield  {author} {\bibinfo {author} {\bibfnamefont {A.}~\bibnamefont
  {Radford}}, \bibinfo {author} {\bibfnamefont {L.}~\bibnamefont {Metz}},\ and\
  \bibinfo {author} {\bibfnamefont {R.}~\bibnamefont {Chintala}},\ }\bibfield
  {title} {\bibinfo {title} {Unsupervised representation learning with deep
  convolutional generative adversarial networks},\ }in\ \href@noop {} {\emph
  {\bibinfo {booktitle} {Proc. of ICLR}}}\ (\bibinfo {year} {2016})\BibitemShut
  {NoStop}%
\bibitem [{\citenamefont {Zinn-Justin}(2002)}]{zinn2002quantum}%
  \BibitemOpen
  \bibfield  {author} {\bibinfo {author} {\bibfnamefont {J.}~\bibnamefont
  {Zinn-Justin}},\ }\href@noop {} {\emph {\bibinfo {title} {Quantum field
  theory and critical phenomena}}},\ Vol.\ \bibinfo {volume} {113}\ (\bibinfo
  {publisher} {Clarendon Press, Oxford},\ \bibinfo {year} {2002})\BibitemShut
  {NoStop}%
\bibitem [{\citenamefont {Bellman}\ and\ \citenamefont
  {Kalaba}(1959)}]{bellman1959mathematical}%
  \BibitemOpen
  \bibfield  {author} {\bibinfo {author} {\bibfnamefont {R.}~\bibnamefont
  {Bellman}}\ and\ \bibinfo {author} {\bibfnamefont {R.}~\bibnamefont
  {Kalaba}},\ }\bibfield  {title} {\bibinfo {title} {A mathematical theory of
  adaptive control processes},\ }\href@noop {} {\bibfield  {journal} {\bibinfo
  {journal} {Proceedings of the National Academy of Sciences}\ }\textbf
  {\bibinfo {volume} {45}},\ \bibinfo {pages} {1288} (\bibinfo {year}
  {1959})}\BibitemShut {NoStop}%
\bibitem [{\citenamefont {Krauth}(2006)}]{krauth2006statistical}%
  \BibitemOpen
  \bibfield  {author} {\bibinfo {author} {\bibfnamefont {W.}~\bibnamefont
  {Krauth}},\ }\href@noop {} {\emph {\bibinfo {title} {Statistical mechanics:
  algorithms and computations}}},\ Vol.~\bibinfo {volume} {13}\ (\bibinfo
  {publisher} {OUP Oxford},\ \bibinfo {year} {2006})\BibitemShut {NoStop}%
\bibitem [{\citenamefont {Chaikin}\ \emph {et~al.}(1995)\citenamefont
  {Chaikin}, \citenamefont {Lubensky},\ and\ \citenamefont
  {Witten}}]{chaikin1995principles}%
  \BibitemOpen
  \bibfield  {author} {\bibinfo {author} {\bibfnamefont {P.~M.}\ \bibnamefont
  {Chaikin}}, \bibinfo {author} {\bibfnamefont {T.~C.}\ \bibnamefont
  {Lubensky}},\ and\ \bibinfo {author} {\bibfnamefont {T.~A.}\ \bibnamefont
  {Witten}},\ }\href@noop {} {\emph {\bibinfo {title} {Principles of condensed
  matter physics}}},\ Vol.~\bibinfo {volume} {10}\ (\bibinfo  {publisher}
  {Cambridge university press Cambridge},\ \bibinfo {year} {1995})\BibitemShut
  {NoStop}%
\bibitem [{\citenamefont {Sethna}(2021)}]{sethna2021statistical}%
  \BibitemOpen
  \bibfield  {author} {\bibinfo {author} {\bibfnamefont {J.}~\bibnamefont
  {Sethna}},\ }\href@noop {} {\emph {\bibinfo {title} {Statistical mechanics:
  entropy, order parameters, and complexity}}},\ Vol.~\bibinfo {volume} {14}\
  (\bibinfo  {publisher} {Oxford University Press, USA},\ \bibinfo {year}
  {2021})\BibitemShut {NoStop}%
\bibitem [{\citenamefont {Delamotte}(2012)}]{delamotte2012introduction}%
  \BibitemOpen
  \bibfield  {author} {\bibinfo {author} {\bibfnamefont {B.}~\bibnamefont
  {Delamotte}},\ }\bibfield  {title} {\bibinfo {title} {An introduction to the
  nonperturbative renormalization group},\ }in\ \href@noop {} {\emph {\bibinfo
  {booktitle} {Renormalization Group and Effective Field Theory Approaches to
  Many-Body Systems}}}\ (\bibinfo  {publisher} {Springer},\ \bibinfo {year}
  {2012})\ pp.\ \bibinfo {pages} {49--132}\BibitemShut {NoStop}%
\bibitem [{\citenamefont {Tanaka}\ \emph {et~al.}(2015)\citenamefont {Tanaka},
  \citenamefont {Kataoka}, \citenamefont {Yasuda},\ and\ \citenamefont
  {Ohzeki}}]{tanaka2015inverse}%
  \BibitemOpen
  \bibfield  {author} {\bibinfo {author} {\bibfnamefont {K.}~\bibnamefont
  {Tanaka}}, \bibinfo {author} {\bibfnamefont {S.}~\bibnamefont {Kataoka}},
  \bibinfo {author} {\bibfnamefont {M.}~\bibnamefont {Yasuda}},\ and\ \bibinfo
  {author} {\bibfnamefont {M.}~\bibnamefont {Ohzeki}},\ }\bibfield  {title}
  {\bibinfo {title} {Inverse renormalization group transformation in bayesian
  image segmentations},\ }\href@noop {} {\bibfield  {journal} {\bibinfo
  {journal} {Journal of the Physical Society of Japan}\ }\textbf {\bibinfo
  {volume} {84}},\ \bibinfo {pages} {045001} (\bibinfo {year}
  {2015})}\BibitemShut {NoStop}%
\bibitem [{\citenamefont {Shiina}\ \emph {et~al.}(2021)\citenamefont {Shiina},
  \citenamefont {Mori}, \citenamefont {Tomita}, \citenamefont {Lee},\ and\
  \citenamefont {Okabe}}]{shiina2021inverse}%
  \BibitemOpen
  \bibfield  {author} {\bibinfo {author} {\bibfnamefont {K.}~\bibnamefont
  {Shiina}}, \bibinfo {author} {\bibfnamefont {H.}~\bibnamefont {Mori}},
  \bibinfo {author} {\bibfnamefont {Y.}~\bibnamefont {Tomita}}, \bibinfo
  {author} {\bibfnamefont {H.~K.}\ \bibnamefont {Lee}},\ and\ \bibinfo {author}
  {\bibfnamefont {Y.}~\bibnamefont {Okabe}},\ }\bibfield  {title} {\bibinfo
  {title} {Inverse renormalization group based on image super-resolution using
  deep convolutional networks},\ }\href@noop {} {\bibfield  {journal} {\bibinfo
   {journal} {Scientific Reports}\ }\textbf {\bibinfo {volume} {11}},\ \bibinfo
  {pages} {1} (\bibinfo {year} {2021})}\BibitemShut {NoStop}%
\bibitem [{\citenamefont {Li}\ and\ \citenamefont {Wang}(2018)}]{li2018neural}%
  \BibitemOpen
  \bibfield  {author} {\bibinfo {author} {\bibfnamefont {S.-H.}\ \bibnamefont
  {Li}}\ and\ \bibinfo {author} {\bibfnamefont {L.}~\bibnamefont {Wang}},\
  }\bibfield  {title} {\bibinfo {title} {Neural network renormalization
  group},\ }\href@noop {} {\bibfield  {journal} {\bibinfo  {journal} {Physical
  review letters}\ }\textbf {\bibinfo {volume} {121}},\ \bibinfo {pages}
  {260601} (\bibinfo {year} {2018})}\BibitemShut {NoStop}%
\bibitem [{\citenamefont {Mehta}\ and\ \citenamefont
  {Schwab}(2014)}]{mehta2014exact}%
  \BibitemOpen
  \bibfield  {author} {\bibinfo {author} {\bibfnamefont {P.}~\bibnamefont
  {Mehta}}\ and\ \bibinfo {author} {\bibfnamefont {D.~J.}\ \bibnamefont
  {Schwab}},\ }\bibfield  {title} {\bibinfo {title} {An exact mapping between
  the variational renormalization group and deep learning},\ }\href@noop {}
  {\bibfield  {journal} {\bibinfo  {journal} {arXiv preprint arXiv:1410.3831}\
  } (\bibinfo {year} {2014})}\BibitemShut {NoStop}%
\bibitem [{\citenamefont {Evenbly}\ and\ \citenamefont
  {White}(2016)}]{evenbly2016entanglement}%
  \BibitemOpen
  \bibfield  {author} {\bibinfo {author} {\bibfnamefont {G.}~\bibnamefont
  {Evenbly}}\ and\ \bibinfo {author} {\bibfnamefont {S.~R.}\ \bibnamefont
  {White}},\ }\bibfield  {title} {\bibinfo {title} {Entanglement
  renormalization and wavelets},\ }\href@noop {} {\bibfield  {journal}
  {\bibinfo  {journal} {Physical review letters}\ }\textbf {\bibinfo {volume}
  {116}},\ \bibinfo {pages} {140403} (\bibinfo {year} {2016})}\BibitemShut
  {NoStop}%
\bibitem [{\citenamefont {Mallat}(2016)}]{mallatReview}%
  \BibitemOpen
  \bibfield  {author} {\bibinfo {author} {\bibfnamefont {S.}~\bibnamefont
  {Mallat}},\ }\bibfield  {title} {\bibinfo {title} {Understanding deep
  convolutional networks},\ }\href@noop {} {\bibfield  {journal} {\bibinfo
  {journal} {Phil. Trans. of Royal Society A}\ }\textbf {\bibinfo {volume}
  {374}} (\bibinfo {year} {2016})}\BibitemShut {NoStop}%
\bibitem [{\citenamefont {Wilson}(1971)}]{wilson1971renormalization}%
  \BibitemOpen
  \bibfield  {author} {\bibinfo {author} {\bibfnamefont {K.~G.}\ \bibnamefont
  {Wilson}},\ }\bibfield  {title} {\bibinfo {title} {Renormalization group and
  critical phenomena. ii. phase-space cell analysis of critical behavior},\
  }\href@noop {} {\bibfield  {journal} {\bibinfo  {journal} {Physical Review
  B}\ }\textbf {\bibinfo {volume} {4}},\ \bibinfo {pages} {3184} (\bibinfo
  {year} {1971})}\BibitemShut {NoStop}%
\bibitem [{\citenamefont {Battle}(1999)}]{battle1999wavelets}%
  \BibitemOpen
  \bibfield  {author} {\bibinfo {author} {\bibfnamefont {G.}~\bibnamefont
  {Battle}},\ }\href@noop {} {\emph {\bibinfo {title} {Wavelets and
  renormalization}}},\ Vol.~\bibinfo {volume} {10}\ (\bibinfo  {publisher}
  {World Scientific},\ \bibinfo {year} {1999})\BibitemShut {NoStop}%
\bibitem [{\citenamefont {Altaisky}(2016)}]{altaisky2016unifying}%
  \BibitemOpen
  \bibfield  {author} {\bibinfo {author} {\bibfnamefont {M.}~\bibnamefont
  {Altaisky}},\ }\bibfield  {title} {\bibinfo {title} {Unifying renormalization
  group and the continuous wavelet transform},\ }\href@noop {} {\bibfield
  {journal} {\bibinfo  {journal} {Physical Review D}\ }\textbf {\bibinfo
  {volume} {93}},\ \bibinfo {pages} {105043} (\bibinfo {year}
  {2016})}\BibitemShut {NoStop}%
\bibitem [{\citenamefont {Kolmogorov}(1962)}]{kolmogorov_1962}%
  \BibitemOpen
  \bibfield  {author} {\bibinfo {author} {\bibfnamefont {A.~N.}\ \bibnamefont
  {Kolmogorov}},\ }\bibfield  {title} {\bibinfo {title} {A refinement of
  previous hypotheses concerning the local structure of turbulence in a viscous
  incompressible fluid at high reynolds number},\ }\href@noop {} {\bibfield
  {journal} {\bibinfo  {journal} {Journal of Fluid Mechanics}\ }\textbf
  {\bibinfo {volume} {13}},\ \bibinfo {pages} {82–85} (\bibinfo {year}
  {1962})}\BibitemShut {NoStop}%
\bibitem [{\citenamefont {Frisch}\ and\ \citenamefont
  {Parisi}(1985)}]{FrischParisi80}%
  \BibitemOpen
  \bibfield  {author} {\bibinfo {author} {\bibfnamefont {U.}~\bibnamefont
  {Frisch}}\ and\ \bibinfo {author} {\bibfnamefont {G.}~\bibnamefont
  {Parisi}},\ }\bibfield  {title} {\bibinfo {title} {Fully developed turbulence
  and intermittency},\ }\href@noop {} {\bibfield  {journal} {\bibinfo
  {journal} {Proceedings of the International Summer School on Turbulence and
  Predictability in Geophysical Fluid Dynamics and Climate Dynamics}\ ,\
  \bibinfo {pages} {84–88}} (\bibinfo {year} {1985})}\BibitemShut {NoStop}%
\bibitem [{\citenamefont {Frisch}(1991)}]{frisch1991global}%
  \BibitemOpen
  \bibfield  {author} {\bibinfo {author} {\bibfnamefont {U.}~\bibnamefont
  {Frisch}},\ }\bibfield  {title} {\bibinfo {title} {From global scaling, a la
  kolmogorov, to local multifractal scaling in fully developed turbulence},\
  }\href@noop {} {\bibfield  {journal} {\bibinfo  {journal} {Proceedings of the
  Royal Society of London. Series A: Mathematical and Physical Sciences}\
  }\textbf {\bibinfo {volume} {434}},\ \bibinfo {pages} {89} (\bibinfo {year}
  {1991})}\BibitemShut {NoStop}%
\bibitem [{\citenamefont {Muzy}\ \emph {et~al.}(1994)\citenamefont {Muzy},
  \citenamefont {Bacry},\ and\ \citenamefont {Arneodo}}]{muzy1994multifractal}%
  \BibitemOpen
  \bibfield  {author} {\bibinfo {author} {\bibfnamefont {J.-F.}\ \bibnamefont
  {Muzy}}, \bibinfo {author} {\bibfnamefont {E.}~\bibnamefont {Bacry}},\ and\
  \bibinfo {author} {\bibfnamefont {A.}~\bibnamefont {Arneodo}},\ }\bibfield
  {title} {\bibinfo {title} {The multifractal formalism revisited with
  wavelets},\ }\href@noop {} {\bibfield  {journal} {\bibinfo  {journal}
  {International Journal of Bifurcation and Chaos}\ }\textbf {\bibinfo {volume}
  {4}},\ \bibinfo {pages} {245} (\bibinfo {year} {1994})}\BibitemShut {NoStop}%
\bibitem [{\citenamefont {Mandelbrot}()}]{mandelbrotmultifractals}%
  \BibitemOpen
  \bibfield  {author} {\bibinfo {author} {\bibfnamefont {B.~B.}\ \bibnamefont
  {Mandelbrot}},\ }\bibfield  {title} {\bibinfo {title} {Multifractals and 1/ƒ
  noise wild self-affinity in physics (1963--1976)},\ }\href@noop {} {\
  }\BibitemShut {NoStop}%
\bibitem [{\citenamefont {Bacry}\ \emph {et~al.}(2001)\citenamefont {Bacry},
  \citenamefont {Delour},\ and\ \citenamefont {Muzy}}]{MRW01}%
  \BibitemOpen
  \bibfield  {author} {\bibinfo {author} {\bibfnamefont {E.}~\bibnamefont
  {Bacry}}, \bibinfo {author} {\bibfnamefont {J.}~\bibnamefont {Delour}},\ and\
  \bibinfo {author} {\bibfnamefont {J.~F.}\ \bibnamefont {Muzy}},\ }\bibfield
  {title} {\bibinfo {title} {Multifractal random walk},\ }\href@noop {}
  {\bibfield  {journal} {\bibinfo  {journal} {Phys. Rev. E}\ }\textbf {\bibinfo
  {volume} {64}},\ \bibinfo {pages} {026103} (\bibinfo {year}
  {2001})}\BibitemShut {NoStop}%
\bibitem [{\citenamefont {Allys}\ \emph {et~al.}(2020)\citenamefont {Allys},
  \citenamefont {Marchand}, \citenamefont {Cardoso}, \citenamefont
  {Villaescusa-Navarro}, \citenamefont {Ho},\ and\ \citenamefont
  {Mallat}}]{Allys_2020}%
  \BibitemOpen
  \bibfield  {author} {\bibinfo {author} {\bibfnamefont {E.}~\bibnamefont
  {Allys}}, \bibinfo {author} {\bibfnamefont {T.}~\bibnamefont {Marchand}},
  \bibinfo {author} {\bibfnamefont {J.-F.}\ \bibnamefont {Cardoso}}, \bibinfo
  {author} {\bibfnamefont {F.}~\bibnamefont {Villaescusa-Navarro}}, \bibinfo
  {author} {\bibfnamefont {S.}~\bibnamefont {Ho}},\ and\ \bibinfo {author}
  {\bibfnamefont {S.}~\bibnamefont {Mallat}},\ }\bibfield  {title} {\bibinfo
  {title} {New interpretable statistics for large-scale structure analysis and
  generation},\ }\href@noop {} {\bibfield  {journal} {\bibinfo  {journal}
  {Physical Review D}\ }\textbf {\bibinfo {volume} {102}} (\bibinfo {year}
  {2020})}\BibitemShut {NoStop}%
\bibitem [{\citenamefont {Zhang}\ and\ \citenamefont
  {Mallat}(2021)}]{zhang2021maximum}%
  \BibitemOpen
  \bibfield  {author} {\bibinfo {author} {\bibfnamefont {S.}~\bibnamefont
  {Zhang}}\ and\ \bibinfo {author} {\bibfnamefont {S.}~\bibnamefont {Mallat}},\
  }\bibfield  {title} {\bibinfo {title} {Maximum entropy models from phase
  harmonic covariances},\ }\href@noop {} {\bibfield  {journal} {\bibinfo
  {journal} {Applied and Computational Harmonic Analysis}\ }\textbf {\bibinfo
  {volume} {53}},\ \bibinfo {pages} {199} (\bibinfo {year} {2021})}\BibitemShut
  {NoStop}%
\bibitem [{\citenamefont {Cheng}\ and\ \citenamefont {M{\'{e}
  }nard}(2021)}]{Cheng_2021}%
  \BibitemOpen
  \bibfield  {author} {\bibinfo {author} {\bibfnamefont {S.}~\bibnamefont
  {Cheng}}\ and\ \bibinfo {author} {\bibfnamefont {B.}~\bibnamefont {M{\'{e}
  }nard}},\ }\bibfield  {title} {\bibinfo {title} {Weak lensing scattering
  transform: dark energy and neutrino mass sensitivity},\ }\bibfield  {journal}
  {\bibinfo  {journal} {Monthly Notices of the Royal Astronomical Society}\
  }\href {https://doi.org/10.1093/mnras/stab2102} {10.1093/mnras/stab2102}
  (\bibinfo {year} {2021})\BibitemShut {NoStop}%
\bibitem [{\citenamefont {Stephane}(1999)}]{stephane1999wavelet}%
  \BibitemOpen
  \bibfield  {author} {\bibinfo {author} {\bibfnamefont {M.}~\bibnamefont
  {Stephane}},\ }\href@noop {} {\bibinfo {title} {A wavelet tour of signal
  processing}} (\bibinfo {year} {1999})\BibitemShut {NoStop}%
\bibitem [{\citenamefont {Ron}\ \emph {et~al.}(2002)\citenamefont {Ron},
  \citenamefont {Swendsen},\ and\ \citenamefont {Brandt}}]{ron2002inverse}%
  \BibitemOpen
  \bibfield  {author} {\bibinfo {author} {\bibfnamefont {D.}~\bibnamefont
  {Ron}}, \bibinfo {author} {\bibfnamefont {R.~H.}\ \bibnamefont {Swendsen}},\
  and\ \bibinfo {author} {\bibfnamefont {A.}~\bibnamefont {Brandt}},\
  }\bibfield  {title} {\bibinfo {title} {Inverse monte carlo renormalization
  group transformations for critical phenomena},\ }\href@noop {} {\bibfield
  {journal} {\bibinfo  {journal} {Physical review letters}\ }\textbf {\bibinfo
  {volume} {89}},\ \bibinfo {pages} {275701} (\bibinfo {year}
  {2002})}\BibitemShut {NoStop}%
\bibitem [{\citenamefont {Bachtis}\ \emph {et~al.}(2022)\citenamefont
  {Bachtis}, \citenamefont {Aarts}, \citenamefont {Di~Renzo},\ and\
  \citenamefont {Lucini}}]{bachtis2022inverse}%
  \BibitemOpen
  \bibfield  {author} {\bibinfo {author} {\bibfnamefont {D.}~\bibnamefont
  {Bachtis}}, \bibinfo {author} {\bibfnamefont {G.}~\bibnamefont {Aarts}},
  \bibinfo {author} {\bibfnamefont {F.}~\bibnamefont {Di~Renzo}},\ and\
  \bibinfo {author} {\bibfnamefont {B.}~\bibnamefont {Lucini}},\ }\bibfield
  {title} {\bibinfo {title} {Inverse renormalization group in quantum field
  theory},\ }\href@noop {} {\bibfield  {journal} {\bibinfo  {journal} {Physical
  Review Letters}\ }\textbf {\bibinfo {volume} {128}},\ \bibinfo {pages}
  {081603} (\bibinfo {year} {2022})}\BibitemShut {NoStop}%
\bibitem [{\citenamefont {Kadanoff}\ \emph {et~al.}(1976)\citenamefont
  {Kadanoff}, \citenamefont {Houghton},\ and\ \citenamefont
  {Yalabik}}]{kadanoff}%
  \BibitemOpen
  \bibfield  {author} {\bibinfo {author} {\bibfnamefont {L.~P.}\ \bibnamefont
  {Kadanoff}}, \bibinfo {author} {\bibfnamefont {A.}~\bibnamefont {Houghton}},\
  and\ \bibinfo {author} {\bibfnamefont {M.~C.}\ \bibnamefont {Yalabik}},\
  }\bibfield  {title} {\bibinfo {title} {Variational approximations for
  renormalization group transformations},\ }\href@noop {} {\bibfield  {journal}
  {\bibinfo  {journal} {Journal of Statistical Physics}\ }\textbf {\bibinfo
  {volume} {14}},\ \bibinfo {pages} {171} (\bibinfo {year} {1976})}\BibitemShut
  {NoStop}%
\bibitem [{\citenamefont {Wilson}\ and\ \citenamefont
  {Fisher}(1972)}]{wilson1972critical}%
  \BibitemOpen
  \bibfield  {author} {\bibinfo {author} {\bibfnamefont {K.~G.}\ \bibnamefont
  {Wilson}}\ and\ \bibinfo {author} {\bibfnamefont {M.~E.}\ \bibnamefont
  {Fisher}},\ }\bibfield  {title} {\bibinfo {title} {Critical exponents in 3.99
  dimensions},\ }\href@noop {} {\bibfield  {journal} {\bibinfo  {journal}
  {Physical Review Letters}\ }\textbf {\bibinfo {volume} {28}},\ \bibinfo
  {pages} {240} (\bibinfo {year} {1972})}\BibitemShut {NoStop}%
\bibitem [{\citenamefont {Fisher}(1974)}]{fisher1974renormalization}%
  \BibitemOpen
  \bibfield  {author} {\bibinfo {author} {\bibfnamefont {M.~E.}\ \bibnamefont
  {Fisher}},\ }\bibfield  {title} {\bibinfo {title} {The renormalization group
  in the theory of critical behavior},\ }\href@noop {} {\bibfield  {journal}
  {\bibinfo  {journal} {Reviews of Modern Physics}\ }\textbf {\bibinfo {volume}
  {46}},\ \bibinfo {pages} {597} (\bibinfo {year} {1974})}\BibitemShut
  {NoStop}%
\bibitem [{\citenamefont {Mallat}(1989{\natexlab{a}})}]{Mallat:89}%
  \BibitemOpen
  \bibfield  {author} {\bibinfo {author} {\bibfnamefont {S.}~\bibnamefont
  {Mallat}},\ }\bibfield  {title} {\bibinfo {title} {A theory for
  multiresolution signal decomposition: the wavelet representation},\
  }\href@noop {} {\ \textbf {\bibinfo {volume} {11}},\ \bibinfo {pages} {674}
  (\bibinfo {year} {July 1989}{\natexlab{a}})}\BibitemShut {NoStop}%
\bibitem [{\citenamefont {Mallat}(1989{\natexlab{b}})}]{Mallat:89b}%
  \BibitemOpen
  \bibfield  {author} {\bibinfo {author} {\bibfnamefont {S.}~\bibnamefont
  {Mallat}},\ }\bibfield  {title} {\bibinfo {title} {Multiresolution
  approximations and wavelet orthonormal bases of \protect{$\text{L}^2$}},\
  }\href@noop {} {\bibfield  {journal} {\bibinfo  {journal} {Trans. Amer. Math.
  Soc.}\ }\textbf {\bibinfo {volume} {315}},\ \bibinfo {pages} {69} (\bibinfo
  {year} {September 1989}{\natexlab{b}})}\BibitemShut {NoStop}%
\bibitem [{\citenamefont {Meyer}(1992)}]{Meyer:92c}%
  \BibitemOpen
  \bibfield  {author} {\bibinfo {author} {\bibfnamefont {Y.}~\bibnamefont
  {Meyer}},\ }\href@noop {} {\emph {\bibinfo {title} {Wavelets and
  Operators}}}\ (\bibinfo  {publisher} {Advanced mathematics. Cambridge
  university press},\ \bibinfo {year} {1992})\BibitemShut {NoStop}%
\bibitem [{\citenamefont {Daubechies}(1992)}]{daubechies1992ten}%
  \BibitemOpen
  \bibfield  {author} {\bibinfo {author} {\bibfnamefont {I.}~\bibnamefont
  {Daubechies}},\ }\href@noop {} {\emph {\bibinfo {title} {Ten lectures on
  wavelets}}}\ (\bibinfo  {publisher} {SIAM},\ \bibinfo {year}
  {1992})\BibitemShut {NoStop}%
\bibitem [{\citenamefont {Kaupu{\v{z}}s}\ \emph {et~al.}(2016)\citenamefont
  {Kaupu{\v{z}}s}, \citenamefont {Melnik},\ and\ \citenamefont
  {Rim{\v{s}}{\=a}ns}}]{kaupuvzs2016corrections}%
  \BibitemOpen
  \bibfield  {author} {\bibinfo {author} {\bibfnamefont {J.}~\bibnamefont
  {Kaupu{\v{z}}s}}, \bibinfo {author} {\bibfnamefont {R.}~\bibnamefont
  {Melnik}},\ and\ \bibinfo {author} {\bibfnamefont {J.}~\bibnamefont
  {Rim{\v{s}}{\=a}ns}},\ }\bibfield  {title} {\bibinfo {title} {Corrections to
  finite-size scaling in the $\varphi$ 4 model on square lattices},\
  }\href@noop {} {\bibfield  {journal} {\bibinfo  {journal} {International
  Journal of Modern Physics C}\ }\textbf {\bibinfo {volume} {27}},\ \bibinfo
  {pages} {1650108} (\bibinfo {year} {2016})}\BibitemShut {NoStop}%
\bibitem [{\citenamefont {Bruna}\ and\ \citenamefont
  {Mallat}(2019)}]{bruna2019multiscale}%
  \BibitemOpen
  \bibfield  {author} {\bibinfo {author} {\bibfnamefont {J.}~\bibnamefont
  {Bruna}}\ and\ \bibinfo {author} {\bibfnamefont {S.}~\bibnamefont {Mallat}},\
  }\bibfield  {title} {\bibinfo {title} {Multiscale sparse microcanonical
  models},\ }\href@noop {} {\bibfield  {journal} {\bibinfo  {journal}
  {Mathematical Statistics and Learning}\ }\textbf {\bibinfo {volume} {1}},\
  \bibinfo {pages} {257} (\bibinfo {year} {2019})}\BibitemShut {NoStop}%
\bibitem [{\citenamefont {Swendsen}\ and\ \citenamefont
  {Wang}(1987)}]{swendsen1987nonuniversal}%
  \BibitemOpen
  \bibfield  {author} {\bibinfo {author} {\bibfnamefont {R.~H.}\ \bibnamefont
  {Swendsen}}\ and\ \bibinfo {author} {\bibfnamefont {J.-S.}\ \bibnamefont
  {Wang}},\ }\bibfield  {title} {\bibinfo {title} {Nonuniversal critical
  dynamics in monte carlo simulations},\ }\href@noop {} {\bibfield  {journal}
  {\bibinfo  {journal} {Physical review letters}\ }\textbf {\bibinfo {volume}
  {58}},\ \bibinfo {pages} {86} (\bibinfo {year} {1987})}\BibitemShut {NoStop}%
\bibitem [{\citenamefont {Wolff}(1989)}]{wolff1989comparison}%
  \BibitemOpen
  \bibfield  {author} {\bibinfo {author} {\bibfnamefont {U.}~\bibnamefont
  {Wolff}},\ }\bibfield  {title} {\bibinfo {title} {Comparison between cluster
  monte carlo algorithms in the ising model},\ }\href@noop {} {\bibfield
  {journal} {\bibinfo  {journal} {Physics Letters B}\ }\textbf {\bibinfo
  {volume} {228}},\ \bibinfo {pages} {379} (\bibinfo {year}
  {1989})}\BibitemShut {NoStop}%
\bibitem [{\citenamefont {Duane}\ \emph {et~al.}(1987)\citenamefont {Duane},
  \citenamefont {Kennedy}, \citenamefont {Pendleton},\ and\ \citenamefont
  {Roweth}}]{duane1987hybrid}%
  \BibitemOpen
  \bibfield  {author} {\bibinfo {author} {\bibfnamefont {S.}~\bibnamefont
  {Duane}}, \bibinfo {author} {\bibfnamefont {A.~D.}\ \bibnamefont {Kennedy}},
  \bibinfo {author} {\bibfnamefont {B.~J.}\ \bibnamefont {Pendleton}},\ and\
  \bibinfo {author} {\bibfnamefont {D.}~\bibnamefont {Roweth}},\ }\bibfield
  {title} {\bibinfo {title} {Hybrid monte carlo},\ }\href@noop {} {\bibfield
  {journal} {\bibinfo  {journal} {Physics letters B}\ }\textbf {\bibinfo
  {volume} {195}},\ \bibinfo {pages} {216} (\bibinfo {year}
  {1987})}\BibitemShut {NoStop}%
\bibitem [{\citenamefont {Goodman}\ and\ \citenamefont
  {Sokal}(1989)}]{goodman1989multigrid}%
  \BibitemOpen
  \bibfield  {author} {\bibinfo {author} {\bibfnamefont {J.}~\bibnamefont
  {Goodman}}\ and\ \bibinfo {author} {\bibfnamefont {A.~D.}\ \bibnamefont
  {Sokal}},\ }\bibfield  {title} {\bibinfo {title} {Multigrid monte carlo
  method. conceptual foundations},\ }\href@noop {} {\bibfield  {journal}
  {\bibinfo  {journal} {Physical Review D}\ }\textbf {\bibinfo {volume} {40}},\
  \bibinfo {pages} {2035} (\bibinfo {year} {1989})}\BibitemShut {NoStop}%
\bibitem [{\citenamefont {Torrie}\ and\ \citenamefont
  {Valleau}(1977)}]{torrie1977nonphysical}%
  \BibitemOpen
  \bibfield  {author} {\bibinfo {author} {\bibfnamefont {G.~M.}\ \bibnamefont
  {Torrie}}\ and\ \bibinfo {author} {\bibfnamefont {J.~P.}\ \bibnamefont
  {Valleau}},\ }\bibfield  {title} {\bibinfo {title} {Nonphysical sampling
  distributions in monte carlo free-energy estimation: Umbrella sampling},\
  }\href@noop {} {\bibfield  {journal} {\bibinfo  {journal} {Journal of
  Computational Physics}\ }\textbf {\bibinfo {volume} {23}},\ \bibinfo {pages}
  {187} (\bibinfo {year} {1977})}\BibitemShut {NoStop}%
\bibitem [{\citenamefont {Marinari}\ and\ \citenamefont
  {Parisi}(1992)}]{marinari1992simulated}%
  \BibitemOpen
  \bibfield  {author} {\bibinfo {author} {\bibfnamefont {E.}~\bibnamefont
  {Marinari}}\ and\ \bibinfo {author} {\bibfnamefont {G.}~\bibnamefont
  {Parisi}},\ }\bibfield  {title} {\bibinfo {title} {Simulated tempering: a new
  monte carlo scheme},\ }\href@noop {} {\bibfield  {journal} {\bibinfo
  {journal} {EPL (Europhysics Letters)}\ }\textbf {\bibinfo {volume} {19}},\
  \bibinfo {pages} {451} (\bibinfo {year} {1992})}\BibitemShut {NoStop}%
\bibitem [{\citenamefont {Wasserman}(2021)}]{wasserman2021}%
  \BibitemOpen
  \bibfield  {author} {\bibinfo {author} {\bibfnamefont {L.}~\bibnamefont
  {Wasserman}},\ }\bibfield  {title} {\bibinfo {title} {All of statistics},\
  }in\ \href@noop {} {\emph {\bibinfo {booktitle} {All of Statistics}}}\
  (\bibinfo  {publisher} {Springer},\ \bibinfo {year} {2021})\BibitemShut
  {NoStop}%
\bibitem [{\citenamefont {Batou}\ and\ \citenamefont
  {Soize}(2013)}]{batou2013calculation}%
  \BibitemOpen
  \bibfield  {author} {\bibinfo {author} {\bibfnamefont {A.}~\bibnamefont
  {Batou}}\ and\ \bibinfo {author} {\bibfnamefont {C.}~\bibnamefont {Soize}},\
  }\bibfield  {title} {\bibinfo {title} {Calculation of lagrange multipliers in
  the construction of maximum entropy distributions in high stochastic
  dimension},\ }\href@noop {} {\bibfield  {journal} {\bibinfo  {journal}
  {SIAM/ASA Journal on Uncertainty Quantification}\ }\textbf {\bibinfo {volume}
  {1}},\ \bibinfo {pages} {431} (\bibinfo {year} {2013})}\BibitemShut {NoStop}%
\bibitem [{\citenamefont {Cardy}(1996)}]{cardy1996scaling}%
  \BibitemOpen
  \bibfield  {author} {\bibinfo {author} {\bibfnamefont {J.}~\bibnamefont
  {Cardy}},\ }\href@noop {} {\emph {\bibinfo {title} {Scaling and
  renormalization in statistical physics}}},\ Vol.~\bibinfo {volume} {5}\
  (\bibinfo  {publisher} {Cambridge university press},\ \bibinfo {year}
  {1996})\BibitemShut {NoStop}%
\bibitem [{\citenamefont {Montanari}\ and\ \citenamefont
  {Semerjian}(2006)}]{montanari2006rigorous}%
  \BibitemOpen
  \bibfield  {author} {\bibinfo {author} {\bibfnamefont {A.}~\bibnamefont
  {Montanari}}\ and\ \bibinfo {author} {\bibfnamefont {G.}~\bibnamefont
  {Semerjian}},\ }\bibfield  {title} {\bibinfo {title} {Rigorous inequalities
  between length and time scales in glassy systems},\ }\href@noop {} {\bibfield
   {journal} {\bibinfo  {journal} {Journal of statistical physics}\ }\textbf
  {\bibinfo {volume} {125}},\ \bibinfo {pages} {23} (\bibinfo {year}
  {2006})}\BibitemShut {NoStop}%
\bibitem [{\citenamefont {Hohenberg}\ and\ \citenamefont
  {Halperin}(1977)}]{hohenberg1977theory}%
  \BibitemOpen
  \bibfield  {author} {\bibinfo {author} {\bibfnamefont {P.~C.}\ \bibnamefont
  {Hohenberg}}\ and\ \bibinfo {author} {\bibfnamefont {B.~I.}\ \bibnamefont
  {Halperin}},\ }\bibfield  {title} {\bibinfo {title} {Theory of dynamic
  critical phenomena},\ }\href@noop {} {\bibfield  {journal} {\bibinfo
  {journal} {Reviews of Modern Physics}\ }\textbf {\bibinfo {volume} {49}},\
  \bibinfo {pages} {435} (\bibinfo {year} {1977})}\BibitemShut {NoStop}%
\bibitem [{\citenamefont {Canet}\ and\ \citenamefont
  {Chat{\'e}}(2007)}]{canet2007non}%
  \BibitemOpen
  \bibfield  {author} {\bibinfo {author} {\bibfnamefont {L.}~\bibnamefont
  {Canet}}\ and\ \bibinfo {author} {\bibfnamefont {H.}~\bibnamefont
  {Chat{\'e}}},\ }\bibfield  {title} {\bibinfo {title} {A non-perturbative
  approach to critical dynamics},\ }\href@noop {} {\bibfield  {journal}
  {\bibinfo  {journal} {Journal of Physics A: Mathematical and Theoretical}\
  }\textbf {\bibinfo {volume} {40}},\ \bibinfo {pages} {1937} (\bibinfo {year}
  {2007})}\BibitemShut {NoStop}%
\bibitem [{\citenamefont {Andreanov}\ \emph {et~al.}(2006)\citenamefont
  {Andreanov}, \citenamefont {Biroli}, \citenamefont {Bouchaud},\ and\
  \citenamefont {Lefevre}}]{andreanov2006field}%
  \BibitemOpen
  \bibfield  {author} {\bibinfo {author} {\bibfnamefont {A.}~\bibnamefont
  {Andreanov}}, \bibinfo {author} {\bibfnamefont {G.}~\bibnamefont {Biroli}},
  \bibinfo {author} {\bibfnamefont {J.-P.}\ \bibnamefont {Bouchaud}},\ and\
  \bibinfo {author} {\bibfnamefont {A.}~\bibnamefont {Lefevre}},\ }\bibfield
  {title} {\bibinfo {title} {Field theories and exact stochastic equations for
  interacting particle systems},\ }\href@noop {} {\bibfield  {journal}
  {\bibinfo  {journal} {Physical Review E}\ }\textbf {\bibinfo {volume} {74}},\
  \bibinfo {pages} {030101} (\bibinfo {year} {2006})}\BibitemShut {NoStop}%
\bibitem [{\citenamefont {Gelman}\ \emph {et~al.}(1997)\citenamefont {Gelman},
  \citenamefont {Gilks},\ and\ \citenamefont {Roberts}}]{gelman1997weak}%
  \BibitemOpen
  \bibfield  {author} {\bibinfo {author} {\bibfnamefont {A.}~\bibnamefont
  {Gelman}}, \bibinfo {author} {\bibfnamefont {W.~R.}\ \bibnamefont {Gilks}},\
  and\ \bibinfo {author} {\bibfnamefont {G.~O.}\ \bibnamefont {Roberts}},\
  }\bibfield  {title} {\bibinfo {title} {Weak convergence and optimal scaling
  of random walk metropolis algorithms},\ }\href@noop {} {\bibfield  {journal}
  {\bibinfo  {journal} {The annals of applied probability}\ }\textbf {\bibinfo
  {volume} {7}},\ \bibinfo {pages} {110} (\bibinfo {year} {1997})}\BibitemShut
  {NoStop}%
\bibitem [{\citenamefont {Frenkel}\ and\ \citenamefont
  {Smit}(2001)}]{frenkel2001understanding}%
  \BibitemOpen
  \bibfield  {author} {\bibinfo {author} {\bibfnamefont {D.}~\bibnamefont
  {Frenkel}}\ and\ \bibinfo {author} {\bibfnamefont {B.}~\bibnamefont {Smit}},\
  }\href@noop {} {\emph {\bibinfo {title} {Understanding molecular simulation:
  from algorithms to applications}}},\ Vol.~\bibinfo {volume} {1}\ (\bibinfo
  {publisher} {Elsevier},\ \bibinfo {year} {2001})\BibitemShut {NoStop}%
\bibitem [{\citenamefont {Berges}\ \emph {et~al.}(2002)\citenamefont {Berges},
  \citenamefont {Tetradis},\ and\ \citenamefont {Wetterich}}]{berges2002non}%
  \BibitemOpen
  \bibfield  {author} {\bibinfo {author} {\bibfnamefont {J.}~\bibnamefont
  {Berges}}, \bibinfo {author} {\bibfnamefont {N.}~\bibnamefont {Tetradis}},\
  and\ \bibinfo {author} {\bibfnamefont {C.}~\bibnamefont {Wetterich}},\
  }\bibfield  {title} {\bibinfo {title} {Non-perturbative renormalization flow
  in quantum field theory and statistical physics},\ }\href@noop {} {\bibfield
  {journal} {\bibinfo  {journal} {Physics Reports}\ }\textbf {\bibinfo {volume}
  {363}},\ \bibinfo {pages} {223} (\bibinfo {year} {2002})}\BibitemShut
  {NoStop}%
\bibitem [{\citenamefont {Matilla}\ \emph {et~al.}(2016)\citenamefont
  {Matilla}, \citenamefont {Haiman}, \citenamefont {Hsu}, \citenamefont
  {Gupta},\ and\ \citenamefont {Petri}}]{matilla2016dark}%
  \BibitemOpen
  \bibfield  {author} {\bibinfo {author} {\bibfnamefont {J.~M.~Z.}\
  \bibnamefont {Matilla}}, \bibinfo {author} {\bibfnamefont {Z.}~\bibnamefont
  {Haiman}}, \bibinfo {author} {\bibfnamefont {D.}~\bibnamefont {Hsu}},
  \bibinfo {author} {\bibfnamefont {A.}~\bibnamefont {Gupta}},\ and\ \bibinfo
  {author} {\bibfnamefont {A.}~\bibnamefont {Petri}},\ }\bibfield  {title}
  {\bibinfo {title} {Do dark matter halos explain lensing peaks?},\ }\href@noop
  {} {\bibfield  {journal} {\bibinfo  {journal} {Physical Review D}\ }\textbf
  {\bibinfo {volume} {94}},\ \bibinfo {pages} {083506} (\bibinfo {year}
  {2016})}\BibitemShut {NoStop}%
\bibitem [{\citenamefont {Gupta}\ \emph {et~al.}(2018)\citenamefont {Gupta},
  \citenamefont {Matilla}, \citenamefont {Hsu},\ and\ \citenamefont
  {Haiman}}]{gupta2018non}%
  \BibitemOpen
  \bibfield  {author} {\bibinfo {author} {\bibfnamefont {A.}~\bibnamefont
  {Gupta}}, \bibinfo {author} {\bibfnamefont {J.~M.~Z.}\ \bibnamefont
  {Matilla}}, \bibinfo {author} {\bibfnamefont {D.}~\bibnamefont {Hsu}},\ and\
  \bibinfo {author} {\bibfnamefont {Z.}~\bibnamefont {Haiman}},\ }\bibfield
  {title} {\bibinfo {title} {Non-gaussian information from weak lensing data
  via deep learning},\ }\href@noop {} {\bibfield  {journal} {\bibinfo
  {journal} {Physical Review D}\ }\textbf {\bibinfo {volume} {97}},\ \bibinfo
  {pages} {103515} (\bibinfo {year} {2018})}\BibitemShut {NoStop}%
\bibitem [{\citenamefont {Bartelmann}\ and\ \citenamefont
  {Schneider}(2001)}]{bartelmann2001weak}%
  \BibitemOpen
  \bibfield  {author} {\bibinfo {author} {\bibfnamefont {M.}~\bibnamefont
  {Bartelmann}}\ and\ \bibinfo {author} {\bibfnamefont {P.}~\bibnamefont
  {Schneider}},\ }\bibfield  {title} {\bibinfo {title} {Weak gravitational
  lensing},\ }\href@noop {} {\bibfield  {journal} {\bibinfo  {journal} {Physics
  Reports}\ }\textbf {\bibinfo {volume} {340}},\ \bibinfo {pages} {291}
  (\bibinfo {year} {2001})}\BibitemShut {NoStop}%
\bibitem [{\citenamefont {Kilbinger}(2015)}]{kilbinger2015cosmology}%
  \BibitemOpen
  \bibfield  {author} {\bibinfo {author} {\bibfnamefont {M.}~\bibnamefont
  {Kilbinger}},\ }\bibfield  {title} {\bibinfo {title} {Cosmology with cosmic
  shear observations: a review},\ }\href@noop {} {\bibfield  {journal}
  {\bibinfo  {journal} {Reports on Progress in Physics}\ }\textbf {\bibinfo
  {volume} {78}},\ \bibinfo {pages} {086901} (\bibinfo {year}
  {2015})}\BibitemShut {NoStop}%
\bibitem [{\citenamefont {Fu}\ \emph {et~al.}(2014)\citenamefont {Fu},
  \citenamefont {Kilbinger}, \citenamefont {Erben}, \citenamefont {Heymans},
  \citenamefont {Hildebrandt}, \citenamefont {Hoekstra}, \citenamefont
  {Kitching}, \citenamefont {Mellier}, \citenamefont {Miller}, \citenamefont
  {Semboloni} \emph {et~al.}}]{fu2014cfhtlens}%
  \BibitemOpen
  \bibfield  {author} {\bibinfo {author} {\bibfnamefont {L.}~\bibnamefont
  {Fu}}, \bibinfo {author} {\bibfnamefont {M.}~\bibnamefont {Kilbinger}},
  \bibinfo {author} {\bibfnamefont {T.}~\bibnamefont {Erben}}, \bibinfo
  {author} {\bibfnamefont {C.}~\bibnamefont {Heymans}}, \bibinfo {author}
  {\bibfnamefont {H.}~\bibnamefont {Hildebrandt}}, \bibinfo {author}
  {\bibfnamefont {H.}~\bibnamefont {Hoekstra}}, \bibinfo {author}
  {\bibfnamefont {T.~D.}\ \bibnamefont {Kitching}}, \bibinfo {author}
  {\bibfnamefont {Y.}~\bibnamefont {Mellier}}, \bibinfo {author} {\bibfnamefont
  {L.}~\bibnamefont {Miller}}, \bibinfo {author} {\bibfnamefont
  {E.}~\bibnamefont {Semboloni}}, \emph {et~al.},\ }\bibfield  {title}
  {\bibinfo {title} {Cfhtlens: cosmological constraints from a combination of
  cosmic shear two-point and three-point correlations},\ }\href@noop {}
  {\bibfield  {journal} {\bibinfo  {journal} {Monthly Notices of the Royal
  Astronomical Society}\ }\textbf {\bibinfo {volume} {441}},\ \bibinfo {pages}
  {2725} (\bibinfo {year} {2014})}\BibitemShut {NoStop}%
\bibitem [{\citenamefont {Liu}\ \emph {et~al.}(2015)\citenamefont {Liu},
  \citenamefont {Pan}, \citenamefont {Li}, \citenamefont {Shan}, \citenamefont
  {Wang}, \citenamefont {Fu}, \citenamefont {Fan}, \citenamefont {Kneib},
  \citenamefont {Leauthaud}, \citenamefont {Van~Waerbeke} \emph
  {et~al.}}]{liu2015cosmological}%
  \BibitemOpen
  \bibfield  {author} {\bibinfo {author} {\bibfnamefont {X.}~\bibnamefont
  {Liu}}, \bibinfo {author} {\bibfnamefont {C.}~\bibnamefont {Pan}}, \bibinfo
  {author} {\bibfnamefont {R.}~\bibnamefont {Li}}, \bibinfo {author}
  {\bibfnamefont {H.}~\bibnamefont {Shan}}, \bibinfo {author} {\bibfnamefont
  {Q.}~\bibnamefont {Wang}}, \bibinfo {author} {\bibfnamefont {L.}~\bibnamefont
  {Fu}}, \bibinfo {author} {\bibfnamefont {Z.}~\bibnamefont {Fan}}, \bibinfo
  {author} {\bibfnamefont {J.-P.}\ \bibnamefont {Kneib}}, \bibinfo {author}
  {\bibfnamefont {A.}~\bibnamefont {Leauthaud}}, \bibinfo {author}
  {\bibfnamefont {L.}~\bibnamefont {Van~Waerbeke}}, \emph {et~al.},\ }\bibfield
   {title} {\bibinfo {title} {Cosmological constraints from weak lensing peak
  statistics with canada--france--hawaii telescope stripe 82 survey},\
  }\href@noop {} {\bibfield  {journal} {\bibinfo  {journal} {Monthly Notices of
  the Royal Astronomical Society}\ }\textbf {\bibinfo {volume} {450}},\
  \bibinfo {pages} {2888} (\bibinfo {year} {2015})}\BibitemShut {NoStop}%
\bibitem [{\citenamefont {Kacprzak}\ \emph {et~al.}(2016)\citenamefont
  {Kacprzak}, \citenamefont {Kirk}, \citenamefont {Friedrich}, \citenamefont
  {Amara}, \citenamefont {Refregier}, \citenamefont {Marian}, \citenamefont
  {Dietrich}, \citenamefont {Suchyta}, \citenamefont {Aleksi{\'c}},
  \citenamefont {Bacon} \emph {et~al.}}]{kacprzak2016cosmology}%
  \BibitemOpen
  \bibfield  {author} {\bibinfo {author} {\bibfnamefont {T.}~\bibnamefont
  {Kacprzak}}, \bibinfo {author} {\bibfnamefont {D.}~\bibnamefont {Kirk}},
  \bibinfo {author} {\bibfnamefont {O.}~\bibnamefont {Friedrich}}, \bibinfo
  {author} {\bibfnamefont {A.}~\bibnamefont {Amara}}, \bibinfo {author}
  {\bibfnamefont {A.}~\bibnamefont {Refregier}}, \bibinfo {author}
  {\bibfnamefont {L.}~\bibnamefont {Marian}}, \bibinfo {author} {\bibfnamefont
  {J.}~\bibnamefont {Dietrich}}, \bibinfo {author} {\bibfnamefont
  {E.}~\bibnamefont {Suchyta}}, \bibinfo {author} {\bibfnamefont
  {J.}~\bibnamefont {Aleksi{\'c}}}, \bibinfo {author} {\bibfnamefont
  {D.}~\bibnamefont {Bacon}}, \emph {et~al.},\ }\bibfield  {title} {\bibinfo
  {title} {Cosmology constraints from shear peak statistics in dark energy
  survey science verification data},\ }\href@noop {} {\bibfield  {journal}
  {\bibinfo  {journal} {Monthly Notices of the Royal Astronomical Society}\
  }\textbf {\bibinfo {volume} {463}},\ \bibinfo {pages} {3653} (\bibinfo {year}
  {2016})}\BibitemShut {NoStop}%
\bibitem [{\citenamefont {Martinet}\ \emph {et~al.}(2018)\citenamefont
  {Martinet}, \citenamefont {Schneider}, \citenamefont {Hildebrandt},
  \citenamefont {Shan}, \citenamefont {Asgari}, \citenamefont {Dietrich},
  \citenamefont {Harnois-D{\'e}raps}, \citenamefont {Erben}, \citenamefont
  {Grado}, \citenamefont {Heymans} \emph {et~al.}}]{martinet2018kids}%
  \BibitemOpen
  \bibfield  {author} {\bibinfo {author} {\bibfnamefont {N.}~\bibnamefont
  {Martinet}}, \bibinfo {author} {\bibfnamefont {P.}~\bibnamefont {Schneider}},
  \bibinfo {author} {\bibfnamefont {H.}~\bibnamefont {Hildebrandt}}, \bibinfo
  {author} {\bibfnamefont {H.}~\bibnamefont {Shan}}, \bibinfo {author}
  {\bibfnamefont {M.}~\bibnamefont {Asgari}}, \bibinfo {author} {\bibfnamefont
  {J.~P.}\ \bibnamefont {Dietrich}}, \bibinfo {author} {\bibfnamefont
  {J.}~\bibnamefont {Harnois-D{\'e}raps}}, \bibinfo {author} {\bibfnamefont
  {T.}~\bibnamefont {Erben}}, \bibinfo {author} {\bibfnamefont
  {A.}~\bibnamefont {Grado}}, \bibinfo {author} {\bibfnamefont
  {C.}~\bibnamefont {Heymans}}, \emph {et~al.},\ }\bibfield  {title} {\bibinfo
  {title} {Kids-450: cosmological constraints from weak-lensing peak
  statistics--ii: Inference from shear peaks using n-body simulations},\
  }\href@noop {} {\bibfield  {journal} {\bibinfo  {journal} {Monthly Notices of
  the Royal Astronomical Society}\ }\textbf {\bibinfo {volume} {474}},\
  \bibinfo {pages} {712} (\bibinfo {year} {2018})}\BibitemShut {NoStop}%
\bibitem [{\citenamefont {Shan}\ \emph {et~al.}(2018)\citenamefont {Shan},
  \citenamefont {Liu}, \citenamefont {Hildebrandt}, \citenamefont {Pan},
  \citenamefont {Martinet}, \citenamefont {Fan}, \citenamefont {Schneider},
  \citenamefont {Asgari}, \citenamefont {Harnois-D{\'e}raps}, \citenamefont
  {Hoekstra} \emph {et~al.}}]{shan2018kids}%
  \BibitemOpen
  \bibfield  {author} {\bibinfo {author} {\bibfnamefont {H.}~\bibnamefont
  {Shan}}, \bibinfo {author} {\bibfnamefont {X.}~\bibnamefont {Liu}}, \bibinfo
  {author} {\bibfnamefont {H.}~\bibnamefont {Hildebrandt}}, \bibinfo {author}
  {\bibfnamefont {C.}~\bibnamefont {Pan}}, \bibinfo {author} {\bibfnamefont
  {N.}~\bibnamefont {Martinet}}, \bibinfo {author} {\bibfnamefont
  {Z.}~\bibnamefont {Fan}}, \bibinfo {author} {\bibfnamefont {P.}~\bibnamefont
  {Schneider}}, \bibinfo {author} {\bibfnamefont {M.}~\bibnamefont {Asgari}},
  \bibinfo {author} {\bibfnamefont {J.}~\bibnamefont {Harnois-D{\'e}raps}},
  \bibinfo {author} {\bibfnamefont {H.}~\bibnamefont {Hoekstra}}, \emph
  {et~al.},\ }\bibfield  {title} {\bibinfo {title} {Kids-450: cosmological
  constraints from weak lensing peak statistics--i. inference from analytical
  prediction of high signal-to-noise ratio convergence peaks},\ }\href@noop {}
  {\bibfield  {journal} {\bibinfo  {journal} {Monthly Notices of the Royal
  Astronomical Society}\ }\textbf {\bibinfo {volume} {474}},\ \bibinfo {pages}
  {1116} (\bibinfo {year} {2018})}\BibitemShut {NoStop}%
\bibitem [{\citenamefont {Laureijs}\ \emph {et~al.}(2011)\citenamefont
  {Laureijs}, \citenamefont {Amiaux}, \citenamefont {Arduini}, \citenamefont
  {Augueres}, \citenamefont {Brinchmann}, \citenamefont {Cole}, \citenamefont
  {Cropper}, \citenamefont {Dabin}, \citenamefont {Duvet}, \citenamefont
  {Ealet} \emph {et~al.}}]{laureijs2011euclid}%
  \BibitemOpen
  \bibfield  {author} {\bibinfo {author} {\bibfnamefont {R.}~\bibnamefont
  {Laureijs}}, \bibinfo {author} {\bibfnamefont {J.}~\bibnamefont {Amiaux}},
  \bibinfo {author} {\bibfnamefont {S.}~\bibnamefont {Arduini}}, \bibinfo
  {author} {\bibfnamefont {J.-L.}\ \bibnamefont {Augueres}}, \bibinfo {author}
  {\bibfnamefont {J.}~\bibnamefont {Brinchmann}}, \bibinfo {author}
  {\bibfnamefont {R.}~\bibnamefont {Cole}}, \bibinfo {author} {\bibfnamefont
  {M.}~\bibnamefont {Cropper}}, \bibinfo {author} {\bibfnamefont
  {C.}~\bibnamefont {Dabin}}, \bibinfo {author} {\bibfnamefont
  {L.}~\bibnamefont {Duvet}}, \bibinfo {author} {\bibfnamefont
  {A.}~\bibnamefont {Ealet}}, \emph {et~al.},\ }\bibfield  {title} {\bibinfo
  {title} {Euclid definition study report},\ }\href@noop {} {\bibfield
  {journal} {\bibinfo  {journal} {arXiv preprint arXiv:1110.3193}\ } (\bibinfo
  {year} {2011})}\BibitemShut {NoStop}%
\bibitem [{\citenamefont {Derrida}\ \emph {et~al.}(2002)\citenamefont
  {Derrida}, \citenamefont {Lebowitz},\ and\ \citenamefont
  {Speer}}]{derrida2002large}%
  \BibitemOpen
  \bibfield  {author} {\bibinfo {author} {\bibfnamefont {B.}~\bibnamefont
  {Derrida}}, \bibinfo {author} {\bibfnamefont {J.}~\bibnamefont {Lebowitz}},\
  and\ \bibinfo {author} {\bibfnamefont {E.}~\bibnamefont {Speer}},\ }\bibfield
   {title} {\bibinfo {title} {Large deviation of the density profile in the
  steady state of the open symmetric simple exclusion process},\ }\href@noop {}
  {\bibfield  {journal} {\bibinfo  {journal} {Journal of statistical physics}\
  }\textbf {\bibinfo {volume} {107}},\ \bibinfo {pages} {599} (\bibinfo {year}
  {2002})}\BibitemShut {NoStop}%
\bibitem [{\citenamefont {Bertini}\ \emph {et~al.}(2015)\citenamefont
  {Bertini}, \citenamefont {De~Sole}, \citenamefont {Gabrielli}, \citenamefont
  {Jona-Lasinio},\ and\ \citenamefont {Landim}}]{bertini2015macroscopic}%
  \BibitemOpen
  \bibfield  {author} {\bibinfo {author} {\bibfnamefont {L.}~\bibnamefont
  {Bertini}}, \bibinfo {author} {\bibfnamefont {A.}~\bibnamefont {De~Sole}},
  \bibinfo {author} {\bibfnamefont {D.}~\bibnamefont {Gabrielli}}, \bibinfo
  {author} {\bibfnamefont {G.}~\bibnamefont {Jona-Lasinio}},\ and\ \bibinfo
  {author} {\bibfnamefont {C.}~\bibnamefont {Landim}},\ }\bibfield  {title}
  {\bibinfo {title} {Macroscopic fluctuation theory},\ }\href@noop {}
  {\bibfield  {journal} {\bibinfo  {journal} {Reviews of Modern Physics}\
  }\textbf {\bibinfo {volume} {87}},\ \bibinfo {pages} {593} (\bibinfo {year}
  {2015})}\BibitemShut {NoStop}%
\bibitem [{\citenamefont {Lin}\ \emph {et~al.}(2017)\citenamefont {Lin},
  \citenamefont {Tegmark},\ and\ \citenamefont {Rolnick}}]{lin2017does}%
  \BibitemOpen
  \bibfield  {author} {\bibinfo {author} {\bibfnamefont {H.~W.}\ \bibnamefont
  {Lin}}, \bibinfo {author} {\bibfnamefont {M.}~\bibnamefont {Tegmark}},\ and\
  \bibinfo {author} {\bibfnamefont {D.}~\bibnamefont {Rolnick}},\ }\bibfield
  {title} {\bibinfo {title} {Why does deep and cheap learning work so well?},\
  }\href@noop {} {\bibfield  {journal} {\bibinfo  {journal} {Journal of
  Statistical Physics}\ }\textbf {\bibinfo {volume} {168}},\ \bibinfo {pages}
  {1223} (\bibinfo {year} {2017})}\BibitemShut {NoStop}%
\bibitem [{\citenamefont {Iso}\ \emph {et~al.}(2018)\citenamefont {Iso},
  \citenamefont {Shiba},\ and\ \citenamefont {Yokoo}}]{iso2018scale}%
  \BibitemOpen
  \bibfield  {author} {\bibinfo {author} {\bibfnamefont {S.}~\bibnamefont
  {Iso}}, \bibinfo {author} {\bibfnamefont {S.}~\bibnamefont {Shiba}},\ and\
  \bibinfo {author} {\bibfnamefont {S.}~\bibnamefont {Yokoo}},\ }\bibfield
  {title} {\bibinfo {title} {Scale-invariant feature extraction of neural
  network and renormalization group flow},\ }\href@noop {} {\bibfield
  {journal} {\bibinfo  {journal} {Physical review E}\ }\textbf {\bibinfo
  {volume} {97}},\ \bibinfo {pages} {053304} (\bibinfo {year}
  {2018})}\BibitemShut {NoStop}%
\bibitem [{\citenamefont {Roberts}\ \emph {et~al.}(2022)\citenamefont
  {Roberts}, \citenamefont {Yaida},\ and\ \citenamefont
  {Hanin}}]{roberts2022principles}%
  \BibitemOpen
  \bibfield  {author} {\bibinfo {author} {\bibfnamefont {D.~A.}\ \bibnamefont
  {Roberts}}, \bibinfo {author} {\bibfnamefont {S.}~\bibnamefont {Yaida}},\
  and\ \bibinfo {author} {\bibfnamefont {B.}~\bibnamefont {Hanin}},\
  }\href@noop {} {\emph {\bibinfo {title} {The Principles of Deep Learning
  Theory: An Effective Theory Approach to Understanding Neural Networks}}}\
  (\bibinfo  {publisher} {Cambridge University Press},\ \bibinfo {year}
  {2022})\BibitemShut {NoStop}%
\bibitem [{\citenamefont {Lee}\ \emph {et~al.}(2019)\citenamefont {Lee},
  \citenamefont {Gommers}, \citenamefont {Waselewski}, \citenamefont
  {Wohlfahrt},\ and\ \citenamefont {O'Leary}}]{lee2019pywavelets}%
  \BibitemOpen
  \bibfield  {author} {\bibinfo {author} {\bibfnamefont {G.}~\bibnamefont
  {Lee}}, \bibinfo {author} {\bibfnamefont {R.}~\bibnamefont {Gommers}},
  \bibinfo {author} {\bibfnamefont {F.}~\bibnamefont {Waselewski}}, \bibinfo
  {author} {\bibfnamefont {K.}~\bibnamefont {Wohlfahrt}},\ and\ \bibinfo
  {author} {\bibfnamefont {A.}~\bibnamefont {O'Leary}},\ }\bibfield  {title}
  {\bibinfo {title} {Pywavelets: A python package for wavelet analysis},\
  }\href@noop {} {\bibfield  {journal} {\bibinfo  {journal} {Journal of Open
  Source Software}\ }\textbf {\bibinfo {volume} {4}},\ \bibinfo {pages} {1237}
  (\bibinfo {year} {2019})}\BibitemShut {NoStop}%
\bibitem [{\citenamefont {Milchev}\ \emph {et~al.}(1986)\citenamefont
  {Milchev}, \citenamefont {Heermann},\ and\ \citenamefont
  {Binder}}]{milchev1986finite}%
  \BibitemOpen
  \bibfield  {author} {\bibinfo {author} {\bibfnamefont {A.}~\bibnamefont
  {Milchev}}, \bibinfo {author} {\bibfnamefont {D.}~\bibnamefont {Heermann}},\
  and\ \bibinfo {author} {\bibfnamefont {K.}~\bibnamefont {Binder}},\
  }\bibfield  {title} {\bibinfo {title} {Finite-size scaling analysis of the
  $\phi$ 4 field theory on the square lattice},\ }\href@noop {} {\bibfield
  {journal} {\bibinfo  {journal} {Journal of statistical physics}\ }\textbf
  {\bibinfo {volume} {44}},\ \bibinfo {pages} {749} (\bibinfo {year}
  {1986})}\BibitemShut {NoStop}%
\bibitem [{\citenamefont {Hasenbusch}(1999)}]{hasenbusch1999monte}%
  \BibitemOpen
  \bibfield  {author} {\bibinfo {author} {\bibfnamefont {M.}~\bibnamefont
  {Hasenbusch}},\ }\bibfield  {title} {\bibinfo {title} {A monte carlo study of
  leading order scaling corrections of 4 theory on a three-dimensional
  lattice},\ }\href@noop {} {\bibfield  {journal} {\bibinfo  {journal} {Journal
  of Physics A: Mathematical and General}\ }\textbf {\bibinfo {volume} {32}},\
  \bibinfo {pages} {4851} (\bibinfo {year} {1999})}\BibitemShut {NoStop}%
\bibitem [{\citenamefont {Tr{\"o}ster}\ \emph {et~al.}(2005)\citenamefont
  {Tr{\"o}ster}, \citenamefont {Dellago},\ and\ \citenamefont
  {Schranz}}]{troster2005free}%
  \BibitemOpen
  \bibfield  {author} {\bibinfo {author} {\bibfnamefont {A.}~\bibnamefont
  {Tr{\"o}ster}}, \bibinfo {author} {\bibfnamefont {C.}~\bibnamefont
  {Dellago}},\ and\ \bibinfo {author} {\bibfnamefont {W.}~\bibnamefont
  {Schranz}},\ }\bibfield  {title} {\bibinfo {title} {Free energies of the
  $\phi$ 4 model from wang-landau simulations},\ }\href@noop {} {\bibfield
  {journal} {\bibinfo  {journal} {Physical Review B}\ }\textbf {\bibinfo
  {volume} {72}},\ \bibinfo {pages} {094103} (\bibinfo {year}
  {2005})}\BibitemShut {NoStop}%
\bibitem [{\citenamefont {Zhong}\ \emph {et~al.}(2018)\citenamefont {Zhong},
  \citenamefont {Barkema}, \citenamefont {Panja},\ and\ \citenamefont
  {Ball}}]{zhong2018critical}%
  \BibitemOpen
  \bibfield  {author} {\bibinfo {author} {\bibfnamefont {W.}~\bibnamefont
  {Zhong}}, \bibinfo {author} {\bibfnamefont {G.~T.}\ \bibnamefont {Barkema}},
  \bibinfo {author} {\bibfnamefont {D.}~\bibnamefont {Panja}},\ and\ \bibinfo
  {author} {\bibfnamefont {R.~C.}\ \bibnamefont {Ball}},\ }\bibfield  {title}
  {\bibinfo {title} {Critical dynamical exponent of the two-dimensional scalar
  $\phi$ 4 model with local moves},\ }\href@noop {} {\bibfield  {journal}
  {\bibinfo  {journal} {Physical Review E}\ }\textbf {\bibinfo {volume} {98}},\
  \bibinfo {pages} {062128} (\bibinfo {year} {2018})}\BibitemShut {NoStop}%
\bibitem [{\citenamefont {Binder}(1992)}]{binder1992finite}%
  \BibitemOpen
  \bibfield  {author} {\bibinfo {author} {\bibfnamefont {K.}~\bibnamefont
  {Binder}},\ }\bibfield  {title} {\bibinfo {title} {Finite size effects at
  phase transitions},\ }in\ \href@noop {} {\emph {\bibinfo {booktitle}
  {Computational methods in field theory}}}\ (\bibinfo  {publisher}
  {Springer},\ \bibinfo {year} {1992})\ pp.\ \bibinfo {pages}
  {59--125}\BibitemShut {NoStop}%
\end{thebibliography}%

\end{document}